\documentclass[12pt,a4paper]{article}
\pdfoutput=1

\usepackage{epsfig}
\usepackage{amsmath}
\usepackage{amssymb}
\usepackage{verbatim}
\usepackage{bm}
\usepackage{dsfont}
\usepackage[footnotesize]{caption}
\usepackage{subfigure}
\usepackage{cancel}
\usepackage{graphics}

\usepackage{color}

\definecolor{darkblue}{rgb}{0,0, 0.5}

\usepackage[pdfpagemode={UseOutlines},bookmarks=true,bookmarksopen=true,
   bookmarksopenlevel=0,bookmarksnumbered=true,hypertexnames=false,
  linkcolor={darkblue},citecolor={darkblue},urlcolor={darkblue},
   pdfstartview={FitV},unicode,breaklinks=true]{hyperref} 
\hypersetup{colorlinks=true}

\usepackage{hyperref}
\usepackage[numbers, sort&compress]{natbib}
\usepackage{hypernat}

\newcommand{\BmL}{$B$$-$$L$\ }

\begin{document}

\begin{flushleft}
DESY 13-050\\
IPMU 13-0091\\
May 2013
\end{flushleft}

\vskip 1cm

\begin{center}
{\LARGE\bf The Gravitational Wave Spectrum from\\[2mm]
Cosmological \boldsymbol{$B$$-$$L$} Breaking}

\vskip 2cm

{\large W.~Buchm\"uller$^1$, V.~Domcke$^1$, K.~Kamada$^1$, K.~Schmitz$^2$}\\[3mm]
$^1${\it Deutsches Elektronen-Synchrotron DESY, 22607 Hamburg, Germany}\\
$^2${\it Kavli IPMU (WPI), University of Tokyo, Kashiwa 277-8583, Japan}
\end{center}

\vskip 1cm

\begin{abstract}
\noindent Cosmological $B$$-$$L$ breaking is a natural and testable mechanism to generate the initial conditions of the hot early universe. If $B$$-$$L$ is broken at the grand unification scale, the false vacuum phase drives hybrid inflation, ending in tachyonic preheating. The decays of heavy $B$$-$$L$ Higgs bosons and heavy neutrinos generate entropy, baryon asymmetry and dark matter and also control the reheating temperature. The different phases in the transition from inflation to the radiation dominated phase produce a characteristic spectrum of gravitational waves. We calculate the complete gravitational wave spectrum due to inflation, preheating and cosmic strings, which turns out to have several features. The production of gravitational waves from cosmic strings has large  uncertainties, with lower and upper bounds provided by Abelian Higgs strings and Nambu-Goto strings, implying $\Omega_\text{GW} h^2 \sim 10^{-13} - 10^{-8}$, much larger than the spectral amplitude predicted by inflation. Forthcoming gravitational wave detectors such as eLISA, advanced LIGO, ET, BBO and DECIGO will reach the sensitivity needed to test the predictions from cosmological  $B$$-$$L$ breaking.

\end{abstract}

\thispagestyle{empty}

\newpage

\section{Introduction \label{sec:introduction}}

Relic gravitational waves (GWs) are a fascinating window to the 
very early universe \cite{Starobinsky:1979ty}. They are generated by quantum fluctuations 
during inflation \cite{Rubakov:1982df}
as well as in the form of classical radiation from cosmic strings
\cite{Vilenkin:1981bx}. Another important source is preheating after inflation
via resonant decay of an oscillating inflaton field \cite{Khlebnikov:1997di}
or violent collisions of bubble-like structures \cite{GarciaBellido:2007dg}
in tachyonic preheating \cite{Felder:2000hj}. 

We have recently proposed a detailed picture of pre- and reheating where
the initial conditions of the hot early universe are generated by the spontaneous
breaking of $B$$-$$L$, the difference of baryon and lepton number
\cite{Buchmuller:2010yy, Buchmuller:2011mw, Buchmuller:2012wn}. The false vacuum phase of unbroken $B$$-$$L$
symmetry yields hybrid inflation with an energy density set by the scale of
grand unification \cite{Copeland:1994vg,Dvali:1994ms}. In the $B$$-$$L$
breaking phase transition ending inflation most of the vacuum energy density
is rapidly transferred to non-relativistic $B$$-$$L$ Higgs bosons, a sizable
fraction also into cosmic strings. The decays of heavy Higgs bosons and heavy 
Majorana neutrinos generate entropy
and baryon asymmetry via thermal and nonthermal leptogenesis \cite{Fukugita:1986hr, Lazarides:1991wu}. 
The temperature evolution during reheating is controlled by the interplay between the $B$$-$$L$ Higgs and the neutrino sector.
The origin of dark matter are thermally produced gravitinos \cite{Bolz:1998ek}.

In this paper we compute the GW spectrum predicted by
cosmological $B$$-$$L$ breaking. It receives contributions from all the
possible sources mentioned above: inflation, cosmic strings and preheating. 
Much work has already been done on the stochastic gravitational background
from inflation (see, e.g.\ Refs.~\cite{Turner:1993vb,Seto:2003kc,Smith:2005mm}). We 
are particularly interested in features of the GW spectrum caused by the 
change of the equation of state in the cosmological evolution. This has previously been
studied in Ref.~\cite{Nakayama:2008wy} with the goal of determining the
reheating temperature of the early universe. Our results are consistent
with those of Ref.~\cite{Nakayama:2008wy}. The main difference is that we can
resort to a time-resolved description of the entire reheating process, studied in Ref.~\cite{Buchmuller:2012wn}. This allows us to gain a better understanding of the connection between features in the GW spectrum and the evolution of the temperature of the thermal bath, pinpointing to which model parameters certain features in the spectrum are related.

A very interesting but also rather uncertain source of GWs are cosmic strings
\cite{Hindmarsh:2011qj}. In the $B$$-$$L$ breaking phase transition local
cosmic strings are formed. The initial state of such a network
can be simulated numerically, and recently the amplitude of the scale-invariant spectrum of GWs produced during the radiation dominated epoch
has been determined \cite{Figueroa:2012kw}. Based on these results we obtain
the GW spectrum for our model, thereby extending the analysis to GWs produced
during reheating and during matter domination. For Abelian Higgs (AH) strings it is usually assumed that
strings lose their energy mostly via radiation of massive particles. In this case we find a GW spectrum which has a very similar shape to that generated by inflation, but is amplified by several orders of magnitude. This result opens up the possibility to measure features in the GW spectrum related to the temperature evolution during reheating. Alternatively, one also considers the possibility that, beyond a certain
 length, strings can be described as Nambu-Goto (NG) strings, which lose their
energy by radiating GWs, see e.g.\ Refs.~\cite{Damour:2001bk,Siemens:2006yp,
Kuroyanagi:2012wm}. We shall also study the implications of
NG strings for the GW spectrum and compare the results with those obtained for
AH~strings.

Tachyonic preheating leads to GWs with a spectrum peaked at very high
frequencies. For certain parameter regimes of hybrid inflation the spectrum
has been determined numerically \cite{Dufaux:2007pt, Dufaux:2008dn}. We shall base our discussion on
analytical estimates for the peak frequencies, which we apply to our model.

Measuring the GW spectrum thus provides a unique possibility to test different aspects of a phase transition in the early universe. Forthcoming space- and ground-based interferometers such as advanced LIGO~\cite{advligopaper}, ET~\cite{etreport}, BBO~\cite{Crowder:2005nr}, DECIGO~\cite{Kawamura:2011zz} and eLISA~\cite{AmaroSeoane:2012km} will reach the sensitivity necessary to probe this scenario. At the same time, millisecond pulsar timing experiments already now put stringent bounds on NG cosmic strings~\cite{Manchester:2012za} and future experiments such as SKA~\cite{Kramer:2004rwa} will further increase this sensitivity. It will however remain a challenge to disentangle the GW spectrum from a phase transition in the early universe from other sources of GWs, due to both astrophysical processes and subsequent cosmological phase transitions, see e.g.\ \cite{Maggiore:1900zz, Binetruy:2012ze}.

The paper is organized as follows. In Sec.~\ref{sec:GWs} we recall some
basic formulas for the production of GWs and the transfer function which
are needed in the subsequent chapters. The main ingredients of our model
for pre- and reheating are described in Sec.~\ref{sec:model}. 
Secs.~\ref{sec:inflation} and \ref{sec:preheating} deal with the production
of GWs during inflation and preheating, and in Secs.~\ref{sec:cosmicstrings}
and \ref{sec:comparison} GWs from cosmic strings are discussed, for the case
of AH strings and NG strings, respectively. Sec.~\ref{sec:reheatingtemp} focuses on probing the reheating temperature by measuring a feature in the GW spectrum. Constraints from the cosmic
microwave background and observational prospects are the topic of Sec.~\ref{sec:observations},
and we conclude in Sec.~\ref{sec:conclusion}. Three appendices deal with the scale factor and temperature evolution during reheating as well as the analytical calculation of the GW background from
NG strings.

\section{Cosmic Gravitational Wave Background}\label{sec:GWs}

In this section we recall some basic formulas which we shall need in our 
calculation of the various contributions to the GW background.
GWs are tensor perturbations of the 
homogeneous background metric. In a flat Friedmann Robertson Walker (FRW) 
background, these perturbations can be parametrized as~\cite{Maggiore:1900zz}
\begin{equation}
 ds^2 = a^2(\tau) (\eta_{\mu \nu} + h_{\mu \nu}) dx^\mu dx^\nu \,.
\label{eq_metric}
\end{equation}
Here $\eta_{\mu \nu} = \text{diag}(-1, 1, 1, 1)$, $a$ is the scale factor and $x^\mu$ are 
conformal coordinates, with $x^i$ denoting the comoving spatial coordinates 
and $\tau = x^0$ the conformal time. These are related to the physical 
coordinates and the cosmic time as $\boldsymbol x_{\text{phys}} = a(\tau)\, \boldsymbol x$ 
and $dt = a(\tau)\, d\tau$, respectively\footnote{Here and in the following, 
Greek letters denote Lorentz indices, $\mu, \nu = 0,1,2,3$, whereas Latin letters 
refer to the spatial indices, $i, j = 1,2,3$, with bold letters indicating 
3-vectors. \smallskip}.

Introducing
\begin{equation}
 \bar h_{\mu \nu} = h_{\mu \nu} - \frac{1}{2} \eta_{\mu \nu} h^{\rho}_{\rho}\,,
\end{equation}
the linearized Einstein equations describing the generation and propagation 
of GWs read
\begin{equation}
 \bar h''_{\mu \nu}(\boldsymbol{x}, \tau) + 2 \frac{a'}{a} 
\bar h'_{\mu \nu}(\boldsymbol{x}, \tau) - \nabla_{\boldsymbol{x}}^2 
\bar h_{\mu \nu}(\boldsymbol{x}, \tau) = 16 \pi G T_{\mu \nu}(\boldsymbol{x},\tau) \,,
\label{eq_einstein_x}
\end{equation}
with a prime denoting the derivative with respect to conformal time; 
$G$ is Newton's constant and $T_{\mu \nu}$ is the anisotropic part of the stress 
energy tensor of the source. The total stress energy tensor is the sum of
$T_{\mu \nu}$ and an isotropic part which determines the background metric.
Outside the source, 
we can choose the transverse traceless (TT) gauge for the GW, 
i.e.\ $h^{0 \mu} = 0$, $h^i_i = 0$,  $\partial^j h_{ij} = 0$,
which implies $\bar h_{\mu \nu} = h_{\mu \nu}$. The mode equation which describes the generation and propagation of these degrees of freedom (DOFs) can be obtained by using an appropriate projection operator 
\cite{Maggiore:1900zz} on the Fourier transform\footnote{Our convention for the  Fourier 
transformation is $h_{ij}(\boldsymbol x, \tau) = \int \frac{d^3 k}{(2\pi)^3}\  
h_{ij}(\boldsymbol k, \tau) \exp(i \boldsymbol{kx})$.} of Eq.~\eqref{eq_einstein_x},
\begin{equation}
 \tilde{h}^{''}_{ij}(\boldsymbol{k}, \tau) + 
\left(k^2 - \frac{a^{''}}{a}\right) \tilde{h}_{ij}(\boldsymbol{k}, \tau) 
= 16 \pi G a \Pi_{ij}(\boldsymbol k, \tau) \,,
\label{eq_einstein_k}
\end{equation}
where $\tilde{h}_{ij} = a h_{ij}$, 
$\Pi_{ij}$ denotes the Fourier transform of the TT part
of the anisotropic stress tensor $T_{\mu  \nu}$, $k = |\boldsymbol k|$, and 
$\boldsymbol k$ is the comoving wavenumber, related to the physical wave number through 
$\boldsymbol k_\text{phys} = \boldsymbol k / a$.

A useful plane wave expansion of GWs is given by
\begin{align}
h_{ij}\left(\boldsymbol{x},\tau\right) = \sum_{A = +,\times}
\int_{-\infty}^{+\infty} \frac{dk}{2\pi} \int d^2\hat{\boldsymbol{k}} \, 
h_A\left(\boldsymbol{k}\right)
e_{ij}^A\big(\hat{k}\big) \, T_k(\tau)
e^{-ik\left(\tau - \hat{\boldsymbol{k}}\boldsymbol{x}\right)}\,.
\label{eq:hFourier}
\end{align}
Here, $\hat{\boldsymbol{k}} = \boldsymbol{k} / k$,
$A = +, \times$ labels the two possible polarization states of
a GW in the TT gauge and
$e_{ij}^{+,\times}$ denote the two corresponding polarization tensors
satisfying the normalization condition $e_{ij}^A e^{ij\,B} = 2 \delta^{AB}$. 
$h_A(\boldsymbol{k})$ are the coefficients of the expansion and the red-shift due to the expansion of the universe is captured in the so-called `transfer function' $T_k(\tau)$.

An analytical expression for $T_k$ can be obtained by studying the homogeneous, i.e. source-free, version of Eq.~\eqref{eq_einstein_k}.
Using the Friedmann equations, we find $a''/a \sim a^2 H^2$. The mode equation describes two distinct regimes. On sub-horizon scales, $k \gg a H$, we can neglect the $a''/a$ term. The solution is thus simply $\tilde h_{ij} \sim \cos(\omega \tau)$ and hence $h_{ij} \sim \cos(\omega \tau)/a$, i.e.\ the modes decay as $1/a$ inside the horizon. On the other hand, on super-horizon scales, $k \ll aH$, we can neglect the $k^2$ term. This yields $2 a' h_{ij}' + a h_{ij}'' = 0$, with the solution
\begin{equation}
 h_{ij}(\tau) = A + B\int^\tau \frac{ d\tau'}{a^2(\tau')} \,,
\end{equation}
where $A$ and $B$ are constants of integration. This solution is a constant plus a decaying mode which can be neglected. Hence on super-horizon scales the amplitude of the mode remains constant, the mode is `frozen'.

With this, we identify the transfer function $T_k$ capturing the effects due to the expansion of the universe as 
\begin{equation}
 T_k(\tau_*, \tau) =  \frac{h^E_{ij}(\boldsymbol{k}, \tau)}{h^E_{ij}(\boldsymbol{k}, \tau_*)} \,,
\label{eq:transferfunction}
\end{equation}
with $h^E_{ij}(\boldsymbol{k}, \tau)$ denoting the envelope of the oscillating function $h_{ij}(\boldsymbol{k}, \tau)$.
For modes present on super-horizon scales, i.e.\ GWs produced by inflation, the reference time $\tau_*$ can be equally replaced by any $\tau < \tau_{k}$, where $\tau_{k}$ denotes the time when a given mode with wavenumber $k$ enters the horizon, $k = a(\tau_{k}) H(\tau_{k})$. To good approximation, the transfer function can then be estimated as (see e.g.~\cite{Smith:2005mm}) 
\begin{equation}
 T_k(\tau_*, \tau) \approx \frac{a(\tau_*)}{a(\tau)} \quad \text{with  } \tau_* = \begin{cases} \tau_i \text{  for sub-horizon sources} \\ \tau_{k} \text{  for super-horizon sources} \end{cases} \,,
\label{eq:approx_transferfunction}
\end{equation}
with $\tau_i$ marking the time when the GW is generated.
Here in the latter case, we assume the amplitude to be constant until $\tau = \tau_{k}$ and then to drop as $1/a$ immediately afterwards. The actual solution to the mode equation yields corrections to both of these assumptions, however as a numerical check reveals the effects compensate each other so that Eq.~\eqref{eq:approx_transferfunction} reproduces the full result very well. We will quantify this statement at the end of Sec.~\ref{sec:inflation} after discussing the transfer function in more detail. For super-horizon sources we will in the following use the more compact notation $T_k(\tau) = T_k(\tau_k, \tau)$.

The GW background is a superposition of GWs propagating with all frequencies
in all directions. An important observable characterizing the GW background
is the ensemble average of the energy density~\cite{Maggiore:1900zz}, which
is expected to be isotropic,
\begin{equation}
 \rho_\text{GW}(\tau)  = \frac{1}{32 \pi G} 
\left \langle \dot h_{ij}\left(\boldsymbol{x},\tau\right) 
\dot h^{ij}\left(\boldsymbol{x},\tau\right) \right \rangle 
 = \int_{-\infty}^\infty d \ln k \, \frac{\partial \rho_\text{GW}(k, \tau)}
{\partial \ln k} \,,
\label{eq:rhoGWtot}
\end{equation}
with the angular brackets denoting the ensemble average and the dot referring 
to the derivative with respect to cosmic time. Alternatively, one also uses
the ratio of the differential energy density to the critical density $\rho_c  = 3 H^2/(8\pi G)$,
\begin{equation}
\label{eq:OmegaGW}
\Omega_\text{GW}(k, \tau) = \frac{1}{\rho_c} \frac{\partial \rho_\text{GW}(k, \tau)}{\partial \ln k} \,,
\end{equation}
where $H$ denotes the Hubble parameter. In the model considered in this paper, the energy density receives contributions of quantum as well as of classical origin,
\begin{equation}
\rho_\text{GW}(\tau) = \rho_\text{GW}^{\rm qu}(\tau) + \rho_\text{GW}^{\rm cl}(\tau)\ .
\end{equation}
The quantum part is due to inflation and therefore stochastic, whereas the
classical part is determined by the contributions to the stress energy tensor
from cosmic strings and from tachyonic preheating,
\begin{equation}
\rho^{\rm cl}_\text{GW}(\tau) = \rho_\text{GW}^{\rm CS} + \rho_\text{GW}^{\rm TP}(\tau)\ .
\end{equation}

For a stochastic GW background the Fourier modes $h_A\left(\boldsymbol{k}\right)$ in Eq.~\eqref{eq:hFourier}
are random variables and their ensemble average is determined by
a time-independent spectral density $S_h(k)$~\cite{Maggiore:1900zz},
\begin{align}
\left<h_A\left(\boldsymbol{k}\right)h_B^*\left(\boldsymbol{k}'\right)\right>
= 2\pi \delta\left(k-k'\right) \frac{1}{4\pi}
\delta^{(2)}\big(\hat{\boldsymbol{k}}-\hat{\boldsymbol{k}}'\big)
\delta_{AB} \frac{1}{2} S_h(k) \,.
\label{eq:Shdef}
\end{align}
This relation reflects the fact that different modes are uncorrelated and
that the background is isotropic.
On sub-horizon scales, $k \gg aH$,  Eqs.~(\ref{eq:hFourier}), \eqref{eq:approx_transferfunction} and
(\ref{eq:Shdef}) yield
\begin{equation}
 \left \langle  h_{ij}\left(\boldsymbol{x},\tau\right) 
 h^{ij}\left(\boldsymbol{x},\tau\right) \right \rangle 
= \frac{1}{\pi}\int_{-\infty}^\infty dk \; S_h(k) \, \frac{a^2(\tau_*)}{a^2(\tau)}\,,
\label{eq:2pointfct_Sh}
\end{equation}
and
\begin{equation}
 \left \langle \dot h_{ij}\left(\boldsymbol{x},\tau\right) 
\dot h^{ij}\left(\boldsymbol{x},\tau\right) \right \rangle 
= \frac{1}{\pi a^2(\tau)}\int_{-\infty}^\infty dk \; k^2 \; S_h(k) \, \frac{a^2(\tau_*)}{a^2(\tau)}\,.
\end{equation}
Comparing this with Eq.~\eqref{eq:rhoGWtot} yields the differential energy density
\begin{align}
\frac{\partial\rho_{\textrm{GW}}\left(k,\tau\right)}{\partial \ln k} =
\frac{a^2(\tau_*)}{16\pi^2 G a^4(\tau)} k^3 S_h(k) \,.
\label{eq:rhoGW}
\end{align}

The classical contribution to the GW energy density is obtained by integrating
Eq.~(\ref{eq_einstein_k}) from the initial time $\tau_i$ of GW production
until today,
\begin{equation}
 h_{ij}(\boldsymbol{k}, \tau) = 
16 \pi G  \frac{1}{a(\tau)} \int_{\tau_i}^{\tau} d\tau' a(\tau')
\mathcal{G}(k,\tau,\tau')\Pi_{ij}(\boldsymbol k, \tau') \,,
\label{eq_hcl_k}
\end{equation}
where $\mathcal{G}(k,\tau,\tau')$ is the retarded Green's function of the
differential operator on the left-hand side of Eq.~(\ref{eq_einstein_k}). 
For sub-horizon modes, i.e. $k\tau \gg 1$, one has 
$\mathcal{G}(k,\tau,\tau') = \sin (k(\tau - \tau')) /k$. It is now
straightforward to evaluate the ensemble average $\langle {\dot h}^2 \rangle$.
Assuming translational invariance and isotropy of the source,
\begin{equation}
\left<\Pi_{ij}(\boldsymbol k, \tau) \Pi^{ij}(\boldsymbol k', \tau')\right> =
(2\pi)^3 \Pi^2(k, \tau, \tau') \delta(\boldsymbol k + \boldsymbol k')\ ,
\label{eq_Pi_isotropy}
\end{equation}
the resulting differential energy density simplifies to
\begin{equation}\label{eq:GWsource}
\frac{\partial\rho_{\textrm{GW}}\left(k,\tau\right)}{\partial \ln k} =
\frac{2G}{\pi}\frac{ k^3}{a^4(\tau)} \int_{\tau_i}^{\tau} d\tau_1
\int_{\tau_i}^{\tau} d\tau_2 
a(\tau_1)a(\tau_2) \cos (k(\tau_1 - \tau_2)) \Pi^2(k,\tau_1,\tau_2)\ .
\end{equation}
Here, in order to perform the ensemble average, we have also averaged the integrand 
over a period $\Delta\tau = 2\pi/k$, assuming ergodicity.

\section{Cosmological \texorpdfstring{\boldsymbol{$B$$-$$L$}}{bl} Breaking}\label{sec:model}

The main goal of this paper is to derive the full spectrum of
GWs whose origin is either directly or
indirectly related to the \BmL phase transition.
In the next chapters, we are going to discuss in turn
all of the relevant sources for GWs.
For now, let us first review how
spontaneous \BmL breaking at the end of hybrid inflation can be
embedded into supersymmetric theories.

%%%%%%%%%%%%%%%%%%%%%%%%%%%%%%%%%%%%%%%%%%%%%%%%%%%%%%%%%%%%%%%%%%%%%%%%%%%%%%

Cosmological \BmL breaking is implemented by the superpotential
\begin{align}\label{BmLsuper}
W_{B-L} = \frac{\sqrt{\lambda}}{2} \Phi \left(v_{B-L}^2 - 2 S_1 S_2\right) \,,
\end{align}
where the chiral superfields $\Phi$, $S_1$ and $S_2$ represent standard
model gauge singlets carrying \BmL charges $0$, $-2$ and $+2$, respectively.
The radial component $\varphi$ of the complex scalar
$\phi = \varphi/\sqrt{2} e^{i\theta} \subset \Phi$ is identified
as the inflaton.
Similarly, the Higgs multiplet $S$ breaking \BmL at the scale
$v_{B-L}$ is contained in the fields $S_{1,2} = S/\sqrt{2} e^{\pm i \Lambda}$.
The actual scalar \BmL breaking Higgs boson $\sigma$ corresponds
in particular to the real part of the complex scalar $s \subset S$.
The parameter $\lambda$ is a dimensionless coupling constant.

%%%%%%%%%%%%%%%%%%%%%%%%%%%%%%%%%%%%%%%%%%%%%%%%%%%%%%%%%%%%%%%%%%%%%%%%%%%%%%

Assuming a canonical K\"ahler potential for $\phi$, the tree-level
scalar potential induced by $W_{B-L}$ is exactly flat in the direction
of the inflaton $\varphi$.
For $\varphi$ larger than some critical value,
$\varphi > \varphi_c = v_{B-L}$, the complex scalars in $S_1$ and $S_2$
are stabilized at their origin, $S_{1,2} = 0$, such that
\BmL is unbroken.
In this phase of unbroken $B$$-$$L$, the energy density of the vacuum
is non-zero, $V_0 = \frac{1}{4}\lambda v_{B-L}^4$, corresponding
to an explicit breaking of supersymmetry and entailing a stage of
hybrid inflation.
The supersymmetric vacuum is stabilized by radiative corrections
at the one-loop level, forcing $\varphi$ to slowly roll down to
$\varphi = 0$.
The corresponding scalar potential for the inflaton field reads
\begin{equation}
V(\varphi) = \frac{\lambda}{4} v_{B-L}^4  + V_{1l}(\varphi/\varphi_c) \ ,
\end{equation}
where for $\varphi \gg\varphi_c$ the one-loop correction is given by
\begin{equation}
V_{1l}(\varphi) \simeq 
\frac{\lambda}{32\pi^2} v_{B-L}^4 \ln(\varphi/\varphi_c)\ .
\end{equation}
Once $\varphi$ passes below $\varphi_c$, the \BmL Higgs boson becomes
tachyonically unstable, i.e.\ it acquires a negative mass squared.
This triggers the sudden end of inflation and the spontaneous
breaking of $B$$-$$L$.
In the true groundstate, we eventually have $\varphi = 0$
and $S_{1,2} = v_{B-L} /\sqrt{2}$.

%%%%%%%%%%%%%%%%%%%%%%%%%%%%%%%%%%%%%%%%%%%%%%%%%%%%%%%%%%%%%%%%%%%%%%%%%%%%%%

The slow-roll parameters $\epsilon$ and $\eta$ of hybrid inflation
as well as the amplitude $\Delta_s^2$ of the scalar metric perturbations
can be readily expressed in terms of $\lambda$ and $v_{B-L}$,
\begin{align}
\epsilon & \approx \frac{\lambda}{16\pi^2} \left|\eta\right| \,, \qquad
\eta \approx - \frac{\lambda \, M_{\textrm{Pl}}^2}{32\pi^3 \, \varphi_*^2}
\approx - \frac{1}{2 N_e^*} \,, \label{eq:epsiloneta}\\ 
\Delta_s^2(k_*) & = \frac{H_{\textrm{inf}}^2}{8\pi^2  \epsilon  M_{\textrm{Pl}}^2}
\approx \frac{64\pi^2}{3} N_e^*
\left(\frac{v_{B-L}}{M_{\textrm{Pl}}}\right)^4 \,. \label{eq:Hinf}
\end{align}
Here, $M_\text{Pl} = (8 \pi G)^{-\frac{1}{2}} = 2.44 \times 10^{18}$~GeV is the reduced Planck mass, $k_* = 0.002\ \textrm{Mpc}^{-1}$ is the chosen pivot scale, which is probed by observations of the CMB, and ${N_e^*\simeq 50}$ denotes the number of e-folds before the end
of inflation, at field value $\varphi_*$ when the pivot scale leaves the
Hubble horizon.
In Eq.~\eqref{eq:Hinf} we have used the slow-roll relation 
$3 M_{\textrm{Pl}}^2 H_{\textrm{inf}}^2 = V_0$, where $H_\text{inf}$ denotes the Hubble parameter during inflation.
The value of $\Delta_s^2$ measured by the PLANCK satellite,
$\Delta_s^2 \simeq 2.18 \times 10^{-9}$ \cite{Ade:2013xla}, fixes $v_{B-L}$ to
a value close to the GUT scale.
More precisely, in a detailed study of hybrid inflation that also
takes into account the production of cosmic strings as well as non-canonical contributions to the K\"ahler potential,
the authors of Ref.~\cite{Nakayama:2010xf} find consistency among
all relevant observations for $v_{B-L}$ values ranging between
$3 \times 10^{15}\,\textrm{GeV}$ and $7 \times 10^{15}\,\textrm{GeV}$
and couplings in the range $10^{-4}\lesssim \sqrt{\lambda}\lesssim 10^{-1}$. Taking into account the recent PLANCK data~\cite{Ade:2013xla}, the upper bound on $v_{B-L}$ comes down to $v_{B-L} \lesssim 6 \times 10^{15}$~GeV.\footnote{For a recent discussion, see e.g.\ Ref.~\cite{Pallis:2013dxa}. \vspace{2mm} }
For definiteness, we shall work with
$v_{B-L} = 5 \times 10^{15}\,\textrm{GeV}$ in the following.
Note that successful inflation only takes place for suitably chosen
initial conditions \cite{Tetradis:1997kp}, which also depend on the
gravitino mass \cite{Nakayama:2010xf}.

Let us now turn to tensor perturbations.
As evident from Eq.~\eqref{eq:epsiloneta}, $\epsilon$
is suppressed by a loop factor as compared to $\eta$.
This results in a very small tensor-to-scalar ratio $r$
and hence a very small amplitude $\Delta_t^2$ of the
tensor metric perturbations,
\begin{align}
\Delta_t^2 &= \frac{2 H_{\textrm{inf}}^2}{\pi^2 M_{\textrm{Pl}}^2}
= \frac{\lambda}{6\pi^2}\left(\frac{v_{B-L}}{M_{\textrm{Pl}}}\right)^4
=  r \, \Delta_s^2 
\simeq r \times 2.18 \times 10^{-9} \,, 
\label{eq:deltatsquared}\\
r &= 16 \,\epsilon \simeq
1.0 \times 10^{-7} \left(\frac{\lambda}{10^{-4}}\right)
\left(\frac{50}{N_e^*}\right) \,. \label{eq:deltatsquared2}
\end{align}
According to the consistency relation $n_t = -r /8$, we then immediately
conclude that our inflationary model always predicts a negligibly
small tensor spectral index $n_t$.
In the calculation of the GW spectrum, we can therefore
neglect any variation of the Hubble scale during inflation.

%%%%%%%%%%%%%%%%%%%%%%%%%%%%%%%%%%%%%%%%%%%%%%%%%%%%%%%%%%%%%%%%%%%%%%%%%%%%%%

The spontaneous breaking of \BmL at the end of hybrid inflation is
accompanied by two important non-perturbative processes.
The first is tachyonic preheating which denotes the transfer of
the initial vacuum energy density $V_0$ into a gas of non-relativistic
\BmL Higgs bosons $\sigma$
along with the non-adiabatic production of all particle
species coupled to the \BmL Higgs field.\footnote{An interesting question in this context, which requires further investigation, concerns the effect of the inflaton on tachyonic preheating, see e.g.\ Ref.~\cite{Martin:2011ib}.}
The second process is the production of topological defects in
the form of cosmic strings. They are characterized by their energy density per unit length $\mu$, which, in the Abelian Higgs model, is given by~\cite{Hill:1987ye}
\begin{equation}
 \mu = 2 \pi v^2_{B-L} \, B\left(\frac{m_S}{m_Z}\right) \,, \quad \text{ with } B(\beta) = 2.4 \left(\ln\frac{2}{\beta} \right)^{-1} \text{ for } \beta < 10^{-2} \,,
\end{equation}
where $m_S$ and $m_Z$ denote the masses of the \BmL Higgs and gauge bosons, respectively.
Preheating as well as cosmic strings act as sources of
GWs and we will discuss their respective contributions
to the GW spectrum in Secs.~\ref{sec:preheating},
\ref{sec:cosmicstrings} and \ref{sec:comparison}.

%%%%%%%%%%%%%%%%%%%%%%%%%%%%%%%%%%%%%%%%%%%%%%%%%%%%%%%%%%%%%%%%%%%%%%%%%%%%%%

The \BmL breaking sector couples to the supersymmetric standard model
(supplemented by three generations of right-handed neutrinos) via Yukawa
terms in the superpotential,
\begin{align}
W \supset W_{\textrm{MSSM}} + W_{n} \,,\quad
W_{n} = h^{\nu}_{ij} \textbf{5}^*_i n_j^c H_u +
\frac{1}{2} h_i^n n_i^c n_i^c S \,,
\end{align}
where $n_i^c$ denote the superfields containing the charge conjugates of
the right-handed neutrinos,
the matrices $h_{ij}^n$ and $h_{ij}^\nu$ encompass Yukawa
coupling constants, and $W_{\text{MSSM}}$ is the superpotential
of the minimal supersymmetric standard model (MSSM).
We assume that the flavour structure of our superpotential
derives from a $U(1)$ flavour symmetry of the Froggatt-Nielsen
type that commutes with $SU(5)$, 
cf.\ Ref.~\cite{Buchmuller:1998zf}.
This is the reason why we arrange all the superfields of our model
in $SU(5)$ multiplets.
In particular, we have $\textbf{5}_i^* = (d^c_i, \, \ell_i)$, $i = 1,2,3$.
Furthermore, we also assume that the colour triplet partners of
the electroweak Higgs doublets $H_u$ and $H_d$ have been projected out.
During the \BmL and the electroweak phase transitions, the
fields $S$ and $H_{u,d}$ acquire vacuum expectation values $v_{B-L}$
and $v_{u,d}$, respectively.
After electroweak symmetry breaking, the superpotential
$W_n$ hence turns into the usual seesaw superpotential
featuring a neutrino Dirac as well as a neutrino Majorana mass term,
thereby providing us with a natural explanation for the smallness of the
standard model neutrino masses.

%%%%%%%%%%%%%%%%%%%%%%%%%%%%%%%%%%%%%%%%%%%%%%%%%%%%%%%%%%%%%%%%%%%%%%%%%%%%%%

After preheating most of the total energy density is stored in
non-relativistic $\sigma$ particles.
These then slowly decay into all three generations of heavy Majorana
neutrinos $N_i$ and sneutrinos $\tilde{N}_i$ via the second operator
in the superpotential $W_n$.
Subsequently, the heavy (s)neutrinos decay in turn into the lepton-Higgs pairs of the MSSM via the first operator in $W_n$.
The (s)neutrino decay products thermalize immediately, thereby giving rise
to a hot thermal bath of MSSM radiation.
This chain of decay and thermalization processes represents the actual
reheating phase of the early universe.
As explained in more detail in 
Refs.~\cite{Buchmuller:2011mw,Buchmuller:2012wn}, it is accompanied
by the generation of a primordial lepton asymmetry in the decay
of the heavy (s)neutrinos as well as the production of a thermal
abundance of gravitinos.
Electroweak sphaleron processes convert the lepton asymmetry into
the baryon asymmetry of the universe.
Our scenario of cosmological \BmL breaking hence
naturally accommodates baryogenesis via leptogenesis.
Moreover, given an appropriate superparticle mass spectrum, the
thermally produced gravitinos either account themselves for the relic
density of dark matter or they generate the dark matter abundance
in the form of MSSM neutralinos in their decays.

The two main quantities controlling the time evolution of the reheating process
are $\Gamma_S^0$ and $\Gamma_{N_1}^0$, i.e.\ the vacuum decay rate of the \BmL Higgs bosons
and its superpartners as well as the vacuum decay rate of the heavy (s)neutrinos of
the first generation,\footnote{For simplicity, we shall assume that the decay of
the \BmL Higgs multiplet into the two heavier (s)neutrino generations is kinematically
forbidden (cf.\ also Ref.~\cite{Buchmuller:2012wn}).}
\begin{align}
\Gamma_S^0 = \frac{m_S}{32\pi}\left(\frac{M_1}{v_{B-L}}\right)^2
\left[1 - \left(\frac{2M_1}{m_S}\right)^2\right]^{1/2} \,,\quad
\Gamma_{N_1}^0 = \frac{\widetilde{m}_1}{4\pi} \left(\frac{M_1}{v_u}\right)^2 \,,
\label{eq:GammaSN1}
\end{align}
with $\widetilde{m}_1$ denoting the effective neutrino mass of the first generation,
\begin{align}
\widetilde{m}_1 = \left[\left(h^\nu\right)^\dagger h^\nu\right]_{11} \frac{v_u^2}{M_1} \,.
\end{align}
According to the Froggatt-Nielsen flavour model that our earlier study in Ref.~\cite{Buchmuller:2012wn}
is based on, the Higgs and (s)neutrino masses, $m_S$ and $M_1$, are expected to differ by some
power of the Froggatt-Nielsen hierarchy parameter $\eta \simeq 1/\sqrt{300}$.
Just as in our previous work, we shall thus assume for definiteness that $m_S =M_1 / \eta^2$.
This reduces the number of free and independent parameters to two,
namely the two neutrino masses $M_1$ and $\widetilde{m}_1$ which then end up being
in one-to-one correspondence to the two decay rates $\Gamma_S^0$ and $\Gamma_{N_1}^0$.
A further important quantity, which can be determined as a function of
$\Gamma_S^0$ and $\Gamma_{N_1}^0$, is the effective (s)neutrino decay rate
$\Gamma_{N_1}^S$,
\begin{align}
\Gamma_{N_1}^S (a) = \gamma^{-1}(a)\, \Gamma_{N_1}^0 \,,\quad
\gamma^{-1}(a) = \left<\frac{M_1}{E_{N_1}}\right>_a^{(S)} \,,
\label{eq:GammaN1S}
\end{align}
which accounts for the fact that the (s)neutrinos which are produced with very
high momenta $p_{N_1} \gg M_1$ in the decay of the \BmL Higgs particles remain relativistic 
up to their decay.
Correspondingly, the factor $\gamma^{-1}$ multiplying $\Gamma_{N_1}^0$ in
Eq.~\eqref{eq:GammaN1S} denotes the time-dependent inverse Lorentz factor
for the heavy (s)neutrinos averaged over the entire (s)neutrino phase space
(cf.\ Ref.~\cite{Buchmuller:2011mw} for an explicit computation of $\gamma^{-1}$).

%%%%%%%%%%%%%%%%%%%%%%%%%%%%%%%%%%%%%%%%%%%%%%%%%%%%%%%%%%%%%%%%%%%%%%%%%%%%%%

\begin{figure}
\begin{center}
\includegraphics[width=12cm]{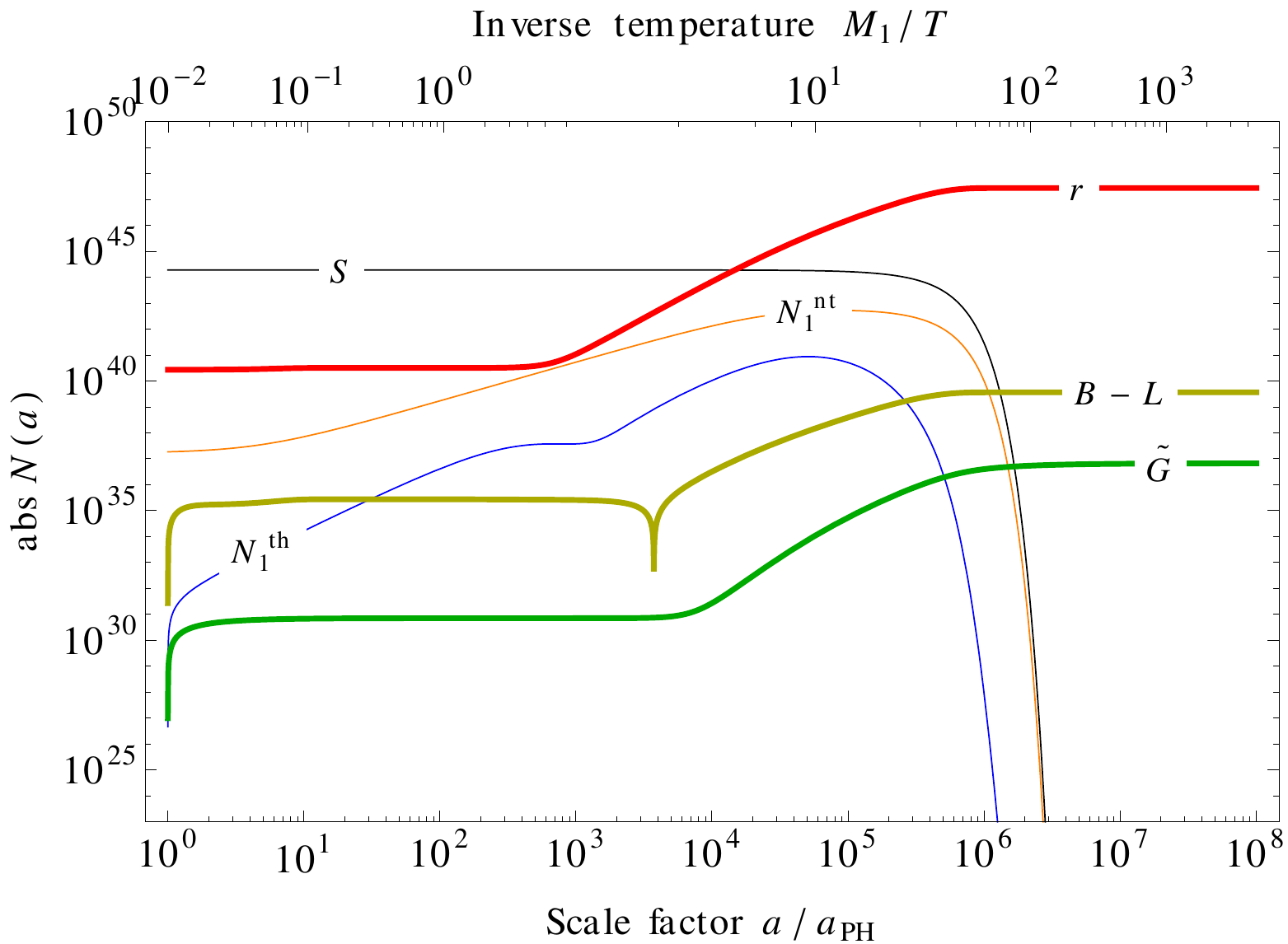}

\vspace*{5mm}

\includegraphics[width=12cm]{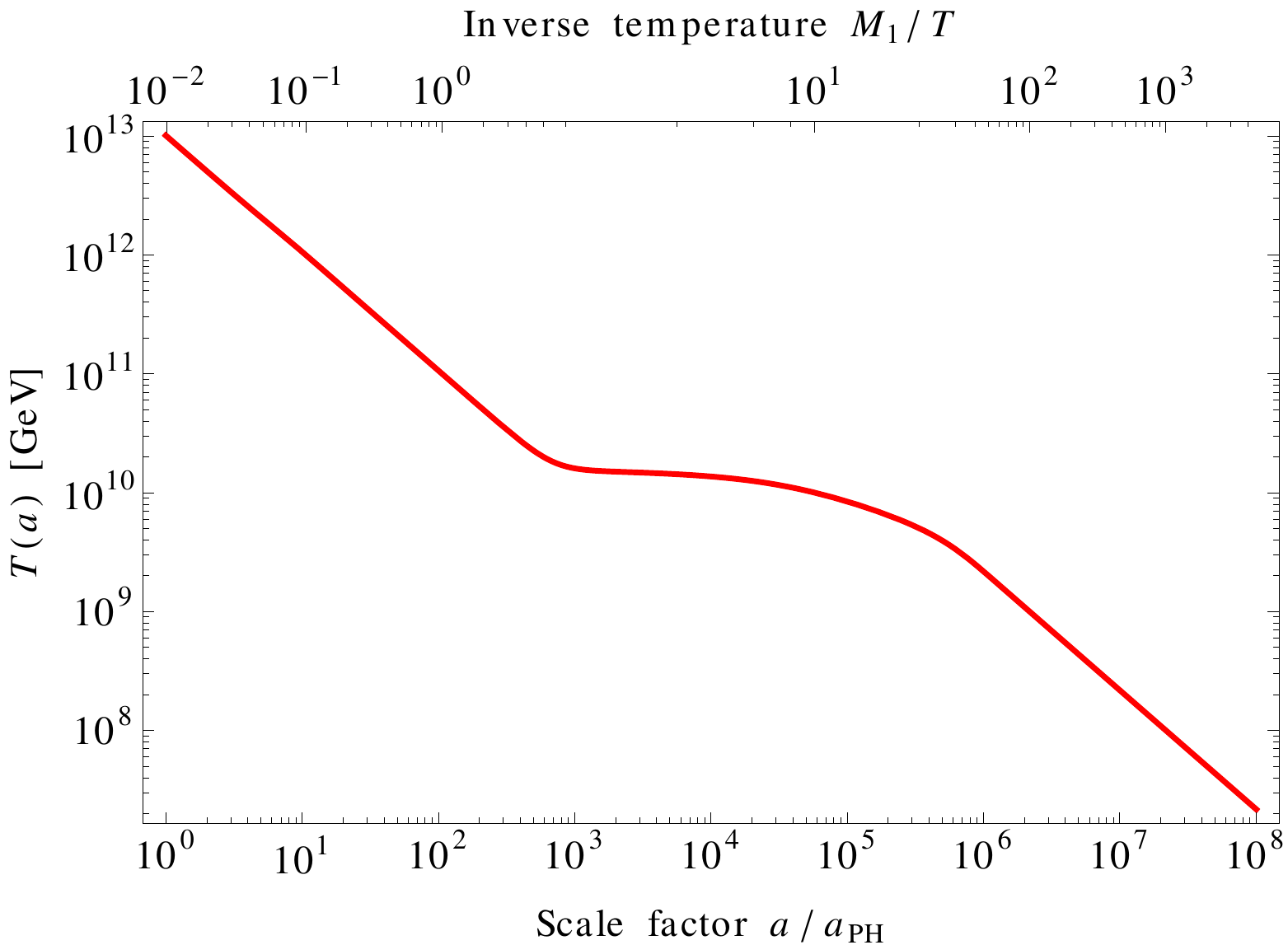}
\caption{{\bf Upper panel:} Comoving number densities of Higgs bosons ($S$), 
thermally and nonthermally produced heavy (s)neutrinos ($N_1^{\mathrm{th}}$, 
$N_1^{\mathrm{nt}}$), radiation ($r$), lepton asymmetry ($B$$-$$L$) and gravitinos ($\widetilde G$). 
{\bf Lower panel:} Emergent plateau of approximately constant
temperature.
Input parameters: Heavy neutrino mass $M_1 =  10^{11}~\mathrm{GeV}$,
effective neutrino mass $\widetilde{m}_1 = 0.04~\mathrm{eV}$.
The $B$$-$$L$ scale is fixed by requiring consistency with hybrid inflation,
$v_{B-L} = 5\times 10^{15}~\mathrm{GeV}$. As in Ref.~\cite{Buchmuller:2012wn}.}
\label{fig:reheat}
\end{center}
\end{figure}

%%%%%%%%%%%%%%%%%%%%%%%%%%%%%%%%%%%%%%%%%%%%%%%%%%%%%%%%%%%%%%%%%%%%%%%%%%%%%%

In order to obtain a detailed and time-resolved picture of
the reheating process, one needs to solve the
set of Boltzmann equations describing the evolution of all
relevant particle species.
Such a study has been performed in Ref.~\cite{Buchmuller:2012wn}.
For completeness, we now recall some of the results of our earlier work
(cf.\ Fig.~\ref{fig:reheat}, upper panel).
A remarkable feature of reheating after the \BmL phase transition
is an approximate plateau in the radiation temperature
around the time when the heavy (s)neutrinos decay
(cf.\ Fig.~\ref{fig:reheat}, lower panel).
This constancy of the temperature over some extended period of
time is a direct consequence of a temporary balance between
entropy production and cosmic expansion.
The temperature at which the plateau
is located represents the characteristic temperature scale
for leptogenesis as well as for the thermal production of gravitinos.
It is typically larger by some $\mathcal{O}(1)$ factor than the
actual reheating temperature $T_{\textrm{RH}}$, which is reached
towards the end of reheating when half of the total energy has been
converted into relativistic particles, cf.\ Sec.~\ref{sec:reheatingtemp}.

\section{Gravitational Waves from Inflation}\label{sec:inflation}

%%%%%%%%%%%%%%%%%%%%%%%%%%%%%%%%%%%%%%%%%%%%%%%%%%%%%%%%%%%%%%%%%%%%%%%%%%%%%%

During inflation quantum fluctuations of the metric are stretched
to ever larger physical scales such that they eventually cross outside
the Hubble horizon.
Outside the horizon, the amplitudes of these metric perturbations
remain preserved and they only begin to evolve again once they re-enter
the Hubble horizon after the end of inflation (cf.\ Sec.~\ref{sec:GWs}).
Inflation hence gives rise to a stochastic background of GWs whose spectrum is directly related to the properties
of the primordial quantum metric fluctuations.
The spectral density $S_h(k)$ for the GWs originating from
inflation is easily calculated.

First, one expands the tensor perturbations $h_{ij}$ into Fourier modes,%
\begin{align}
h_{ij}\left(\mathbf{x},\tau\right) =
\sum_{A = +,\times} \int \frac{d^3k}{\left(2\pi\right)^3}
\,\varphi_k^A (\tau) \, e_{ij}^A \,e^{i\mathbf{k}\mathbf{x}} \,.
\label{eq_fouriermode}
\end{align}
For each wavenumber $k$, we thus have two modes $\varphi_k^+$ and $\varphi_k^\times$.
After rescaling these fields in order to render them
canonically normalized, their two-point function is given
by the usual expression for free scalar fields in an inflationary background,
\begin{align}
\phi_k^A(\tau) = \frac{M_{\textrm{Pl}} \; \varphi_k^A(\tau) }{\sqrt{2} \left(2\pi\right)^{3/2}} \,
 \,, \quad
\left<\phi_k^A(\tau) \phi_{k'}^B (\tau)\right> =
\frac{H_{\textrm{inf}}^2}{2k^3} \,\delta^{AB}
\delta^{(3)}\left(\mathbf{k} + \mathbf{k}'\right) \,, \quad
k \ll aH \,.
\end{align}
The evolution of the modes $\phi_k^A$ after the end of inflation
is accounted for by the transfer function $T_k$ 
(cf.\ Eq.~\eqref{eq:transferfunction}),
which enables us to write down an expression for the two-point function that is valid at all times,
\begin{align}
\left<\phi_k^A(\tau) \phi_{k'}^B (\tau)\right> =
\frac{H_{\textrm{inf}}^2}{2k^3} \, T_k^2(\tau) \,\delta^{AB}
\delta^{(3)}\left(\mathbf{k} + \mathbf{k}'\right) \,.
\end{align}
The correlation function of the tensor perturbations $h_{ij}$
is correspondingly given by
\begin{align}
\left<h_{ij}\left(\mathbf{x},\tau\right)
h^{ij}\left(\mathbf{x},\tau\right)\right>  =
 \int_{-\infty}^{+\infty} dk \,
\frac{H_{\textrm{inf}}^2}{\pi^2 k M_{\textrm{Pl}}^2} \,T_k^2(\tau)
= \frac{1}{\pi} \int_{-\infty}^{+\infty} dk \,
S_h(k) \,T_k^2(\tau) \,,
\label{eq:spectraltransfer}
\end{align}
from which we can readily read off the spectral density $S_h$,
\begin{align}
S_h(k) = \frac{H_{\textrm{inf}}^2}{\pi k M_{\textrm{Pl}}^2} \,.
\end{align}
Note that Eq.~\eqref{eq:spectraltransfer} extends Eq.~\eqref{eq:2pointfct_Sh}
which is only valid for sub-horizon modes.
According to Eq.~\eqref{eq:rhoGW}, we hence obtain for
today's energy spectrum of GWs from inflation
\begin{align}
\Omega_{\textrm{GW}}(k,\tau_0) = \frac{k^3}{6 a_0^2 H_0^2}
\frac{H_{\textrm{inf}}^2}{\pi^2 k M_{\textrm{Pl}}^2} \,T_k^2(\tau_0) =
\frac{\Delta_t^2}{12} \frac{k^2}{a_0^2 H_0^2} \,T_k^2(\tau_0) \,,
\label{eq:OmegaGWInf}
\end{align}
where  $a_0 = a(\tau_0)$ denotes the value of the scale factor today.
%

%%%%%%%%%%%%%%%%%%%%%%%%%%%%%%%%%%%%%%%%%%%%%%%%%%%%%%%%%%%%%%%%%%%%%%%%%%%%%%

To good approximation, the transfer function $T_k$ corresponds to the
ratio of the scale factor at the time when the mode with wavenumber
$k$ crosses back inside the Hubble horizon to the present value of the
scale factor, $T_k \simeq a\left(\tau_k\right) /a_0$, cf.\ Eq.~\eqref{eq:approx_transferfunction}.
$T_k$ therefore depends on the expansion history of the universe and
exhibits a different functional dependence on $k$ depending on whether a
given mode re-enters the Hubble horizon
during matter domination, radiation domination or reheating.

It is instructive to compute the transfer function $T_k$ analytically for the corresponding three
intervals of $k$ values,
\begin{align}
k\ \in\ [k_0,k_{\textrm{eq}}),\,\ [k_{\textrm{eq}},k_{\textrm{RH}}),\,\
[k_{\textrm{RH}},k_{\textrm{PH}}) \,\ , \quad k_i = a_i H(a_i) \,,
\label{eq:kintervals}
\end{align}
where the subscript $i = 0,\textrm{eq},\textrm{RH},\textrm{PH}$ labels
the boundaries of the three epochs between preheating and 
today.\footnote{As a consequence of the late-time
acceleration of the universe, perturbation modes only re-enter the
Hubble horizon until $a_{k_{\textrm{min}}} \simeq
\left(\Omega_m /
\left(2\Omega_{\Lambda}\right)\right)^{1/3} a_0$.
At later times, the physical wavelengths of the modes grow faster than
the Hubble horizon, similarly as during inflation, so that they
cross outside the Hubble horizon again.
The smallest wavenumber that ever crosses inside the Hubble horizon
after inflation consequently corresponds to $k_{\textrm{min}}
\simeq (3/2)^{1/2}\left(2\Omega_\Lambda\right)^{1/6} (\Omega_m)^{1/3} k_0
\simeq 0.86\, k_0$.
By comparison, the mode with wavenumber $k_0$ re-enters the Hubble horizon slightly
earlier at $a \simeq \Omega_m a_0$. }
$k_{\textrm{eq}}$ and $k_{\textrm{RH}}$
stand for the wavenumbers of the modes that re-enter the Hubble
horizon at the time of radiation-matter equality at a redshift
of roughly $3300$ and close to the end of reheating,
when half of the non-relativistic \BmL Higgs bosons and its superpartners have decayed.
$k_{\textrm{PH}}$ is the wavenumber of the mode that has just grown
to the size of the Hubble horizon by the end of inflation and which begins to
move inside the horizon once the expansion of the universe becomes matter
dominated in the course of preheating.
Metric fluctuations with $k > k_{\textrm{PH}}$
are never stretched to horizon-size scales and thus always remain at the
quantum level.

%%%%%%%%%%%%%%%%%%%%%%%%%%%%%%%%%%%%%%%%%%%%%%%%%%%%%%%%%%%%%%%%%%%%%%%%%%%%%%

\medskip
\noindent \textbf{$\boldsymbol{k}_\text{eq}, \boldsymbol{k}_{\text{RH}}$ and $\boldsymbol{k}_\text{PH}$:}
As a preparation for our computation of $T_k$, let us now
determine in turn $k_{\textrm{eq}}$, $k_{\textrm{RH}}$ and $k_{\textrm{PH}}$.
After the end of the reheating process, the comoving entropy density of radiation is
conserved, $a^3 s_r = \textrm{const.}$, such that the Friedmann equation
takes the following form,
\begin{align}
H(a(\tau)) =  H_0 \left[\Omega_{\Lambda}
+ \Omega_m\left(\frac{a_0}{a(\tau)}\right)^3
+ \frac{g_*(\tau)}{g_*^0}\left(\frac{g_{*,s}^0}{g_{*,s}(\tau)}\right)^{4/3}
\Omega_r\left(\frac{a_0}{a(\tau)}\right)^4\right]^{1/2} \,.
\label{eq:HFriedmann}
\end{align}
Here, $\Omega_{\Lambda}$,  $\Omega_m$ and $\Omega_r$ denote the ratios
of the different energy densities to the critical density today,
while $g_*(\tau)$ and $g_{*,s}(\tau)$ are the effective sums of relativistic
DOFs entering the radiation energy and entropy densities, $\rho_r$ and $s_r$,
at conformal time $\tau$, respectively.
A mode with wavenumber $k$ crosses the Hubble horizon at time $\tau_k$
and scale factor $a(\tau_k)$, the latter of which is determined by the relation
\begin{align}
k = a(\tau_k) \, H(a(\tau_k)) \,.
\label{eq:kaH}
\end{align}
The boundary wavenumber $k_{\textrm{eq}}$ can now be calculated
by means of Eqs.~\eqref{eq:HFriedmann} and \eqref{eq:kaH}.
In doing so, we actually only have to take into account the
dominating contributions to the total energy density
at the time of radiation-matter equality, i.e.\ only the second
and the third term on the right-hand side of Eq.~\eqref{eq:HFriedmann}.%
\footnote{Similarly, if we were interested in the wavenumber of the mode
crossing the Hubble horizon at the time of matter-vacuum equality,
we would only have to consider the first and the second term.}
This yields
\begin{align}
k_{\textrm{eq}} =
\left(\frac{g_{*,s}^{\textrm{eq}}}{g_{*,s}^0}\right)^{2/3}
\left(\frac{2g_{*}^0}{g_{*}^{\textrm{eq}}}\right)^{1/2}
\frac{\Omega_m}{{\Omega_r}^{1/2}} \, k_0 =
7.33 \times 10^{-2} \, a_0 \,\Omega_m h^2 \,\textrm{Mpc}^{-1}\,,
\label{eq:keq}
\end{align}
with $g_{*,s}^{\rm eq} = 3.91$, $g_{*,s}^0 =3.91$ and $g_*^{\rm eq} = 3.36$.
The present value of $g_*$ is sensitive to the mass spectrum of the light standard
model neutrinos.
If all neutrinos are non-relativistic at present, we have $g_*^0 = 2$.
On the other hand, if the lightest neutrino has not yet turned non-relativistic, $g_*^0$
is slightly larger, $g_*^0 = 2.45$.
Note that the numerical value of $k_{\textrm{eq}}$ is, however, not affected by this
subtlety as it only depends on $\Omega_r / g_*^0$.

%%%%%%%%%%%%%%%%%%%%%%%%%%%%%%%%%%%%%%%%%%%%%%%%%%%%%%%%%%%%%%%%%%%%%%%%%%%%%%

As compared to $k_{\textrm{eq}}$, the computation of the boundary wavenumber
$k_{\textrm{RH}}$ is complicated by two effects.
First, according to our definition of $T_{\textrm{RH}}$ as the temperature when half of the total energy budget has been converted into relativistic particles, the MSSM DOFs
do not account for half of the total energy density at $a = a_{\textrm{RH}}$.
Instead, the energy density of the non-relativistic \BmL Higgs bosons and its
superpartners at $a = a_{\textrm{RH}}$ is balanced by MSSM radiation and
the heavy, relativistic but nonthermal (s)neutrinos together,
\begin{align}
\rho_{\textrm{tot}} \left(a_{\textrm{RH}}\right) = 
2 \, \rho_S \left(a_{\textrm{RH}}\right) =
2 \left[\rho_r \left(a_{\textrm{RH}}\right) + \rho_N \left(a_{\textrm{RH}}\right) +
\rho_{\tilde{N}} \left(a_{\textrm{RH}}\right) \right] \,. \label{eq:defaR}
\end{align}
In the following, we shall quantify the contribution of the thermal bath to
the total energy density at $a = a_{\textrm{RH}}$ by the factor
$\alpha_{\textrm{RH}} = \rho_{\textrm{tot}} \left(a_{\textrm{RH}}\right)/
\rho_r \left(a_{\textrm{RH}}\right) \geq 2$.
Second, the comoving radiation entropy density $a^3 s_r$ is only conserved once all
non-relativistic particles and heavy (s)neutrinos have decayed, i.e.\ only
for sufficiently late times after $a = a_{\textrm{RH}}$.
In order to quantify the amount of entropy production after $a = a_{\textrm{RH}}$,
we introduce the dilution factor
\begin{align}
D = \frac{\left(a^3 s_r\right)_{a \gg a_{\textrm{RH}}}}
{\left(a^3 s_r\right)_{a = a_{\textrm{RH}}}} \geq 1\,.
\label{eq_Delta}
\end{align}
Based on $D$, we may also define two further useful quantities:
the would-be reheating temperature $\widetilde{T}_{\textrm{RH}}$ as well as
the would-be radiation energy density at reheating $\tilde{\rho}_r^{\textrm{RH}}$,
\begin{align}
\widetilde{T}_{\textrm{RH}} = D^{1/3} \, T_{\textrm{RH}} \,, \quad
\tilde{\rho}_r^{\textrm{RH}} = D^{4/3} \, \rho_r \left(a_{\textrm{RH}}\right) \,,
\label{eq:defTwb}
\end{align}
which represent the temperature and the energy density the thermal bath would have had at $a=a_{\textrm{RH}}$
if, extrapolating back in time from the present epoch, no entropy production
took place as long as $a > a_{\textrm{RH}}$.
Both, $\alpha_{\textrm{RH}}$ and $D$, need to be determined by solving
the Boltzmann equations and depend on the parameters of our model.
We will discuss this parameter dependence in Sec.~\ref{sec:reheatingtemp}.
For now, we simply state that, taking both effects quantified by the two factors
$\alpha_{\textrm{RH}}$ and $D$ into account, one finds the following
expression for $k_{\textrm{RH}}$,
\begin{align}
k_{\textrm{RH}} =  \left(\frac{\alpha_{\textrm{RH}}}{2}\right)^{1/2} D^{-1/3}
\left(\frac{g_{*,s}^0}{g_{*,s}^{\textrm{RH}}}\right)^{1/3}
\left(\frac{2 g_{*}^{\textrm{RH}}}{g_{*}^0}\right)^{1/2} {\Omega_r}^{1/2}
\frac{T_{\textrm{RH}}}{T_{\gamma}^0} \,k_0 \,,
\label{eq:kRH}
\end{align}
with $g^{\rm RH}_* = g^{\rm RH}_{*,s} = 915/4$.
Given the above relations, one easily sees that
\begin{align}
\frac{\alpha_{\textrm{RH}}}{2} =
R \, D^{4/3} \,, \quad
R = \frac{\rho_S\left(a_{\textrm{RH}}\right)}{\tilde{\rho}_r^{\textrm{RH}}} \,.
\label{eq:defR}
\end{align}
The ratio $R$ can be
shown to take a constant value across the entire parameter space,
$R \simeq 0.41$ (cf.\ Appendix~\ref{app:scalefactor}). Physically, the quantity $1/R$ corresponds to the increase of $(a/a_{\text{PH}})^4 \, (\rho_r + \rho_N + \rho_{\tilde N})$ after $a = a_\text{RH}$ due to the decay of the remaining \BmL Higgs bosons.
Because $R$ is a constant, it is possible to rewrite Eq.~\eqref{eq:kRH} in terms of the would-be reheating
temperature $\widetilde{T}_{\textrm{RH}}$ or, equivalently, in terms of the
effective temperature
$T_* = R^{1/2}\, \widetilde{T}_{\textrm{RH}} \simeq 0.64\, \widetilde{T}_{\textrm{RH}}$,
\begin{align}
k_{\textrm{RH}} = & \:
R^{1/2}\left(\frac{g_{*,s}^0}{g_{*,s}^{\textrm{RH}}}\right)^{1/3}
\left(\frac{2 g_{*}^{\textrm{RH}}}{g_{*}^0}\right)^{1/2} {\Omega_r}^{1/2}
\frac{\widetilde{T}_{\textrm{RH}}}{T_{\gamma}^0} \,k_0  \\
= & \: \left(\frac{g_{*,s}^0}{g_{*,s}^{\textrm{RH}}}\right)^{1/3}
\left(\frac{2 g_{*}^{\textrm{RH}}}{g_{*}^0}\right)^{1/2} {\Omega_r}^{1/2}
\frac{T_*}{T_{\gamma}^0} \,k_0 =
2.75 \times 10^{14} \, a_0 \,\textrm{Mpc}^{-1} \left(\frac{T_*}{10^7 \,\textrm{GeV}}\right) \,.
\label{eq:kRHTeff}
\end{align}
Here, $T_*$ is defined such that it appears in our final expression for $k_{\textrm{RH}}$
in exactly the same way as the actual reheating temperature $T_{\textrm{RH}}$ would appear
in $k_{\textrm{RH}}$ if one were to perform a more naive calculation, neglecting the two correction
factors $\alpha_{\textrm{RH}}$ and $D$.
Put differently, if one tried to deduce the reheating temperature from a measurement
of $k_{\textrm{RH}}$ making use of the naive formula for $k_{\textrm{RH}}$,
i.e.\ Eq.~\eqref{eq:kRH} with $\alpha_{\textrm{RH}} = 2$ and $D = 1$,
one would end up with
the effective temperature $T_*$.
As we will see shortly, the GW spectrum exhibits a kink just at
$k = k_{\textrm{RH}}$.
This is why we shall refer to $T_*$ as the `effective kink temperature'.
Finally, we emphasize that the distinction between $T_{\textrm{RH}}$
and $T_*$ is crucial since the reheating temperature $T_{\textrm{RH}}$
turns out to be sensitive to the properties of the
Higgs sector as well as of the neutrino sector, while the effective kink temperature $T_*$
is solely determined by the properties of the Higgs sector, cf.\ Sec.~\ref{sec:reheatingtemp} for a more
detailed discussion.

%%%%%%%%%%%%%%%%%%%%%%%%%%%%%%%%%%%%%%%%%%%%%%%%%%%%%%%%%%%%%%%%%%%%%%%%%%%%%%

The energy density at the end of preheating is given by the vacuum
energy density during inflation, $V_0 = \lambda v_{B-L}^4/4$, which
yields for the wavenumber $k_{\textrm{PH}}$,
\begin{align}
\frac{k_{\textrm{PH}}}{k_{\textrm{RH}}} = & \: C_{\textrm{RH}}
\left(\frac{V_0}{\rho^\text{RH}_{\textrm{tot}}}\right)^{1/6} =
\frac{C_{\textrm{RH}}}{\alpha_{\textrm{RH}}^{1/6}}
\left(\frac{30\lambda}{4\pi^2g_{*}^{\textrm{RH}}}\right)^{1/6}
\left(\frac{v_{B-L}}{T_{\textrm{RH}}}\right)^{2/3} \label{eq_kPHkRH} \\ = & \: 
\frac{C_{\textrm{RH}}}{R^{1/6}}
\left(\frac{15\lambda}{4\pi^2g_{*}^{\textrm{RH}}}\right)^{1/6}
\left(\frac{v_{B-L}}{\widetilde{T}_{\textrm{RH}}}\right)^{2/3} =
C_{\textrm{RH}} R^{1/6}
\left(\frac{15\lambda}{4\pi^2g_{*}^{\textrm{RH}}}\right)^{1/6}
\left(\frac{v_{B-L}}{T_*}\right)^{2/3}\,. \nonumber
\end{align}
Here, we have introduced $C_{\textrm{RH}} =
\big(a_{\textrm{PH}}H_{\textrm{PH}}^{2/3}\big)/
\big(a_{\textrm{RH}}H_{\textrm{RH}}^{2/3}\big)$ in order to account for
the complicated evolution of the Hubble parameter during reheating.
Making the simplifying assumption that the universe is strictly matter
dominated throughout the entire reheating process, one has
$C_{\textrm{RH}} = 1$.
Hence, the actual value of $C_{\textrm{RH}}$ ought to be larger than $1$.
In fact, as we demonstrate in Appendix~\ref{app:scalefactor}, $C_{\textrm{RH}}$
turns out to take a constant value in the entire parameter space, $C_{\textrm{RH}} \simeq 1.13$.
All of the four wavenumbers $k_0$, $k_{\textrm{eq}}$, $k_{\textrm{RH}}$
and $k_{\textrm{PH}}$ can be translated into frequencies
$f = k/\left(2\pi a_0\right)$ at which GW experiments
could observe the corresponding modes,
\begin{align}
f_0 = & \:
3.58 \times 10^{-19} \,\textrm{Hz} \left(\frac{h}{0.70}\right) \,, \\
f_{\textrm{eq}} = & \:
1.57 \times 10^{-17} \,\textrm{Hz}
\left(\frac{\Omega_mh^2}{0.14}\right) \,, \\
f_{\textrm{RH}} = & \:
4.25 \times 10^{-1} \,\textrm{Hz}
\left(\frac{T_*}{10^7\,\textrm{GeV}}\right) \,,  \label{eq_fRH} \\
f_{\textrm{PH}} = & \:
1.93 \times 10^4 \,\textrm{Hz}
\left(\frac{C_{\textrm{RH}}}{1.13}\right)
\left(\frac{R}{0.41}
\frac{\lambda}{10^{-4}}\right)^{1/6}
\left(\frac{10^{-15} \, v_{B-L}}{5 \,\textrm{GeV}}\right)^{2/3}
\left(\frac{T_*}{10^7\,\textrm{GeV}}\right)^{1/3} \,.
\label{eq:fPH}
\end{align}

%%%%%%%%%%%%%%%%%%%%%%%%%%%%%%%%%%%%%%%%%%%%%%%%%%%%%%%%%%%%%%%%%%%%%%%%%%%%%%

\medskip
\noindent \textbf{Transfer function:}
\noindent For modes with wavenumbers well inside one of the three intervals
in Eq.~\eqref{eq:kintervals}, the transfer function $T_k$ can be
readily computed analytically.
Approximating $T_k$ by the ratio $a\left(\tau_k\right)/a_0$
(cf.~Eq.~\eqref{eq:approx_transferfunction}) and
making again use of the Friedmann equation as well as the conservation
of the comoving entropy density for $a \gg a_{\textrm{RH}}$, one obtains
\begin{align}
T_k \simeq {\Omega_r}^{1/2}
\bigg(\frac{g_{*}^k}{g_{*}^0}\bigg)^{1/2}
\left(\frac{g_{*,s}^0}{g_{*,s}^k}\right)^{2/3} \frac{k_0}{k}
\begin{cases}
\frac{1}{\sqrt{2}}\, k_{\textrm{eq}} / k \,, & k_0 \ll k \ll k_{\textrm{eq}} \\
1 \,, & k_{\textrm{eq}} \ll k \ll k_{\textrm{RH}} \\
\sqrt{2} \,R^{1/2}\,C_{\textrm{RH}}^3 \, k_{\textrm{RH}} / k \,, &
k_{\textrm{RH}} \ll k \ll k_{\textrm{PH}}
\end{cases} \,.
\label{eq:Tkresult}
\end{align}
As long as a mode with wavenumber $k$ re-enters into the Hubble horizon
during radiation domination, $g_{*}^k$ and $g_{*,s}^k$ denote the usual
values of $g_{*}(\tau)$ and $g_{*,s}(\tau)$ at time $\tau_k$. On the other hand,
during reheating and matter domination $g_{*}^k$
and $g_{*,s}^k$ correspond to
$g_{*}^{\textrm{RH}}$ and $g_{*,s}^{\textrm{RH}}$ as well as to
$g_{*}^{\textrm{eq}}$ and $g_{*,s}^{\textrm{eq}}$,
respectively.
Inserting now our result for $T_k$ in Eq.~\eqref{eq:Tkresult}
into Eq.~\eqref{eq:OmegaGWInf}, we arrive at the following
expression for the energy spectrum of the GWs
from inflation, (see Fig.~\ref{fig_infl_vs_cs})
\begin{align}
\Omega_{\textrm{GW}} (k) = \frac{\Delta_t^2}{12} \, \Omega_r \,
\frac{g_{*}^k}{g_{*}^0}
\left(\frac{g_{*,s}^0}{g_{*,s}^k}\right)^{4/3}
\begin{cases}
\frac{1}{2}\, \left(k_{\textrm{eq}} / k\right)^2 \,, & k_0 \ll k \ll k_{\textrm{eq}} \\
1 \,, & k_{\textrm{eq}} \ll k \ll k_{\textrm{RH}} \\
2\,R\, C_{\textrm{RH}}^6 \, \left(k_{\textrm{RH}} / k\right)^2 \,, &
k_{\textrm{RH}} \ll k \ll k_{\textrm{PH}}
\end{cases} \,.
\label{eq:OmegaGWinfresult}
\end{align}
Evidently, the energy spectrum $\Omega_{\textrm{GW}}$ decreases
like $k^{-2}$ at its edges and features a plateau in its center.
In the context of cosmological \BmL breaking, the height of the
plateau is controlled by the coupling $\lambda$,
which determines the self-interaction 
of the \BmL breaking Higgs field, as well as by the \BmL breaking scale
(cf.\ Eqs.~\eqref{eq:deltatsquared}, \eqref{eq:deltatsquared2}),
\begin{align}
\Omega^\text{pl}_{\textrm{GW}} h^2
&= \frac{\lambda}{72\pi^2}\left(\frac{v_{B-L}}{M_{\textrm{Pl}}}\right)^4
\Omega_r h^2\ \bigg(\frac{g_{*}^k}{g_{*}^0}\bigg) 
\left(\frac{g_{*,s}^0}{g_{*,s}^k}\right)^{4/3}\nonumber \\
&= 3.28 \times 10^{-22} \left(\frac{\lambda}{10^{-4}}\right)
\left(\frac{v_{B-L}}{5\times 10^{15}\ \textrm{GeV}}\right)^4
\left(\frac{\Omega_r}{8.5\times 10^{-5}}\right)\bar{g}^k \ , \label{eq_Omegaplateau_inflation}
\end{align}
where $\bar{g}^k = (4 g_{*}^k/427)(427/(4 g_{*,s}^k))^{4/3}$ is a ratio
of energy and entropy DOFs. Note that a further effect modifying the GW spectrum to a similar extent as the change of the DOFs during radiation domination is a contribution to the stress energy tensor due to freely streaming neutrinos, cf.\ Ref.~\cite{Watanabe:2006qe}. Since this effect is only relevant for frequencies below $10^{-10}$~Hz, i.e.\ outside the range accessible in near-future experiments, we omit it in this paper.

%%%%%%%%%%%%%%%%%%%%%%%%%%%%%%%%%%%%%%%%%%%%%%%%%%%%%%%%%%%%%%%%%%%%%%%%%%%%%%

In order to describe the two kinks in the energy spectrum
at wavenumbers around $k_{\textrm{eq}}$ and $k_\textrm{RH}$ more accurately, let us
rewrite the transfer function in Eq.~\eqref{eq:Tkresult} as
\begin{align}
T_k = \Omega_m
\bigg(\frac{g_{*}^k}{g_{*}^{\textrm{eq}}}\bigg)^{1/2}
\left(\frac{g_{*,s}^{\textrm{eq}}}{g_{*,s}^k}\right)^{2/3}
\left(\frac{k_0}{k}\right)^2
T_1\left(k/k_{\textrm{eq}}\right)
T_2\left(k/k_{\textrm{RH}}\right) \,,
\label{eq:T1T2}
\end{align}
where $T_1$ and $T_2$ denote two auxiliary
transfer functions accounting for the transition from matter
to radiation domination and from radiation domination
to reheating, respectively.
We have determined both functions numerically by solving
the equation of motion for the Fourier modes $\varphi_k^A$
in appropriate $k$ ranges.
Our numerical results are reasonably well described by the
two fit functions
\begin{align}
\label{eq:T1T1result}
T_1(x) \simeq & \: c_1^{(0)} \left(1+c_1^{(1)}x + c_1^{(2)}x^2\right)^{+1/2}
\,, \quad c_1^{(0)} \simeq 0.73 \,,\:\: c_1^{(1)} \simeq \phantom{-}1.64 \,,\:\:
c_1^{(2)} \simeq 3.87 \,, \\
T_2(x) \simeq & \: \phantom{c_1^{(0)}}
\left(1+c_2^{(1)}x + c_2^{(2)}x^2\right)^{-1/2}
\,, \quad  \phantom{c_1^{(0)} \simeq 0.73 \,,}\:\:\:  c_2^{(1)} \simeq -0.38 \,,\:\: c_2^{(2)} \simeq 1.04 \,.
\label{transferfktT2}
\end{align}

%%%%%%%%%%%%%%%%%%%%%%%%%%%%%%%%%%%%%%%%%%%%%%%%%%%%%%%%%%%%%%%%%%%%%%%%%%%%%%

At first sight, it might appear surprising that we obtain a coefficient
$c_1^{(0)}$ different from $1$.
This is, however, merely a
consequence of our decision to approximate $T_k$ by
$a\left(\tau_k\right)/a_0$ in our analytical calculation.
Alternatively, we could have also determined $T_k$ directly in terms
of the analytical solution of the mode equation for modes
re-entering the Hubble horizon during matter domination.
Instead of Eq.~\eqref{eq:T1T2}, the starting point of our numerical
computations would then have been~\cite{Turner:1993vb},%
\footnote{This is the expression for $T_k$ usually given in the literature~(see e.g.\ Ref.~\cite{Nakayama:2008wy}),
except for the fact that in our expression $g_{*}^k$ and $g_{*,s}^k$
are divided by $g_{*}^{\textrm{eq}}$ and $g_{*,s}^{\textrm{eq}}$, while
they are divided by $g_{*}^0$ and $g_{*,s}^0$ in the standard
expression. Our result thus coincides with the one in the literature in
the case of massless neutrinos and is smaller than the standard result by
a factor of
$\left(g_{*}^0/g_{*}^{\textrm{eq}}\right)^{1/2} \simeq 0.77$ in
the case of three non-relativistic neutrino species in the present epoch.}
\begin{align}
T_k = \Omega_m
\bigg(\frac{g_{*}^k}{g_{*}^{\textrm{eq}}}\bigg)^{1/2}
\left(\frac{g_{*,s}^{\textrm{eq}}}{g_{*,s}^k}\right)^{2/3}
\left(\frac{3 j_1(2 k/ H_0)}{2k/H_0}\right)
T_1\left(k/k_{\textrm{eq}}\right)
T_2\left(k/k_{\textrm{RH}}\right) \,,
\label{eq:TkAlt}
\end{align}
where $j_1$ denotes the spherical Bessel function of the first kind of
order $1$.
In the limit of a large argument, $j_1(x)$ is well approximated by $1/x$.
By comparing Eqs.~\eqref{eq:T1T2} and \eqref{eq:T1T1result} with
Eq.~\eqref{eq:TkAlt}, the coefficient $c_1^{(0)}$ can hence be identified
with the factor $3/4$ multiplying $\left(k/H_0\right)^{-2}$ in the
large-$k$ limit.
By contrast, the coefficient $c_1^{(1)}$ is inaccessible by analytical calculations.
It is slightly larger than the value usually given in the literature,
$c_1^{(1)} \simeq 1.57$ \cite{Nakayama:2008wy},
which is mainly due to our slightly larger reference wavenumber $k_{\textrm{eq}}$,
cf.\ Eq.~\eqref{eq:keq}.
The same is true for $c_1^{(2)}$, which is often given
as $c_1^{(2)} \simeq 3.42$ \cite{Nakayama:2008wy}.
Analytically, the product $\big[c_1^{(0)}\big]^2 c_1^{(2)}$ is
expected to be exactly $2$, which is well reproduced by our numerical
results, $\big[c_1^{(0)}\big]^2 c_1^{(2)} \simeq 2.05$ (cf.\ Eq.~\eqref{eq:OmegaGWinfresult}).

%%%%%%%%%%%%%%%%%%%%%%%%%%%%%%%%%%%%%%%%%%%%%%%%%%%%%%%%%%%%%%%%%%%%%%%%%%%%%%

The authors of Ref.~\cite{Nakayama:2008wy} also compute the transfer function $T_2$
for the particular case of reheating via ordinary inflaton decay.
The coefficients in their fit function,
$c_2^{(1)} \simeq -0.32$ and $c_2^{(2)} \simeq 0.99$,
are very close to our numerical results, which illustrates that
the kink in the GW spectrum at $k = k_{\textrm{RH}}$
turns out to have the same shape, regardless of whether reheating
proceeds as in our case or as in the standard scenario.
The reason for this insensitivity is clear:
the shape of the GW spectrum is solely controlled
by the evolution of the scale factor $a(t)$, which, in turn,
remains qualitatively unaffected when reheating via the decay of non-relativistic
inflaton particles into arbitrary relativistic DOFs is traded for reheating via
the decay of non-relativistic \BmL Higgs bosons into
relativistic (s)neutrinos, cf.\ Appendix~\ref{app:scalefactor}.
As discussed below Eq.~\eqref{eq:kRHTeff}, it is rather
the position of the kink and its dependence on the reheating
temperature
that distinguishes between reheating via inflaton decay
and reheating after cosmological \BmL breaking.
In Sec.~\ref{sec:reheatingtemp}, we will explore this connection between $k_{\textrm{RH}}$
and the temperature of the thermal bath at different times in more detail.

%%%%%%%%%%%%%%%%%%%%%%%%%%%%%%%%%%%%%%%%%%%%%%%%%%%%%%%%%%%%%%%%%%%%%%%%%%%%%%

Before concluding this section, we finally remark that analytically the
coefficient $c_2^{(2)}$ is expected to correspond to
$\left(2 \,R \,C_{\textrm{RH}}^6\right)^{-1} \simeq 0.59$ (cf.\ Eq.~\eqref{eq:OmegaGWinfresult}).
We have checked explicitly that the discrepancy between this expectation
and our numerical result can be entirely attributed to the uncertainty
of our analytical calculation related to the fact that we approximate $T_h$
by the ratio of the scale factor $a(\tau_k)/a_0$.
The deviation of $\left[c_2^{(2)}\right]^{1/2}$ from
$\left(2 \,R \,C_{\textrm{RH}}^6\right)^{-1/2} \simeq 0.77$
is hence at the same level as the deviation of $c_1^{(0)}$ from $1$.
This implies that, for modes entering the Hubble horizon during matter domination,
the approximation $T_k \approx a(\tau_k)/a_0$ apparently
overestimates the actual value of the transfer function by roughly $20\,\%$ to $30\,\%$.
At the same time, as evident from the numerical result for
$\big[c_1^{(0)}\big]^2 c_1^{(2)}$,
we find good agreement between our numerical and our analytical calculation
for modes entering the Hubble horizon during radiation domination.

%%%%%%%%%%%%%%%%%%%%%%%%%%%%%%%%%%%%%%%%%%%%%%%%%%%%%%%%%%%%%%%%%%%%%%%%%%%%%%

\section{Gravitational Waves from Preheating}\label{sec:preheating}

As mentioned in Sec.~\ref{sec:model}, the phase transition at the end of hybrid inflation is accompanied by a non-perturbative process called tachyonic preheating~\cite{Felder:2000hj}. The $B$$-$$L$ Higgs field develops a tachyonic mass, which leads to an exponential growth of its long-wavelength fluctuations around its vacuum expectation value (vev). This process results in the conversion of nearly all of the vacuum energy of inflation into non-relativistic Higgs particles, while at the same time leading to a non-vanishing vev of the \BmL Higgs field and hence a breaking of the $U(1)_{B-L}$ symmetry accompanied by the formation of cosmic strings. The process is rather rapid and violent, involving the collision of `bubble'-like structures with different Higgs vevs and hence leads to the production of GWs~\cite{GarciaBellido:2007dg}.

The process of tachyonic preheating forms a classical, sub-horizon source for GWs which is active only for a short time. The resulting GW spectrum can be obtained by calculating the solution to the mode equation, Eq.~\eqref{eq_hcl_k}, and inserting it into Eq.~\eqref{eq:rhoGWtot}\footnote{Note that Eq.~\eqref{eq_Pi_isotropy} cannot be employed here, because translational invariance does not hold for the scales we are interested in during the preheating phase governed by bubble collisions.}. The anisotropic stress tensor $\Pi_{ij}$ entering Eq.~\eqref{eq_hcl_k} is determined by the dynamics of preheating and vanishes after the end of preheating, allowing the GWs to propagate freely for $\tau \gg \tau_{\text{PH}}$. The remaining challenge is thus to calculate $\Pi_{ij}$ during preheating. This task can be performed numerically, see e.g.\ 
Ref.~\cite{Dufaux:2007pt} for a detailed description of the method and an application to preheating after chaotic inflation as well as 
Ref.~\cite{Dufaux:2008dn} for an application to tachyonic preheating after hybrid inflation. The following discussion will be based on analytical estimates supported by the results of these simulations~\cite{Felder:2006cc, GarciaBellido:2007dg,  Dufaux:2007pt, Dufaux:2008dn}.

GWs from tachyonic preheating are expected to yield a spectrum which is strongly peaked at a typical (physical) scale $R_{\text{PH}}$ associated with the preheating process. The corresponding comoving wave number describing the position of this peak in today's spectrum is readily obtained by redshifting this scale,\begin{equation}
 k_{\text{PH}} = a_{\text{PH}}\, R^{-1}_{\text{PH}} = \frac{a_{\text{PH}}}{a_{\text{RH}}} \, \frac{a_{\text{RH}}}{a_0} \, R^{-1}_{\text{PH}} \, a_0 = \frac{a_{\text{PH}}}{a_{\text{RH}}} \, D^{-1/3} \left( \frac{g_{*,s}^0}{g_{*,s}^{\text{RH}}} \right)^{1/3} \frac{T^0_{\gamma}}{T_{\text{RH}}}\,  R^{-1}_{\text{PH}}  \, a_0\,,
\label{eq_k_tp}
\end{equation}
with $D$ accounting for the deviation from an adiabatic expansion after $a_\text{RH}$, see Eq.~\eqref{eq_Delta}.
The corresponding amplitude of the GW spectrum can be estimated by using the picture of bubble collisions, which implies that the fraction of energy converted into GWs at preheating is given by $\rho_\text{GW}/\rho_c \sim (R_{\text{PH}} H_{\text{PH}})^2$ \cite{Dufaux:2007pt}. This quantity corresponds to the integrated GW wave spectrum  (cf.~Eq.~\eqref{eq:OmegaGW}),
\begin{equation}
\int_{-\infty}^{\infty} d \ln k \; \Omega_\text{GW}(k, \tau) = \frac{\rho_\text{GW}(\tau)}{\rho_c(\tau)} \,.
\label{eq:Omegatot}
\end{equation}
Hence, for a strongly peaked spectrum, we can estimate the amplitude of this peak at $\tau = \tau_\text{PH}$ as
\begin{equation}
  \Omega_\text{GW}^{\text{PH}}(k_{\text{PH}}) \simeq\ c_{\text{PH}} (R_{\text{PH}} H_{\text{PH}})^2 \,.
 \end{equation}
Here $c_{\text{PH}}$ is a model dependent numerical factor,  e.g.\ $c_{\text{PH}} = 0.05$ for the model considered in Ref.~\cite{GarciaBellido:2007dg}.
Analogously to the stochastic GW background from inflation, this result can be redshifted to today (cf.~Eqs.~\eqref{eq:OmegaGWInf} and \eqref{eq:Tkresult}), yielding
\begin{equation}
\begin{split}
 \Omega_\text{GW} (k_{\text{PH}})   h^2 & \simeq \ 
c_{\text{PH}} \, (R_{\text{PH}} H_{\text{PH}})^2\ 
\frac{a_{\text{PH}}}{a_{\text{RH}}}\ \Omega_r h^2 \
\frac{g_*^{\text{RH}}}{g_*^0} \left( \frac{g_{*,s}^{0}}{g_{*,s}^{\text{RH}}} 
\right)^{4/3} \, \left(2 C_\text{RH}^3 R \right) \\
 & \simeq\ 1.5 \times 10^{-5}\ c_{\text{PH}} \ (R_{\text{PH}} H_{\text{PH}})^2\
\frac{a_{\text{PH}}}{a_{\text{RH}}} 
\left(\frac{\Omega_r h^2/ g_*^0}{1.237\times 10^{-5}}\right) \ .
\label{eq_Omega_tp}
\end{split}
\end{equation}
Estimating $H_{\text{PH}}$, $R_{\text{PH}}$ and $a_{\text{PH}}/a_{\text{RH}}$ in the context of our model, Eqs.~\eqref{eq_k_tp} and \eqref{eq_Omega_tp} enable us to predict the characteristic features of the GW spectrum due to preheating.

Tachyonic preheating is a very rapid process, and we can therefore to very good approximation express $H_{\text{PH}}$ as the Hubble parameter at the end of inflation,
\begin{equation}
 H_{\text{PH}} \simeq H_{\rm inf} = \left(\frac{\lambda}{12}\right)^{1/2} \frac{v^2_{B-L}}{M_\text{Pl}} \,.
\label{eq_HPH}
\end{equation}
An estimate of $R_{\text{PH}}$ can be obtained by studying the preheating process. For tachyonic preheating associated with the breaking of a local $U(1)$ symmetry, there are two typical scales, one associated with the dynamics of the scalar field, cf.\ Ref.~\cite{Dufaux:2008dn}, and the other associated with the presence of the gauge field, cf.\ Ref.~\cite{Dufaux:2010cf}. For the former, there are two distinct possibilities, depending on what triggers the onset of tachyonic preheating: the inflaton crossing the critical point with a significant velocity or quantum diffusion, triggered by the growth of quantum fluctuations around the critical point of the scalar potential. For the range of model parameters of interest here (cf. Ref.~\cite{Buchmuller:2012wn}), the inflaton velocity is the parameter governing the onset of preheating and hence~\cite{Dufaux:2008dn}
\begin{equation}
 \left(R^{(s)}_{\text{PH}}\right)^{-1} = (\lambda \, v_{B-L} \, |\dot \phi_c|)^{1/3} = 0.15 \, \lambda^{5/6} \, v_{B-L}^{2/3} \, M_\text{Pl}^{1/3} \,.
\end{equation}
The typical scale associated with the gauge field is given by the mass $m_Z$ of the gauge boson~\cite{Dufaux:2010cf},
\begin{equation}
 \left(R^{(v)}_{\text{PH}}\right)^{-1} \sim m_Z = 2 \sqrt{2} \, g \, v_{B-L} \,.
\end{equation}

We now exploit the equation of state of the universe during reheating, which implies
\begin{equation}
  \frac{a_{\text{PH}}}{a_{\text{RH}}} =  
C_{\text{RH}} \; \left( \frac{H_{\text{RH}}}{H_{\text{PH}}} \right)^{2/3}  \,,
\end{equation}
with $H_\text{PH}$ given by Eq.~\eqref{eq_HPH} and $C_\text{RH}$ as introduced below Eq.~\eqref{eq_kPHkRH}. The Hubble rate at $a_\text{RH}$ is approximately equal to the decay rate of the non-relativistic particles of the Higgs sector, $ H_\text{RH} \simeq  0.58 \; \Gamma_S^0$, cf.\ Eq.~\eqref{eq:GammaSN1} and Appendix~\ref{app:scalefactor}.
For the parameter point quoted below, cf.\ Eq.~\eqref{eq:exampleparameterpoint}, this implies $a_\text{PH}/a_\text{RH} \simeq 2 \times 10^{-6}$.
Using the Friedmann equation and $\alpha_\text{RH}$ as defined in Sec.~\ref{sec:inflation}, this implies
\begin{equation}
 T_\text{RH} = 0.76 \; \alpha_\text{RH}^{-1/4} \left( \frac{90}{\pi^2 g_*^\text{RH}} \right)^{1/4} \sqrt{ \Gamma_S M_\text{Pl}} \,.
\end{equation}
With this, the positions and amplitudes of the peaks in the GW spectrum associated with preheating are given by
\begin{small}
\begin{align}
  f_{\text{PH}}^{(s)} & \simeq 6.3 \times 10^6 \text{ Hz }  \left(  \frac{10^{-11} M_1}{\text{ GeV}} \right)^{1/3} \left(\frac{5~\text{ GeV}}{10^{-15}\, v_{B-L}} \right)^{2}  \left( \frac{10^{-13} \, m_S}{3~\text{ GeV}} \right)^{7/6} ,  \nonumber \\
\Omega_\text{GW}^{(s)} \left(f_{\text{PH}}^{(s)}\right) h^2 & \simeq 3.6 \times 10^{-16}\  \frac{c_{\text{PH}}}{0.05} 
\left(  \frac{10^{-11} M_1}{\text{ GeV}} \right)^{4/3} \left(\frac{5~\text{ GeV}}{10^{-15}\, v_{B-L}} \right)^{-2}  \left( \frac{10^{-13} \, m_S}{3~\text{ GeV}} \right)^{-4/3} , \nonumber \\
  f_{\text{PH}}^{(v)} & \simeq 7.5 \times 10^{10} \text{ Hz }  g \left(  \frac{10^{-11} M_1}{\text{ GeV}} \right)^{1/3} \left( \frac{10^{-13} \, m_S}{3~\text{ GeV}} \right)^{-1/2} , \nonumber \\
\Omega_\text{GW}^{(v)} \left(f_{\text{PH}}^{(v)} \right) h^2 & \simeq 2.6 \times 10^{-24} \frac{1}{g^2} \, \frac{c_{\text{PH}}}{0.05} \left( \frac{10^{-11} M_1}{\text{ GeV}} \right)^{4/3} \left(\frac{5~\text{ GeV}}{10^{-15}\, v_{B-L}} \right)^{2}  \left( \frac{10^{-13} \, m_S}{3~\text{ GeV}} \right)^{2}  ,
\label{eq_pred_preheating}
\end{align}
\end{small}
\hspace{-8pt} where we have used $C_\text{RH} = 1.13$ and $R=0.41$, cf.\ Sec.~\ref{sec:inflation},  and set $D = 2$ and $\alpha_\text{RH} = 2$, cf.\ Appendix~\ref{app:temperatures}.
Note that $f_{\text{PH}}^{(s)}$ and $f_{\text{PH}}^{(v)}$ are related to
microscopic quantities of the preheating process and are therefore much larger
than the frequency $f_{\text{PH}}$, cf. Eq.~\eqref{eq:fPH}, which is
the relevant quantity for inflation, determined by the Hubble parameter
at preheating.

In Fig.~\ref{fig_infl_vs_cs}, we depict the peaks of the GW spectrum due to preheating (in red) for $c_\text{PH} = 0.05$ and a typical parameter point,
\begin{equation}
  g^2 = 1/2\,, \quad v_{B-L} = 5 \times 10^{15}~\text{GeV} \,,  \quad M_1 = 10^{11}~\text{GeV}\,, \quad  m_S = 3 \times 10^{13}~\text{GeV} \,,
\label{eq:exampleparameterpoint}
\end{equation}
together with the contributions from inflation (cf.\ Sec.~\ref{sec:inflation}) and from AH cosmic strings (cf.\ Sec.~\ref{sec:cosmicstrings}). The frequencies and corresponding amplitudes of the two peaks are given by Eq.~\eqref{eq_pred_preheating}. The shape of the peaks in Fig.~\ref{fig_infl_vs_cs} is parametrized by
\begin{equation}
\Omega^{(i)}_\text{GW} h^2 = \Omega_\text{GW}^{(i)}\left(f_{\text{PH}}^{(i)}\right) \times \left( \frac{f}{f_{\text{PH}}^{(i)}}\right)^2 \text{Exp}\left[1 - \left(f/f_{\text{PH}}^{(i)}\right )^2 \right] \,,
\label{eq_preheating_fit}
\end{equation}
with $i = s,v$; this is motivated by the results found in Ref.~\cite{GarciaBellido:2007dg} for the scalar peak. We do stress however that for the purpose of this paper, we are mainly interested in the position of the peaks. A precise quantitative description of the shape of the spectrum at these frequencies, in particular for the peak corresponding to the vector boson, requires a more detailed study.

\section{GWs from the Cosmic String Network \texorpdfstring{\\}{break} in the Abelian Higgs Model}\label{sec:cosmicstrings}

So far, we have discussed the gravitational wave spectrum due to inflation and due to the processes accompanying tachyonic preheating. We now discuss a third source, the emission of GWs from cosmic strings in the scaling regime. Cosmic strings are produced during the \BmL breaking phase transition ending inflation. As the universe evolves, the cosmic string network enters into a scaling regime, i.e.\ the characteristic scale of the string network remains constant relative to the horizon size $H^{-1}$. This entails a constant fraction of energy stored in the cosmic string network and thus a
continuous emission of energy, which occurs at least partly in the form of GWs.
In this section, we review the calculation of the resulting GW background in the Abelian Higgs (AH) model following Ref.~\cite{Figueroa:2012kw}. An alternative approach, based on the Nambu-Goto (NG) model of cosmic strings will be discussed in Sec.~\ref{sec:comparison}.

Starting point of this discussion is Eq.~\eqref{eq:GWsource}, which, for a classical, sub-horizon source, connects the GW spectrum to the unequal time correlator of the source. For a scaling network of cosmic strings in the AH model, we can now proceed and evaluate this expression by exploiting general properties of the unequal time correlator of a scaling source discussed in Ref.~\cite{Durrer:1998rw}. Introducing the dimensionless variable $x = k \tau$, one can express $\Pi^2(k, \tau, \tau')$ as
\begin{equation}
\Pi^2(k, \tau, \tau') = \frac{4 v_{B-L}^4}{\sqrt{\tau \tau'}} C^T(x, x')\ , 
\label{eq_Pi_AH}
\end{equation}
where $C^T(x, x')$ is essentially local in time \cite{Durrer:1998rw},
\begin{equation}
C^T(x,  x') \sim \delta(x - x') \widetilde C(x) \ .
\label{eq_C_AH}
\end{equation}
As the numerical simulation in Ref.~\cite{Durrer:1998rw} shows,  $\widetilde C$ is a function that falls off rapidly for $x \gg 1$, i.e.\ for modes well inside the horizon. Inserting Eqs.~\eqref{eq_Pi_AH} and \eqref{eq_C_AH} into Eq.~\eqref{eq:GWsource} 
yields
\begin{equation}
 \Omega_\text{GW}(k) =  \frac{ k^2}{3 \pi^2 H_0^2 a_0^2} 
\left(\frac{v_{B-L}}{M_{\rm Pl}}\right)^4 \int_{x_i}^{x_0} dx \frac{a^2(x/k)}{a_0^2 \, x} \widetilde C(x) \,. 
\label{eq_OmegaGW_2}
\end{equation}
As a result of the rapid decrease of $\widetilde C(x)$ for $x \gg 1$, this integral is dominated by its lower bound and basically insensitive to the upper bound for $x_0 \gg x_i \gtrsim 1$,
\begin{equation}
 \int_{x_i}^{x_0} dx \frac{a^2(x/k)}{a_0^2 \, x} \widetilde C(x) \simeq \int_{x_i}^{\infty} dx \frac{a^2(x/k)}{a_0^2 \,  x} \widetilde C(x) \ .
\label{eq_intC}
\end{equation}
For scales which entered the Hubble horizon after the $B$$-$$L$ phase transition, $x_i = k \, \tau_k$ is an ${\cal O}(1)$ constant. 
Hence, the $k$-dependence of Eq.~\eqref{eq_intC} can be traced back to $a(x/k)$. For radiation domination, we have $a(\tau) \simeq \sqrt{\Omega_r} H_0 \tau a_0^2$, where we have neglected the change in the effective number of DOFs.
This yields
\begin{align}
\int_{x_i}^{\infty} dx \frac{a^2(x/k)}{a_0^2 \, x} \widetilde C(x) 
\simeq \frac{\Omega_r H_0^2 a_0^2}{2 k^2} F^r \ ,
\end{align}
where $F^r$ is a constant, and therefore a flat spectrum, 
$\Omega_\text{GW} \propto k^0$. On the other hand,
for matter domination one has $a(x/k) \propto k^{-2}$, which yields $\Omega_\text{GW} \propto k^{-2}$. 

For scales which entered the horizon at very early times before the cosmic string network reached its scaling regime, the lower boundary in Eq.~\eqref{eq_intC} refers to the onset of scaling, thus becoming larger than one would expect without taking this effect into account. Note that $x_i$ is now also $k$-dependent. The qualitative effect of this is a suppression of the spectrum at these frequencies. However, in the model we are discussing here we expect scaling to set in well before the end of reheating, and hence this effect only influences the spectrum at very large frequencies which are currently experimentally inaccessible. We will thus omit it in the following discussion.

In summary, we can express today's spectrum of GWs from a scaling network of cosmic strings as\footnote{Note that in Eq.~\eqref{eq_masterformula}, the normalization of the `$1/k^2$-flanks' was obtained by matching to the plateau value for $k = k_{\text{RH}}$ and $k = k_{\text{eq}}$, respectively. However, since close to these points the dominant component is not much larger than the other components, a more detailed knowledge of $\widetilde C(x)$ is necessary to determine the spectrum at these points. This could lead to a slight shift in the normalization of the `flanks', see also Eq.~\eqref{eq:OmegaGWinfresult}.  \label{fn_tildeC}}
\begin{equation}
\Omega_\text{GW}(k) \simeq  \Omega_\text{GW}^\text{pl} 
\left\{\begin{array}{cl} 
(k_{\rm eq}/k)^2, & \quad k_0 < k < k_{\rm eq} \\ 
1, & \quad k_{\rm eq} < k < k_{\rm RH}  \\
(k_{\rm RH}/k)^2 , & \quad  k_{\rm RH} < k  < k_{\text{PH}} \end{array}\right. \ .
\label{eq_masterformula}
\end{equation}
Here $k_{\rm eq}$, $k_{\rm RH}$ and $k_\text{PH}$ are given by Eqs.~\eqref{eq:keq}, \eqref{eq:kRH} and \eqref{eq:fPH}, and the amplitude of the plateau $\Omega_\text{GW}^\text{pl}$ can be estimated using
the result of the numerical simulations in Ref.~\cite{Figueroa:2012kw},
\begin{equation}
\begin{split}
\Omega_\text{GW}^\text{pl} h^2 &= \frac{1}{6\pi^2} F^r 
\left(\frac{v_{B-L}}{M_{\rm Pl}}\right)^4 \Omega_r h^2 \\
&= 4.0 \times 10^{-14} \frac{F^r}{F^r_{\text{FHU}}} 
\left( \frac{v_{B-L}}{5 \times 10^{15} \text{GeV}} \right)^4 
\left( \frac{\Omega_r h^2} {4.2 \times 10^{-5}} \right) \,,
\end{split}
\label{eq_omega_plat}
\end{equation}
where $F^r_{\text{FHU}} = 4.0\times 10^3$ is the numerical constant determined
in Ref.~\cite{Figueroa:2012kw} for global cosmic strings. The corresponding
constant for local strings is expected to have the same order of magnitude
\cite{hindmarsh}.
\begin{figure}[t!]
\begin{center}
\includegraphics[width=0.8\textwidth]{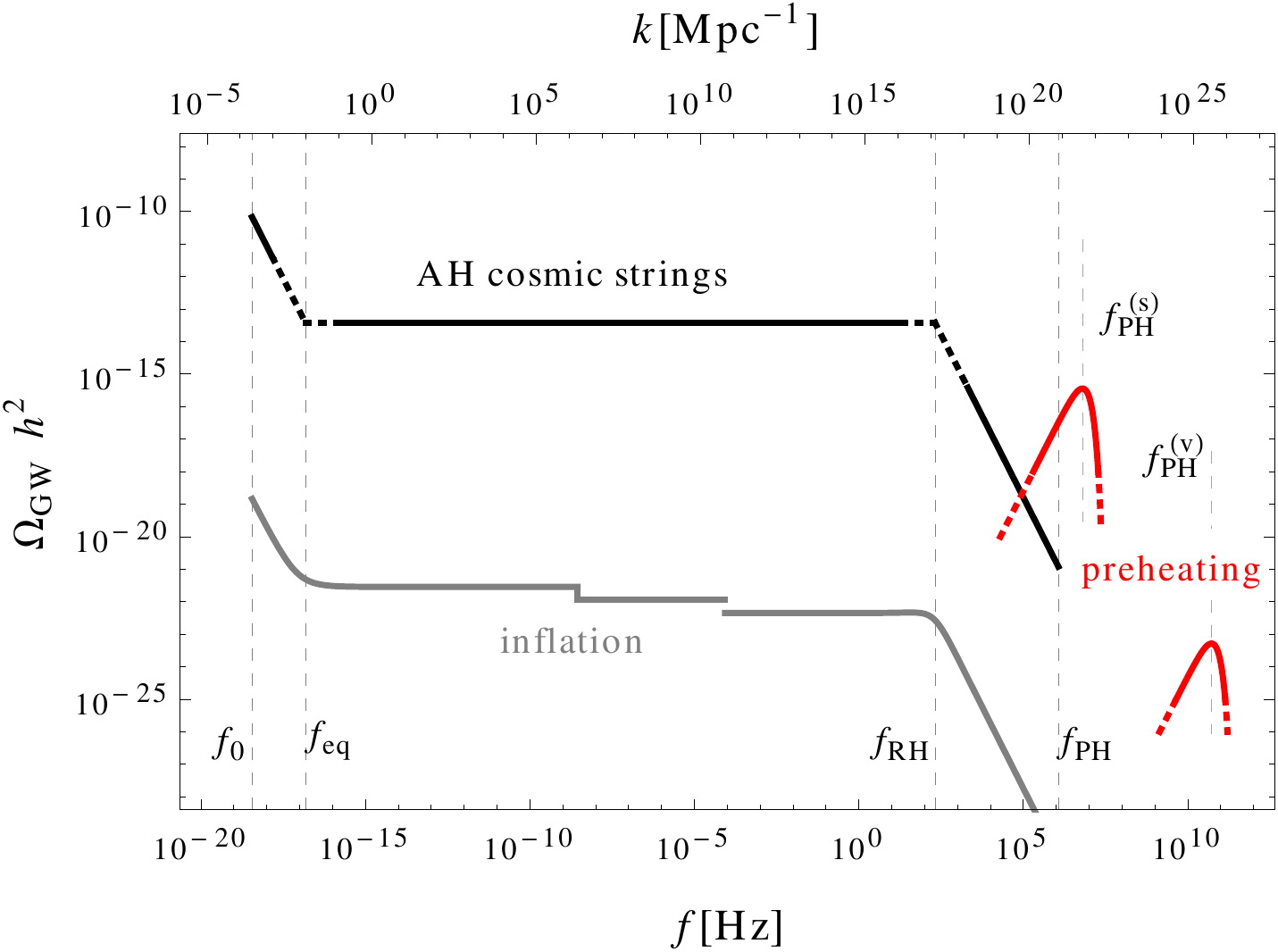}
\caption{GW spectrum today due to inflation (gray), preheating (red) and  AH cosmic strings (black) for $M_1 = 10^{11}$~GeV, $v_{B-L} = 5 \times 10^{15}$~GeV and $m_S = 3 \times 10^{13}$~GeV. $f_0$, $f_\text{eq}$, $f_\text{RH}$ and $f_\text{PH}$ denote the frequencies associated with a horizon sized wave today, at matter-radiation equality, at reheating and at preheating, respectively. $f_{\text{PH}}^{(s)}$ and $f_{\text{PH}}^{(v)}$ denote the positions of the peaks in the GW spectrum associated with the scalar and the vector boson present at preheating. The dashed segments in the spectrum indicate the uncertainties due to the breakdown of the analytical approximations. The GW spectrum from inflation is based on the analytical approximations as well as the numerical values for the transfer function, cf.\ Eqs.~\eqref{eq:OmegaGWinfresult}, \eqref{eq:T1T1result}, and~\eqref{transferfktT2}; the `steps' in the plateau are determined by the changes in the number of DOFs at the QCD scale and at a SUSY scale of 1~TeV. The GW spectrum from preheating is given by Eqs.~\eqref{eq_pred_preheating} and \eqref{eq_preheating_fit}, with $c_\text{PH} = 0.05$ and $g^2 = 1/2$. The GW spectrum from AH cosmic strings is determined by Eqs.~\eqref{eq_masterformula} and \eqref{eq_omega_plat}, with $F^r = F^r_\text{FHU}$.
 } \label{fig_infl_vs_cs}
\end{center}
\end{figure}

Eq.~\eqref{eq_masterformula} strikingly resembles the result found for the stochastic GW background from inflation, cf.\ Eq.~\eqref{eq:OmegaGWinfresult}, up to an overall normalization factor, cf.\ Fig.~\ref{fig_infl_vs_cs}. Note, however, that the origin is quite different. In the case of inflation, the GWs can be traced back to vacuum fluctuations of the space-time metric which remain frozen outside the horizon. After horizon re-entry, they propagate according to the source-free wave equation in FRW space. The amplitude of the resulting stochastic background today is determined only by the redshift of these modes after entering the horizon. On the other hand, the GWs from cosmic strings stem from a classical source, which is active until today. Only the nature of the unequal time correlator, with its rapid decrease for $x \gg 1$, effectively removes the impact of the source when the corresponding mode is well within the horizon. In more physical terms, this implies that the dominant source for GWs from 
AH cosmic 
strings are Hubble-sized structures of the cosmic string network. This explains why the wavenumbers associated with the horizon at $a_{\text{RH}}$ and $a_{\text{eq}}$ play crucial roles in the GW spectrum from cosmic strings, although the GW modes associated with cosmic strings never actually `cross' the horizon.
For cosmic strings the height of the plateau is enhanced by a very large 
numerical factor. On the contrary, GWs from inflation are suppressed by a
small Yukawa coupling. This explains the enormous difference in amplitude
between GWs from inflation and cosmic strings. 

Note that, contrary to inflation, the height of the plateau for the GW background from cosmic strings does not directly translate into a tensor contribution to the CMB scalar power spectrum. This can be traced back to the very different mechanism responsible for generating GWs in inflation and from cosmic strings, in particular concerning correlation properties on super-horizon scales. Determining the effect of GWs from cosmic strings on the CMB requires a specific numerical simulation~\cite{Hindmarsh:2011qj, Ade:2013xla}.

\section{GWs from Cosmic String Loops \texorpdfstring{\\}{break} in the Nambu-Goto Model}\label{sec:comparison}

Calculating the spectrum of gravitational waves produced by cosmic strings is a very difficult task due to the huge range of length and energy scales  involved \cite{Vilenkintextbook, Hindmarsh:2011qj}. A lot of important work has been done  in the case of global as well as local strings in both the Abelian Higgs (AH) model and the Nambu-Goto (NG) model. The former treats cosmic strings in a field theory setup and is based on solving the equations of motion of a $U(1)$ field theory. In this case the string width is found to be governed by the Higgs mass. The latter on the other hand approximates cosmic strings as one-dimensional objects, i.e.\ as infinitely thin strings. Numerical simulations have been carried out for both the AH~\cite{Vincent:1997cx, Moore:2001px, Hindmarsh:2008dw, Bevis:2006mj} and the NG model~\cite{Albrecht:1989mk, Vanchurin:2005pa, Olum:2006ix, BlancoPillado:2011dq} for the initial phase after string formation. In the AH setup, it has been shown how AH strings approach the scaling regime by 
radiating scalar and vector bosons. However, the time range over which AH strings can be simulated is limited, and predictions for the time evolution of the network depend on additional assumptions. On the contrary, in the NG setup, the scaling regime is reached by the formation of string loops, which lose energy and shrink by emitting GWs. Consequently, the resulting network of Hubble-sized cosmic strings is similar in AH and NG simulations (leading to similar predictions for anisotropies in the CMB), but there are serious discrepancies and a number of unresolved problems concerning the production of cosmic strings, the time evolution of the string network and the dominant energy loss mechanism of cosmic strings. 

In particular the last point has quite dramatic consequences for the expected GW background. Since, by assumption, NG strings do not radiate massive or massless particles, their contribution to the GW background is much larger than the contribution from AH strings. After having studied the GW background obtained from the cosmic string network
in Sec.~\ref{sec:cosmicstrings}, we will now turn to the additional contribution due to the decay of cosmic string loops in the NG model.\footnote{In particular, we focus here on GWs from `cusps' on cosmic string loops, which form the dominant contribution for NG strings. Further, subdominant contributions are expected from `kinks' on cosmic string loops~\cite{Damour:2001bk, Siemens:2006yp}, kinks on infinite strings~\cite{Kawasaki:2010yi} as well as from the string network.}
One may expect that sufficiently long strings, whose width is small compared to the curvature radius, are described by the NG action for elementary strings. Comparing both results will yield an estimate of the theoretical uncertainties involved.

The theory of GW emission by NG strings has 
been developed in Refs.~\cite{Damour:2001bk, Siemens:2006yp}.
In the following we extend the discussion in Ref.~\cite{Kuroyanagi:2012wm}
to our model.
The time evolution of the NG string network is determined by the three 
parameters $\Gamma$, $\alpha$ and $p$ \cite{Hindmarsh:2011qj}. They
describe the total energy loss rate
\begin{equation} 
\frac{d}{dt}E = \Gamma G\mu^2 \ ,
\end{equation}
with $\Gamma = \mathcal{O}(50)$, 
the average size of loops at the time $t_i$ of formation, characterized 
by the parameter $\alpha$, which yields for the total length at time $t$,
\begin{equation}
l(t,t_i) = \alpha t_i - \Gamma G \mu (t - t_i) \ ,
\end{equation}
and the reconnection probability $p$ of crossing strings, with $p=1$
for field theory strings.
The lifetime of a loop formed at time $t_i$ is 
$\tau_\text{loop}(t_i) = t|_{l=0} - t_i = \alpha t_i/(\Gamma G \mu)$. We shall
restrict our discussion to the case of short-lived string loops, with  
$\tau_\text{loop}(t_i) \ll t_i$, i.e. $\alpha \ll \Gamma G \mu$. 

The value of the parameter $\alpha$, i.e., the size of string loops at 
formation, is still a matter of debate 
and in the literature values ranging from $0.1$ to $10^{-16}$ are considered
(see, e.g., Refs.~\cite{Hindmarsh:2011qj,Kuroyanagi:2012wm,Lorenz:2010sm,BlancoPillado:2011dq}).
Following Refs.~\cite{Hindmarsh:2011qj,Kuroyanagi:2012wm}, we treat 
$\alpha$ as a free parameter. As we shall see in Sec.~9, the rather large 
values found 
in the numerical simulations in Ref.~\cite{BlancoPillado:2011dq} would be 
inconsistent with constraints from millisecond pulsar timing measurements for
the model under consideration. 

Given $\alpha$ and assuming
that in the scaling regime the string network only loses energy by
gravitational radiation, one can derive an expression for
the rate $R$ at which an
observer detects GWs with frequency $f$ and amplitude $h$, emitted at
redshift $z$ \cite{Kuroyanagi:2012wm},
\begin{equation} \label{eq:d2rdzdh}
\frac{d^2 R}{dz dh}(f,h,z) \simeq \frac{3}{4} 
\frac{\theta_m^2}{(1+z)(\alpha+\Gamma G \mu)h}\frac{1}{\alpha \gamma^2 t^4(z)}
\frac{dV(z)}{dz}\Theta(1-\theta_m) \ . 
\end{equation}
Here the beaming angle $\theta_m$ of the GW burst depends on the loop size 
$l$ and the GW amplitude $h$,
\begin{align}
\theta_m(f,l,z) & = ((1+z)fl)^{-1/3}\ , \label{angle}\\
l(f,h,z) &= 
\left((1+z)^{1/3}\frac{h r(z) f^{4/3}}{\kappa G\mu}\right)^{3/2}\ , \label{length}\\ 
h(f,l,z) &=\frac{\kappa G\mu l}{((1+z)fl)^{1/3} f r (z)} \label{amplitude}\ ,
\end{align}
with $\kappa \simeq 2.7$ a numerical constant; $\gamma = \xi/t$ is the ratio
of the correlation length $\xi \propto H^{-1}$ of the string network divided by the cosmic time, with
$\gamma_r^2 \simeq 0.1$ and $\gamma_m^2 \simeq 0.3$ for radiation domination and
matter domination, respectively, cf.\ Ref.~\cite{Kuroyanagi:2012wm}. The step function provides a
low-frequency cutoff for the emitted burst, $f > 1/(l(1+z))$.
The distance $r(z)$ to the emitting string loop, 
the time $t(z)$ of emission and the change of volume with redshift are
given by
\begin{align}
r(z) &=\int_0^z \frac{dz'}{H(z')} = \frac{1}{H_0}\varphi_r(z) \ , \label{eq_phi_r}\\
t(z) &=\int_z^\infty \frac{dz'}{(1+z')H(z')} = \frac{1}{H_0} \varphi_t(z)\ , \label{eq_phi_t}\\
\frac{dV}{dz} &= \frac{1}{H_0^3} \varphi_V(z)\ , \quad 
\varphi_V(z) = \frac{4 \pi\varphi_r^2(z)}{(1+z)^3} \frac{H_0}{H(z)} \ .
\end{align}

The spectral amplitude of the GW background is now obtained by
integrating the rate over all redshifts and GW amplitudes \cite{Kuroyanagi:2012wm},
\begin{align}\label{eq:theintegral}
\Omega_{\rm GW}(f) = \frac{2 \pi^2 f^3}{3 H_0^2} 
\int_{0}^{h_*} dh \int_0^{z_{\rm PH}} dz  h^2 \frac{d^2R}{dz dh}\ . 
\end{align}
Here $z_{\rm PH}$ is the maximal redshift given by the time of preheating.
Amplitudes smaller than $h_*$ corresponds to bursts that are detected 
within time intervals smaller than their oscillation period. Bursts with
larger amplitudes cannot be resolved and are therefore not counted
\cite{Siemens:2006yp}. The maximal GW amplitude depends on the frequency and 
is defined by the condition
\begin{align}\label{eq:hmax}
\int_{h_*}^{\infty} dh \int_0^{z_{\rm PH}} dz  \frac{d^2R}{dz dh} = f\ .
\end{align}

In order to understand the parameter dependence of the GW spectrum it is
instructive to evaluate the integral in Eq.~\eqref{eq:theintegral} approximately
analytically. Details of this calculation are described in Appendix~B. 
In the different epochs after preheating, i.e.\ early matter domination, 
radiation domination, matter domination and $\Lambda$-domination, bounded
by the redshifts $z_{\rm PH} > z_{\rm RH} > z_{\rm eq} > z_{\Lambda} > 0$,
one easily finds approximate expressions for the integrand, starting from
\begin{equation}\label{eq:hzdep}
\frac{H(z)}{H_0} \simeq 
\begin{cases}
\sqrt{\Omega_\Lambda +\Omega_{m}(1+z)^3+\Omega_{r}(1+z)^4} \ , 
\quad 0<z<z_{\rm RH} \\
(H_{\rm RH}/H_0) ((1+z)/(1+z_{\rm RH}))^{3/2}\ , \quad z_\text{RH} < z < z_{\rm PH}
\end{cases} \ ,
\end{equation}
where we have neglected the change in the DOFs.
For each epoch
the integration has to be carried out in a range $z_c < z < z_m$, which is
determined by the requirement to have small enough string loops and 
large enough frequencies: $l(f,h,z) \leq \alpha t(z)$ and
$\theta_m(f,l,z) \leq 1$. The inequalities are saturated at $z_c$ and
$z_m$, respectively. For a given frequency $f$, the restricted range in
$z$ leads to lower and upper bounds on the amplitude $h$. Taking also
the general upper bound $h_*$ into account, one obtains for each epoch
a frequency range in which GWs are emitted: $f_c < f < f_1$ for 
$\Lambda$-domination, $f_c < f < f_2$ for matter domination, 
$f_{\rm eq}^{(\text{NG})} < f < f_3$ for radiation domination and    
$f_3 < f < f_{\rm PH}^{(\text{NG})}$ for the first matter dominated phase.
\begin{figure}
\begin{center}
\includegraphics[width = 0.7\textwidth]{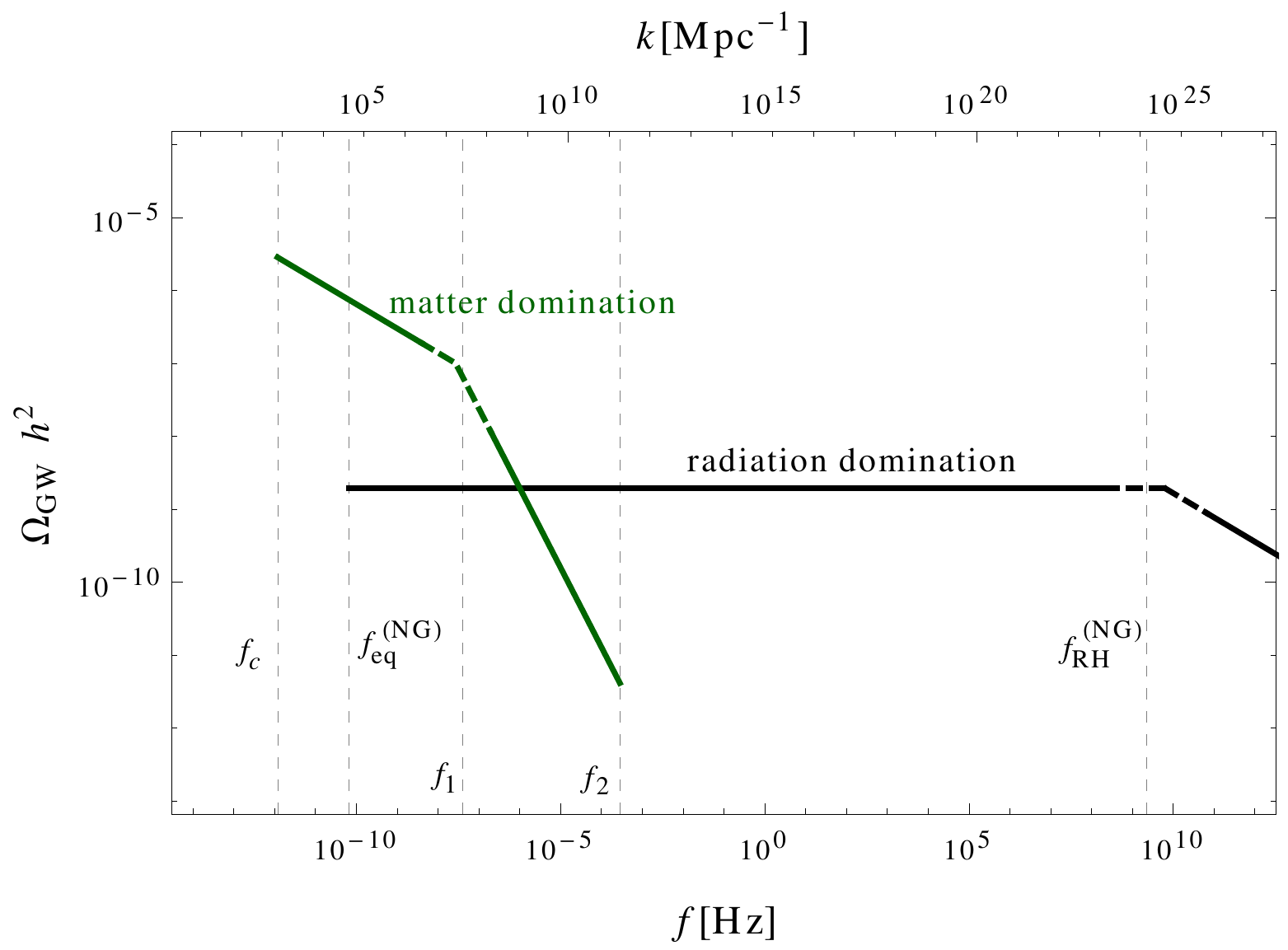}
\caption{Contributions from different epochs to the GW spectrum for NG 
strings for $\alpha = 10^{-6}$; $G\mu = 2.0\times 10^{-7}$ as obtained e.g.\ by the parameter choice in Eq.~\eqref{eq:exampleparameterpoint}. $f_c$, $f_\text{eq}^\text{(NG)}$, $f_1$, $f_2$ and $f_\text{RH}^{(\text{NG})}$ mark the bounding values of the intervals calculated in Eq.~\eqref{eq:theresult}. $f_3 \simeq 2.5 \times 10^{39}$~Hz is outside the physically relevant frequency range.
}
\label{fig:NGstring10-6}
\end{center}
\end{figure}
The frequencies $f_i$ satisfy the relations ($\bar{f}_i = \alpha f_i/H_0$),
\begin{equation}
\begin{split}
\bar{f}_c = \Omega_m^{1/2} &< \bar{f}_{\rm eq}^{(\text{NG})} = 
\frac{\Omega_m}{\Omega_r^{1/2}}
< \bar{f}_1 = \zeta\ \Omega_m^{11/10} 
< \bar{f}_2 = \zeta\ \frac{\Omega_m^{11/5}}{\Omega_r^{11/10}} \\
& < \bar{f}_{\rm RH}^{(\text{NG})} = \Omega_r^{1/2} (1+z_{\rm RH})
< \bar{f}_3 = \zeta\  \Omega_r^{11/10} (1 + z_\text{RH})^{11/5} \ ,
\end{split}
\end{equation}
where
\begin{align}
\zeta = \left(\frac{3\pi^2 \varphi_0^2}{\gamma^2\Gamma G\mu}\right)^{3/5} \ ,
\end{align}
and the constant $\varphi_0$ is defined in Eq.~\eqref{eq:phi0}.
The frequencies $f_c$, $f_{\rm eq}^{(\text{NG})}$ and $f_{\rm RH}^{(\text{NG})}$ essentially agree, up to a
factor $1/\alpha$, with the corresponding frequencies defined in
Sec.~\ref{sec:inflation}
(cf.~Eqs.~\eqref{eq:keq} and \eqref{eq:kRH})\footnote{For simplicity, we
  have neglected  small changes of the equation of state during the
  various epochs in this section. The difference between $f_c$ and
  $f_0$ is due to the approximation \eqref{approx3} for $\varphi_t(z)$.}.  

Using the described approximations for the integrand $d^2R/(dzdh)$,
one obtains after a straightforward calculation, cf.\ Appendix~\ref{app:GWBNG}, the GW spectral amplitude
\begin{small}
\begin{equation}
\frac{\Omega_{\rm GW}(f)}{\Omega_{\rm GW}^{\rm pl}} \simeq 
\begin{cases}
\dfrac{\gamma_r^2}{\gamma_{\rm
    m}^2}\left(\dfrac{1}{5}\dfrac{\Omega_{m}^{5/3}}{\Omega_{\rm
      r}} +\dfrac{32}{25}\dfrac{\Omega_{m}^{7/6}}{\Omega_{\rm
      r}}\right)\left(\dfrac{H_0}{\alpha f}\right)^{1/3}, & \hspace{3mm}f_c\phantom{^{(\text{NG}}}\hspace{-3mm} < f < f_1\\ 
\dfrac{32}{25} \left(\dfrac{6\pi \varphi_0^{2}}{\gamma_m^2\Gamma G
    \mu}
\right)^\frac{5}{11}\dfrac{\gamma_r^2}{\gamma_m^2}\dfrac{\Omega_{\rm
    m}^2}{\Omega_{r} }\left(\dfrac{H_0}{\alpha f}\right)^\frac{12}{11} ,
& \hspace{3mm}f_1\phantom{^{(\text{NG}}}\hspace{-3mm} < f < f_2  \\
1 \ ,& f_{\rm eq}^{\rm (NG)} < f <  f_{\rm RH}^{\rm (NG)} \\ 
\left(1 + \dfrac{32}{25}\dfrac{\gamma_r^2}{\gamma_m^2}\right)\Omega_{\rm
  r}^{1/6}(1+z_{\rm RH})^{1/3} \left(\dfrac{H_0}{\alpha f}\right)^{1/3}, & 
f_{\rm RH}^{\rm (NG)} < f < f_3 \\ 
\dfrac{32}{25 } \left(\dfrac{6\pi \varphi_0^{2}}{\gamma_m^2\Gamma G
    \mu} \right)^{\hspace{-1mm}\frac{5}{11}}\dfrac{\gamma_r^2}{\gamma_m^2}
\dfrac{\Omega_{m}^2}{\Omega_{r}}(1+z_{\rm RH})^2
\left(\dfrac{H_0}{\alpha f}\right)^\frac{12}{11}, & \hspace{3mm}f_3\phantom{^{(\text{NG}}}\hspace{-3mm} < f \ 
\end{cases} \,,
\label{eq:theresult}
\end{equation}
\end{small}
where the height of the plateau is given by
\begin{equation}
\Omega_{\rm GW}^{\rm pl}h^2 = 
\frac{5\pi^3\kappa^2 G\mu}{\Gamma\gamma_r^2}\Omega_r h^2
= 5 \times 10^{-9} 
\left( \frac{G\mu}{5\times 10^{-7}} \right) 
\left( \frac{\Omega_r h^2} {4.2 \times 10^{-5}} \right) \,.
\end{equation}
The various contributions are displayed in Fig.~\ref{fig:NGstring10-6}.
In the frequency interval  $f_c < f < f_1$ the contribution from matter
domination (second term in the parenthesis) dominates over the one from 
vacuum domination. The frequency intervals $f_1 < f < f_2$ as well as
$f_{\rm eq}^{\rm (NG)} < f < f_{\rm RH}^{\rm (NG)}$ receive contributions only
from a single epoch, matter domination and radiation domination, respectively.
In the frequency interval $f_{\rm RH}^{\rm (NG)} < f < f_3$ radiation domination
and early matter domination contribute, and for $f_3 < f$ only early
matter domination is relevant. For small values of $\alpha$, this last
interval can be neglected since the frequencies exceed $f_{\rm PH}^{(v)}$,
the maximally possible frequency for a vortex of finite width. 

\begin{figure}
\begin{center}
\includegraphics[width = 0.7\textwidth]{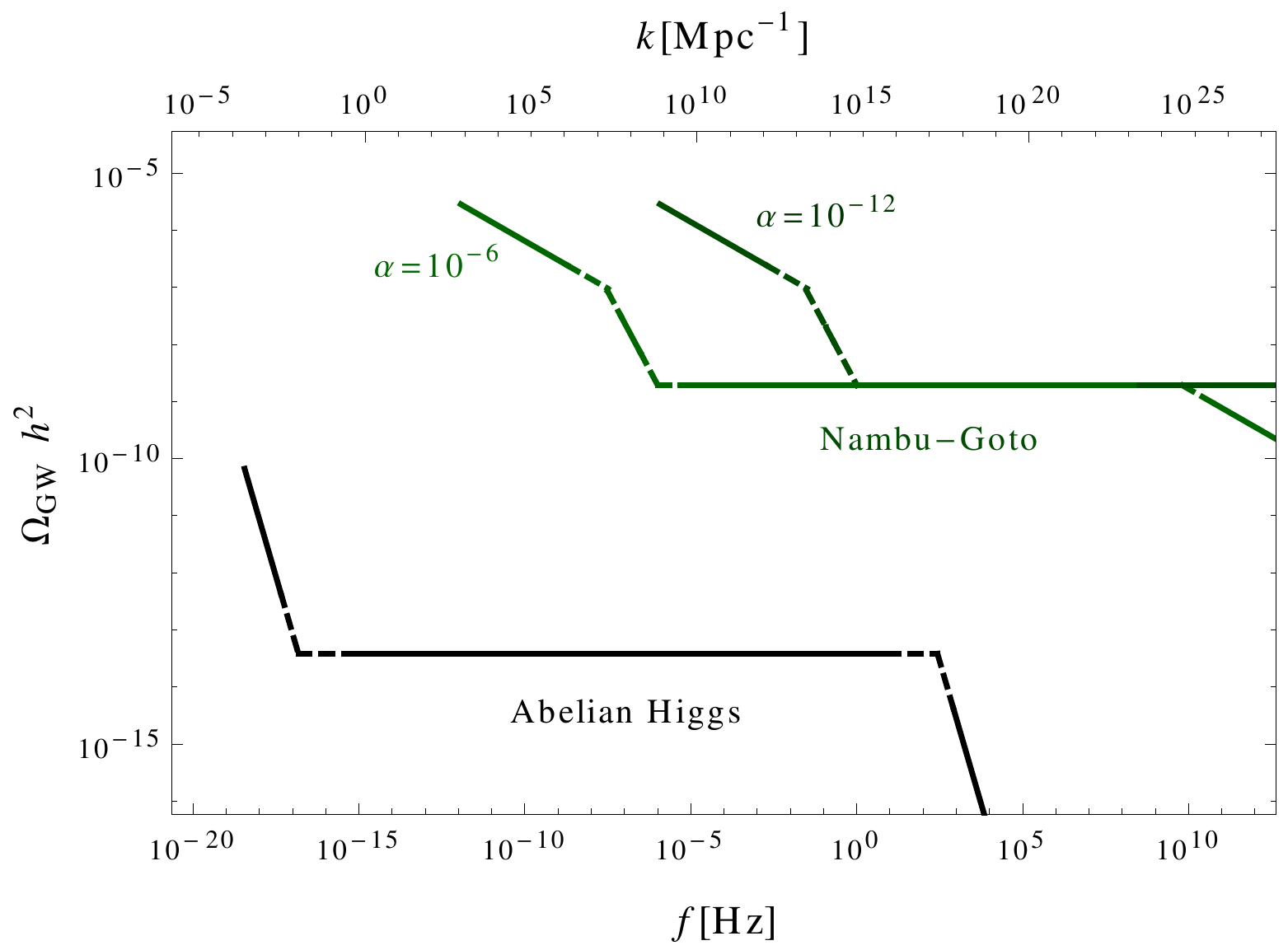}
\caption{Comparison of the GW spectra predicted by AH strings and NG strings
for two values of $\alpha$. The AH curve is obtained as in Fig.~\ref{fig_infl_vs_cs}, for the NG curves
$G\mu = 2.0\times 10^{-7}$, cf.\ Fig.~\ref{fig:NGstring10-6}. }
\label{fig:comparison}
\end{center}
\end{figure}

In Fig.~\ref{fig:comparison} the predictions for the GW spectrum of AH strings
and NG strings are compared for two different values of $\alpha$. In
both approaches the radiation dominated epoch leads to a plateau for
intermediate frequencies. Compared to AH strings the boundaries are
shifted to higher frequencies by a factor $1/\alpha$ for NG
strings. This is due to the fact that the maximal loop size $\alpha
t_i$ is shorter that the horizon size $\sim t_i$ when the string loops
are formed.  Also the frequency dependence for small and
large frequencies is different, again a consequence of the different mechanisms of
gravitational radiation. Most
striking is the difference in normalization by five orders of magnitude!
The reason are the different assumptions on how the string network loses
energy and stays in the scaling regime. Whereas the energy loss of AH
strings is mainly due to massive radiation, NG strings dump all their energy
into GWs. Hence, these two cases provide lower and
upper bounds on the GW background produced by cosmic
strings, and it is conceivable that the true answer corresponds to
intermediate values. To find the answer to this important question is
clearly a theoretical challenge!

%%%%%%%%%%%%%%%%%%%%%%%%%%%%%%%%%%%%%%%%%%%%%%%%%%%%%%%%%%%%%%%%%%%%%%%%%%%%%%

\section{Probing the Temperature during Reheating}
\label{sec:reheatingtemp}

%%%%%%%%%%%%%%%%%%%%%%%%%%%%%%%%%%%%%%%%%%%%%%%%%%%%%%%%%%%%%%%%%%%%%%%%%%%%%%

The transition between matter domination and radiation domination during
reheating may result in a characteristic kink in the GW
spectrum at a frequency $f_{\textrm{RH}}$ (cf.\ Eq.~\eqref{eq_fRH}),
or correspondingly at a wavenumber $k_{\textrm{RH}}$ (cf.\ Eq.~\eqref{eq:kRHTeff}).
As we have seen in Secs.~\ref{sec:inflation}, \ref{sec:cosmicstrings} and \ref{sec:comparison},
at wavenumbers around $k = k_{\textrm{RH}}$, the GW spectrum receives contributions
from the GWs amplified during inflation as well as from the GWs emitted by
cosmic strings.
Here, the stochastic GW background originating from inflation always features
a kink at $k = k_{\textrm{RH}}$.
Its shape is accurately described the transfer function $T_2$ (cf.\ Eq.~\eqref{eq:T1T1result})
which we determined by means of a numerical calculation in Sec.~\ref{sec:inflation}.
The amplitude of the entire stochastic GW background is, however, heavily suppressed
by the small value of the Yukawa coupling $\lambda$ (cf.\ Eq.~\eqref{eq_Omegaplateau_inflation}),
rendering the inflationary signal clearly subdominant to the amplitude of the GWs produced in the
decay of cosmic strings.
Whether or not an observable kink is present in the GW spectrum at
$k = k_{\textrm{RH}}$, hence, depends on the nature of the cosmic strings.

%%%%%%%%%%%%%%%%%%%%%%%%%%%%%%%%%%%%%%%%%%%%%%%%%%%%%%%%%%%%%%%%%%%%%%%%%%%%%%

In Sec.~\ref{sec:cosmicstrings} we found that
the decay of the AH string network induces a GW spectrum which is similar in shape to
the inflationary GW background and which therefore exhibits a kink around $k = k_{\textrm{RH}}$
as well.
As opposed to this, no characteristic feature in the GW spectrum seems to be associated with
the wavenumber $k_{\textrm{RH}}$ in the case of NG strings
(cf.\ Sec.~\ref{sec:comparison} and, in particular, Fig.~\ref{fig:comparison}).
Since the reheating process represents an arguably early stage in the evolution
of the cosmic string network, the AH picture of strings of a non-negligible thickness
likely still applies during reheating.
Solely based on this notion, we would, thus, expect a kink to appear in the
GW spectrum at $k = k_{\textrm{RH}}$.
At late times, cosmic strings are, however, likely best described by the NG model,
which predicts an additional contribution to the GW spectrum due to the gravitational
radiation from cosmic string loops.
As we have seen in the previous section, the GW bursts during the era of radiation
domination are expected
to yield a flat spectrum exceeding the AH contribution by many orders of magnitude
in large regions of parameter space (cf.\ Fig.~\ref{fig:comparison}).
In particular, the kink at $k = k_{\textrm{RH}}$
predicted by the AH model might be obscured by the NG contribution to the GW spectrum.

%%%%%%%%%%%%%%%%%%%%%%%%%%%%%%%%%%%%%%%%%%%%%%%%%%%%%%%%%%%%%%%%%%%%%%%%%%%%%%

In this section, we shall meanwhile assume that owing to a fortunate conjunction of circumstances,
i.e.\ a fortunate interplay of parameter values or some other effect not accounted
for in our analysis, the kink at $k = k_{\textrm{RH}}$ is \textit{not} wiped out due to the
late-time contribution to the GW spectrum from cosmic string loops.
As we expect the GW signal due to AH strings to exceed the inflationary background
by many orders of magnitude, this assumption then leads us to the exciting
observation that future GW experiments
(cf.\ Sec.~\ref{sec:observations}) might in fact be able to observe this kink in the GW
spectrum and hence directly probe the reheating process after inflation.
In this section, we shall now discuss the possible insights into the
reheating process that one might gain from measuring the position of the
kink, i.e.\ from determining the wavenumber $k_{\textrm{RH}}$.

%%%%%%%%%%%%%%%%%%%%%%%%%%%%%%%%%%%%%%%%%%%%%%%%%%%%%%%%%%%%%%%%%%%%%%%%%%%%%%

In scenarios of reheating solely based on the decay of the inflaton $\phi$, 
the wavenumber $k_{\textrm{RH}}$ is typically estimated in terms of the inflaton
decay temperature $T_\phi$~\cite{Nakayama:2008wy},
\begin{align}
k_{\textrm{RH}} \simeq \left(\frac{g_{*,s}^0}{g_{*,s}^{\textrm{RH}}}\right)^{1/3}
\left(\frac{2 g_{*}^{\textrm{RH}}}{g_{*}^0}\right)^{1/2} {\Omega_r}^{1/2}
\frac{T_\phi}{T_{\gamma}^0} \,k_0
\simeq 1.95 \times 10^{14} \,\textrm{Mpc}^{-1} \left(\frac{T_\phi}{10^7 \,\textrm{GeV}}\right) \,,
\label{eq:kRHTphi}
\end{align}
where $T_\phi$ is defined such that at $T = T_\phi$ the Hubble rate $H$ has just dropped
to the value of the inflaton vacuum decay rate $\Gamma_\phi^0$,
\begin{align}
H(a_\phi) = \Gamma_\phi^0 \,,\quad T_\phi = T(a_\phi) \,.
\label{eq:Tphidef}
\end{align}
Studies estimating $k_{\textrm{RH}}$ as in Eq.~\eqref{eq:kRHTphi}
then often refer to $T_\phi$ as the `reheating temperature'.
Likewise, one may also relate $k_{\textrm{RH}}$ to the temperature
at matter-radiation equality during reheating, i.e.\ the temperature that we
refer to as the reheating temperature $T_{\textrm{RH}}$.
In the standard scenario of reheating, the temperature $T_{\textrm{RH}}$ is defined such that
\begin{align}
\rho_{\textrm{tot}} \left(a_{\textrm{RH}}\right) =
2 \,\rho_\phi\left(a_{\textrm{RH}}\right) = 2 \,\rho_r\left(a_{\textrm{RH}}\right) \,,\quad
T_{\textrm{RH}} = T\left(a_{\textrm{RH}}\right) \,.
\label{eq:TRHdefphi}
\end{align}
Consequently, it is only marginally smaller than $T_\phi$ and hence, instead of
using Eq.~\eqref{eq:kRHTphi}, one may equally estimate $k_{\textrm{RH}}$ as
\begin{align}
k_{\textrm{RH}} \simeq \left(\frac{g_{*,s}^0}{g_{*,s}^{\textrm{RH}}}\right)^{1/3}
\left(\frac{2 g_{*}^{\textrm{RH}}}{g_{*}^0}\right)^{1/2} {\Omega_r}^{1/2}
\frac{T_{\textrm{RH}}}{T_{\gamma}^0} \,k_0
\simeq 1.95 \times 10^{14} \,\textrm{Mpc}^{-1}
\left(\frac{T_{\textrm{RH}}}{10^7 \,\textrm{GeV}}\right) \,.
\label{eq:kRHtRH}
\end{align}

%%%%%%%%%%%%%%%%%%%%%%%%%%%%%%%%%%%%%%%%%%%%%%%%%%%%%%%%%%%%%%%%%%%%%%%%%%%%%%

Quantitatively, Eqs.~\eqref{eq:kRHTphi} and \eqref{eq:kRHtRH} surely represent
reasonable estimates of the wavenumber $k_{\textrm{RH}}$.
As apparent from our discussion in Sec.~\ref{sec:inflation}, they,
however, miss some crucial features of the reheating
process on the conceptional side.
In order to obtain a more accurate estimate of $k_{\textrm{RH}}$, one also has to take
into account that at $a = a_\phi$ (or $a = a_{\textrm{RH}}$) only a fraction (only half)
of the total energy is contained in radiation as well as that after $a = a_\phi$
(or $a = a_{\textrm{RH}}$) still some amount of entropy is produced.
This was the reason why, for our scenario of reheating, we introduced the correction
factors $\alpha_{\textrm{RH}}$ and $D$ in Sec.~\ref{sec:inflation}.
Recall that the appearance of the factors $\alpha_{\textrm{RH}}^{1/2}$ and $D^{-1/3}$
in Eq.~\eqref{eq:kRH} then gave rise to an effective kink temperature $T_*$
replacing $T_\phi$ or $T_{\textrm{RH}}$ in our estimate for
$k_{\textrm{RH}}$ (cf.\ Eq.~\eqref{eq:kRHTeff}).
The distinction between these different temperatures is important
since they potentially all exhibit different dependencies on the fundamental
parameters of the underlying particle physics model.
Thus, if one is to deduce information about particle physics parameters
from a measurement of $k_{\textrm{RH}}$, it is of great importance
to know exactly which temperature one is actually probing.

%%%%%%%%%%%%%%%%%%%%%%%%%%%%%%%%%%%%%%%%%%%%%%%%%%%%%%%%%%%%%%%%%%%%%%%%%%%%%%

Having solved the full set of Boltzmann equations governing the reheating
process, we fortunately possess a detailed and time-resolved picture of
the evolution of the radiation temperature during reheating (cf.\ Fig.~\ref{fig:reheat}).
This allows us not only to derive a numerical estimate for $k_{\textrm{RH}}$ that is
preciser than those in  Eqs.~\eqref{eq:kRHTphi} and \eqref{eq:kRHtRH} (cf.\ Eq.~\eqref{eq:kRHTeff}),
but also to pinpoint to which temperature $k_{\textrm{RH}}$ is actually
related conceptionally.
Put differently, we might say:
typically, the reheating process is merely characterized by only one temperature
scale that roughly reflects the temperature evolution during reheating, namely
the reheating temperature $T_{\textrm{RH}}$.
By contrast, we are now able to identify several key temperatures marking crucial points
in the reheating process, which enables us to obtain a much better understanding
of the connection between the kink in the GW spectrum and the temperature of
the thermal bath during reheating.
At this point, it is important to note that the discussion in this section does not rely on
any particular details of our reheating scenario.
Instead, it rather applies to any scenario of reheating that is characterized
by the successive decay of two nonthermal species (corresponding to
$S$ and $N_1$ in our scenario).
The kink in the GW spectrum in the context of a two-stage scenario
of reheating has not been treated in the literature before, which is 
why we dedicate an entire section to it.

%%%%%%%%%%%%%%%%%%%%%%%%%%%%%%%%%%%%%%%%%%%%%%%%%%%%%%%%%%%%%%%%%%%%%%%%%%%%%%

\begin{figure}
\begin{center}
\begin{minipage}{\textwidth}
\centering 
\includegraphics[width=13 cm]{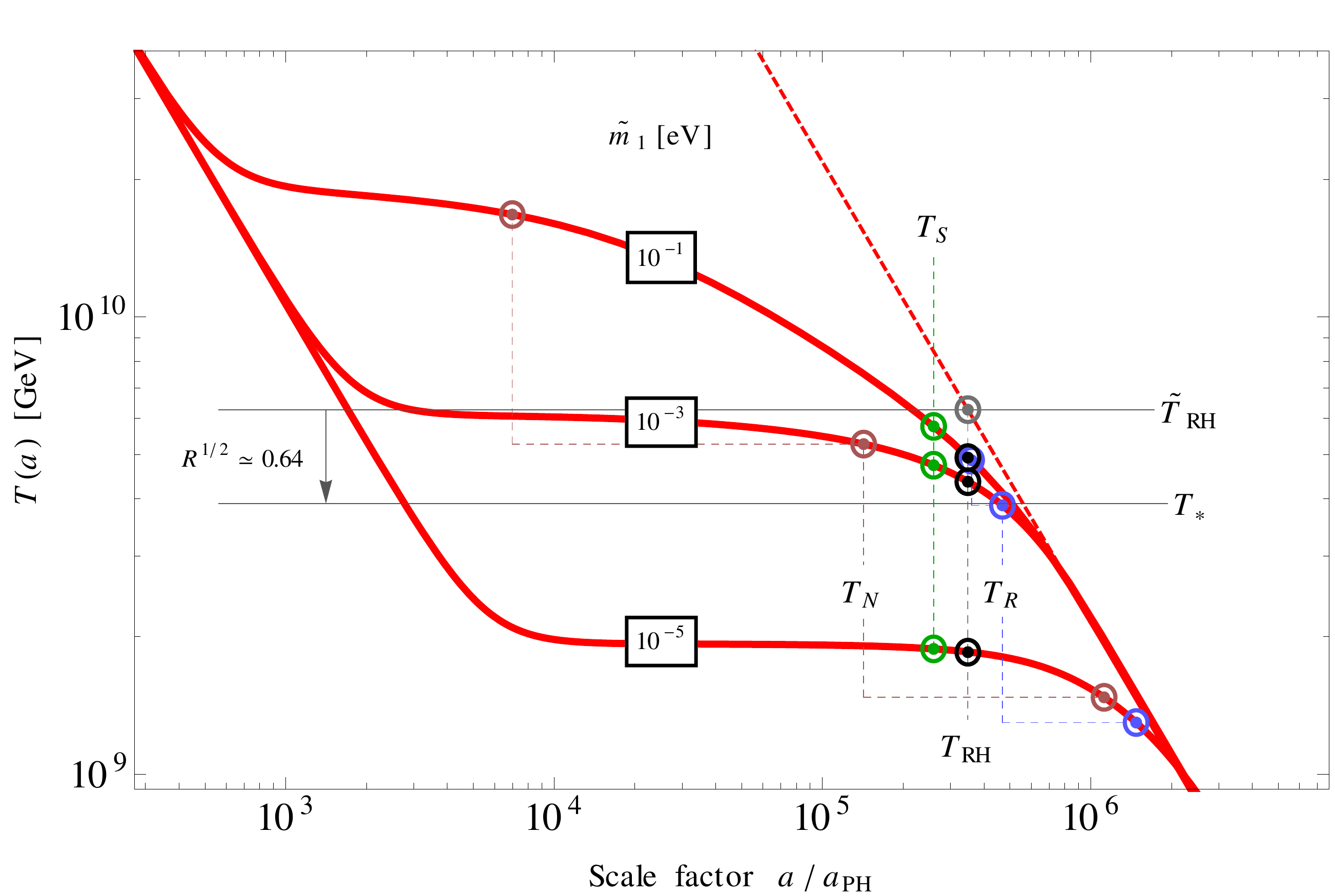}
\includegraphics[width=13 cm]{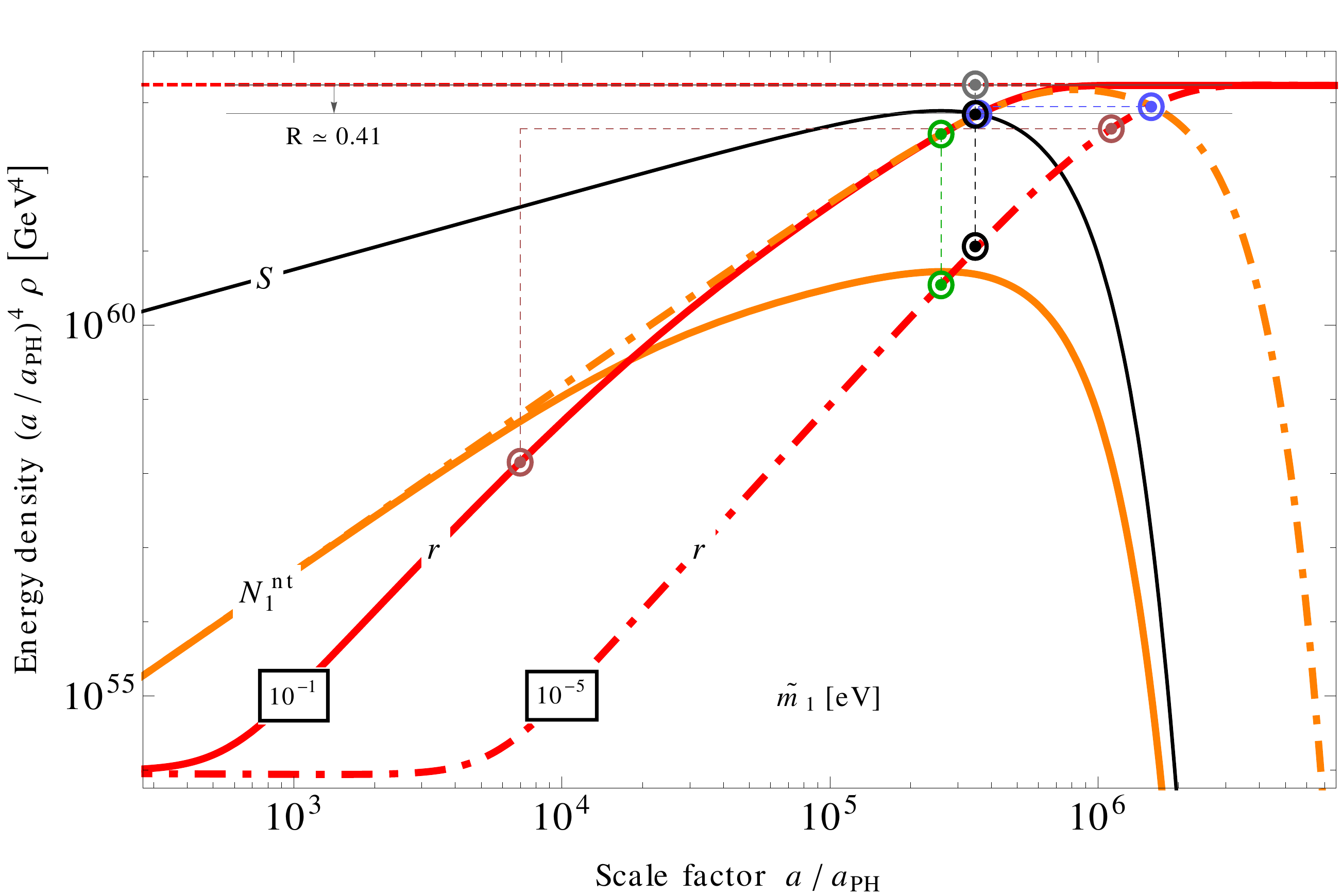}
\caption{Evolution of the radiation temperature \textbf{(upper panel)}
as well as of the \BmL Higgs $(S)$, nonthermal (s)neutrino $(N_1^\text{nt})$ and radiation $(r)$
energy densities \textbf{(lower panel)} as functions of the scale factor $a$
for three different values of $\widetilde{m}_1$, with all other free parameters
kept fixed, $M_1 = m_S/300 = 10^{11}\,\textrm{GeV}$, $v_{B-L} = 5 \times 10^{15}\,\textrm{GeV}$.
The ratio $\Gamma_{N_1}^S(a_\text{RH}) / \Gamma_S^0$ consequently takes the values
$4.4 \times 10^{-2}$, $2.1 \times 10^0$, $1.5 \times 10^2$ for $\widetilde{m}_1 = 10^{-5},\, 10^{-3},\,10^{-1}\,\textrm{eV}$,
respectively.
The colored markers indicate the values of the various benchmark
temperatures defined in Sec.~\ref{sec:reheatingtemp} as labeled in the upper panel.}
\label{fig:Trhocompare}
\end{minipage}
\end{center}
\end{figure}

%%%%%%%%%%%%%%%%%%%%%%%%%%%%%%%%%%%%%%%%%%%%%%%%%%%%%%%%%%%%%%%%%%%%%%%%%%%%%%

Equipped with the three decay rates $\Gamma_S^0$, $\Gamma_{N_1}^0$
and $\Gamma_{N_1}^S$ introduced in Sec.~\ref{sec:model} as well as the energy densities
of all species involved in the reheating process, we are able to
define several benchmark temperatures characterizing the temperature evolution
during reheating: the reheating temperature $T_{\textrm{RH}}$, the would-be reheating
temperature $\widetilde{T}_{\textrm{RH}}$, the effective kink temperature $T_*$,
the neutrino decay temperature $T_N$, the \BmL Higgs decay temperature $T_S$ and
the radiation domination temperature $T_R$.
The three temperatures $T_{\textrm{RH}}$, $\widetilde{T}_{\textrm{RH}}$ and
$T_*$ are closely related to the wavenumber $k_{\textrm{RH}}$, which is why we
devote most of our attention to them in this section.
By contrast, there is no direct connection between $k_{\textrm{RH}}$ and
any of the three remaining temperatures, $T_N$, $T_S$ and $T_R$.
On the other hand, $T_N$, $T_S$ and $T_R$ also represent important benchmark
temperatures, which are all occasionally employed as `the reheating temperature'.
We shall therefore incorporate the temperatures $T_N$, $T_S$ and $T_R$ into our
discussion as well.
Appendix~\ref{app:temperatures} provides more information on them, including their formal definitions.
The differences between all six benchmark temperatures
as well as their dependencies on the effective neutrino mass
$\widetilde{m}_1$ are illustrated in Fig.~\ref{fig:Trhocompare}.
Here, the values of $\widetilde{m}_1$ are chosen
as to show how the hierarchy among the various temperatures changes when
the ratio of the two decay rates $\Gamma_{N_1}^S$ and $\Gamma_S^0$ is varied
between small and large values.

%%%%%%%%%%%%%%%%%%%%%%%%%%%%%%%%%%%%%%%%%%%%%%%%%%%%%%%%%%%%%%%%%%%%%%%%%%%%%%

\medskip\noindent\textbf{Reheating temperature:}
The physical scale corresponding to the wavenumber $k_{\textrm{RH}}$, which marks
the position of the kink in the GW spectrum, enters the Hubble horizon at $a=a_{\textrm{RH}}$,
when only half of the total energy is still accounted for by
non-relativistic \BmL Higgs particles,
while the other half is already contained in relativistic DOFs, i.e.\
heavy nonthermal (s)neutrinos and thermal MSSM radiation
(cf.\ Eq.~\eqref{eq:defaR}),
\begin{align}
\rho_{\textrm{tot}} \left(a_{\textrm{RH}}\right) = 
2 \, \rho_S \left(a_{\textrm{RH}}\right) =
2 \left[\rho_r\left(a_{\textrm{RH}}\right) + \rho_N\left(a_{\textrm{RH}}\right) + \rho_{\tilde{N}}\left(a_{\textrm{RH}}\right)  \right] \,,\quad
T_{\textrm{RH}} = T\left(a_{\textrm{RH}}\right) \,.
\end{align}
In this paper, we refer to the temperature at this time as the reheating
temperature $T_{\textrm{RH}}$.
Note that in our reheating scenario, as opposed to the standard case
(cf.\ Eq.~\eqref{eq:TRHdefphi}), thermal radiation only makes up
a fraction $\alpha_{\textrm{RH}}^{-1} < \frac{1}{2}$ of the total energy at $a = a_{\textrm{RH}}$.

%%%%%%%%%%%%%%%%%%%%%%%%%%%%%%%%%%%%%%%%%%%%%%%%%%%%%%%%%%%%%%%%%%%%%%%%%%%%%%

\medskip\noindent\textbf{Would-be reheating temperature:}
The continuing decay of the \BmL Higgs particles and heavy (s)neutrinos
after $a=a_{\textrm{RH}}$ results in the production of further entropy,
which effectively dilutes the number densities of all species present
during reheating.
This dilution may be quantified in terms of the factor $D$,
which we introduced in Eq.~\eqref{eq_Delta}.
By means of the factor $D$, we are also able to determine the would-be
reheating temperature $\widetilde{T}_{\textrm{RH}} = D^{1/3}\, T_{\textrm{RH}}$
(cf.\ Eq.~\eqref{eq:defTwb}
and the geometrical construction of $\widetilde{T}_{\textrm{RH}}$
in Fig.~\ref{fig:Trhocompare}).
Thanks to our semi-analytical computation of the scale factor as well as
the energy densities of relativistic and non-relativistic DOFs in
Appendix~\ref{app:scalefactor} and employing the Frogatt-Nielsen relation $M_1 = \eta^2 \, m_S$, we are able to compute
$\widetilde{T}_{\textrm{RH}}$ as a function of $M_1$ without having to solve
any Boltzmann equations numerically,
\begin{align}
\widetilde{T}_{\textrm{RH}} \simeq 6.1 \times 10^9 \,\textrm{GeV}
\left(\frac{M_1}{10^{11}\,\textrm{GeV}}\right)^{1.5} \,.
\label{eq:Twbres}
\end{align}
As opposed to the actual reheating temperature, the would-be reheating
temperature does not depend on $\widetilde{m}_1$, since it is insensitive
to the details of the intermediate stage of reheating at which a sizable
fraction of the total energy is contained in heavy (s)neutrinos.
Instead, it only depends on the asymptotic value of the temperature
after all heavy (s)neutrinos have decayed as well as on $a_\text{RH}/a_\text{PH}$ (cf.\ Fig.~\ref{fig:Trhocompare}),
both of which, as evident from the discussion in Appendix~\ref{app:scalefactor}, do
not vary with $\widetilde{m}_1$.

%%%%%%%%%%%%%%%%%%%%%%%%%%%%%%%%%%%%%%%%%%%%%%%%%%%%%%%%%%%%%%%%%%%%%%%%%%%%%%

\medskip\noindent\textbf{Effective kink temperature:}
In our estimate for $k_{\textrm{RH}}$ (cf.\ Eq.~\eqref{eq:kRH}),
the two correction factors $\alpha_{\textrm{RH}}$ and $D$
appear in the combination $(\alpha_{\textrm{RH}}/2)^{1/2}\,D^{-1/3}$,
which is the same as $R^{1/2}D^{1/3}$ according to Eq.~\eqref{eq:defR}.
Comparing our estimate to the standard expression for $k_{\textrm{RH}}$ in terms of $T_{\textrm{RH}}$,
the temperature to which the wavenumber $k_{\textrm{RH}}$ is effectively related,
hence, turns out be $T_*$, the effective kink temperature, rather than $T_{\textrm{RH}}$,
\begin{align}
T_* = R^{1/2} \, \widetilde{T}_{\textrm{RH}} = R^{1/2}\, D^{1/3} \, T_{\textrm{RH}} \,.
\end{align}
The lower panel of Fig.~\ref{fig:Trhocompare} illustrates
how the correction factor $R$ can be constructed geometrically from the energy
density of the \BmL Higgs particles at $a = a_{\textrm{RH}}$ and the
asymptotic value of the radiation energy density.
As shown in Appendix~\ref{app:scalefactor}, the factor $R$ is a constant,
$R \simeq 0.41$, such that $T_*$ ends up being proportional to
$\widetilde{T}_{\textrm{RH}}$,
\begin{align}
T_* \simeq 3.9 \times 10^9 \,\textrm{GeV}
\left(\frac{M_1}{10^{11}\,\textrm{GeV}}\right)^{1.5} \,.
\end{align}
In Fig.~\ref{fig:Trhocompare}, this step from $\widetilde{T}_{\textrm{RH}}$ to
$T_*$ corresponds to lowering the value of $\widetilde{T}_{\textrm{RH}}$ by the
factor $R^{1/2}$ in the upper panel, with $R$ taken from the lower panel.
Together with Eq.~\eqref{eq:kRHTeff}, this result for the effective kink temperature
now eventually provides us with our master formula for the wavenumber
$k_{\textrm{RH}}$ as a function of $M_1$,
\begin{align}
k_{\textrm{RH}} \simeq 2.75 \times 10^{14} \,\textrm{Mpc}^{-1}
\left(\frac{T_*}{10^7\,\textrm{GeV}}\right)
\simeq 1.1 \times 10^{17} \,\textrm{Mpc}^{-1}
\left(\frac{M_1}{10^{11}\,\textrm{GeV}}\right)^{1.5} \,.
\end{align}

%%%%%%%%%%%%%%%%%%%%%%%%%%%%%%%%%%%%%%%%%%%%%%%%%%%%%%%%%%%%%%%%%%%%%%%%%%%%%%

The wavenumber $k_{\textrm{RH}}$, or correspondingly $f_{\textrm{RH}}$, can
be determined from a measurement of the amplitude of the GW spectrum at two frequencies
$f_- \ll f_{\textrm{RH}}$ and $f_+ \gg f_{\textrm{RH}}$ by means of the following relation,
\begin{align}
f_{\textrm{RH}} = \left[\tilde{c}_2^{(2)}
\frac{\Omega_{\textrm{GW}}(f_+)}{\Omega_{\textrm{GW}}(f_-)}\right]^{1/2} f_+ \,.
\label{eq_measurefRH}
\end{align}
Here, $\tilde{c}_2^{(2)}$ is the analogue of the coefficient $c_2^{(2)} \simeq 1.04$ in
the transfer function $T_2$ (cf.\ Eq.~\eqref{eq:T1T1result}) for the kink
in the GW spectrum induced by the decay of the AH string network.
Its numerical value can, in principle, be computed from the function
$\widetilde{C}$ (cf.\ Eq.~\eqref{eq_OmegaGW_2} and Footnote~\ref{fn_tildeC}).
In absence of a more detailed knowledge of $\widetilde{C}$, we however
only know that $\tilde{c}_2^{(2)}$ ought to be of $\mathcal{O}(1)$.
A measurement of $f_{\textrm{RH}}$ would then allow us to infer the \BmL Higgs
decay rate $\Gamma_S^0$ which, in the context of our Froggatt-Nielsen flavour
model, would be equivalent to determining the \BmL Higgs mass $m_S$ and the
(s)neutrino mass $M_1$,
\begin{align}
m_S \simeq 2.1 \times 10^{13}\,\textrm{GeV}
\left(\frac{f_{\textrm{RH}}}{100\,\textrm{Hz}}\right)^{0.67} \,,\quad
M_1 \simeq 7.1 \times 10^{10}\,\textrm{GeV}
\left(\frac{f_{\textrm{RH}}}{100\,\textrm{Hz}}\right)^{0.67} \,.
\end{align}
Again, we stress that these results are independent of $\widetilde{m}_1$ and, hence,
independent of the (s)neutrino decay rate $\Gamma_{N_1}^0$.
In turn, had we relied on the assumption that $k_{\textrm{RH}}$ was directly proportional
to $T_{\textrm{RH}}$ instead of $T_*$, we would have erroneously arrived at the
conclusion that $k_{\textrm{RH}}$ would be a function of both neutrino masses, $M_1$
and $\widetilde{m}_1$.

%%%%%%%%%%%%%%%%%%%%%%%%%%%%%%%%%%%%%%%%%%%%%%%%%%%%%%%%%%%%%%%%%%%%%%%%%%%%%%

\begin{figure}
\begin{center}
\includegraphics[width=0.95\textwidth]{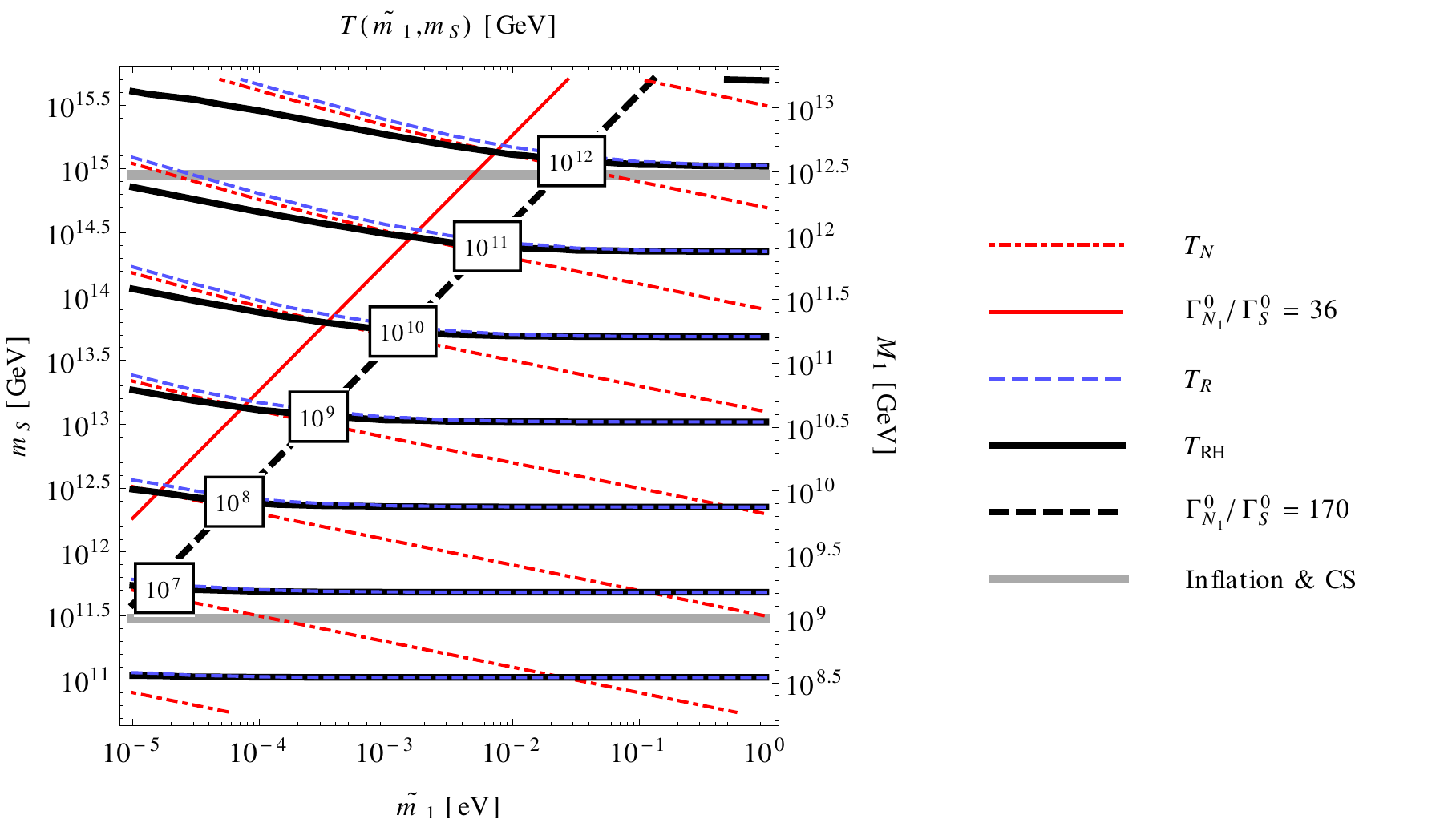}
\caption{Neutrino decay temperature $T_N$, radiation domination temperature $T_R$
and reheating temperature $T_{\textrm{RH}}$ as functions of the effective neutrino
mass $\widetilde{m}_1$ and the heavy (s)neutrino mass $M_1 = \eta^2 m_S$.
The bounds on $M_1$ marked by the gray horizontal lines derive from constraints
on the coupling constant $\lambda$ imposed by the requirement of successful hybrid
inflation and the upper bound on the cosmic string tension (cf.\ Ref.~\cite{Buchmuller:2012wn}).}
\label{fig:TRH}
\end{center}
\end{figure}

%%%%%%%%%%%%%%%%%%%%%%%%%%%%%%%%%%%%%%%%%%%%%%%%%%%%%%%%%%%%%%%%%%%%%%%%%%%%%%

\begin{table}
\begin{center}
\begin{tabular}{c||ccc|ccc|cc}
$T_i$ & $T_{i,0}^{(-)}\,\left[10^9\,\textrm{GeV}\right]$  & $p_i^{(-)}$ & $q_i^{(-)}$ &
$T_{i,0}^{(+)}\,\left[10^9\,\textrm{GeV}\right]$ & $p_i^{(+)}$ & $q_i^{(+)}$ &
$c_i$ & $d_i$ \\\hline\hline
$T_N$ & $9.4$ & $1.25$ & $0.25$ & $2.9$ & $1.2$ & $0.3$ & $36$ & $0.46$\\
$T_S$ & $5.6$ & $1.5$ & $0.0$ & $3.1$ & $1.3$ & $0.2$ & $300$ & $3.0$ \\
$T_{\textrm{RH}}$ & $4.8$ & $1.5$ & $0.0$ & $3.0$ & $1.3$ & $0.2$ & $170$ & $1.7$\\
$T_R$ & $4.7$ & $1.5$ & $0.0$ & $2.5$ & $1.2$ & $0.3$ & $170$ & $1.7$
\end{tabular}
\caption{Numerical values to be used in the fit formulas for
$T_N$, $T_S$, $T_{\textrm{RH}}$ and $T_R$ in Eq.~\eqref{eq:Ti}.}
\label{tab:Tfit}
\end{center}
\end{table}

%%%%%%%%%%%%%%%%%%%%%%%%%%%%%%%%%%%%%%%%%%%%%%%%%%%%%%%%%%%%%%%%%%%%%%%%%%%%%%

In order to determine the temperatures $T_N$, $T_S$, $T_{\textrm{RH}}$ and $T_R$
as functions of $M_1$ and $\widetilde{m}_1$, we have solved the set of Boltzmann
equations governing the reheating process (cf.\ Ref.~\cite{Buchmuller:2012wn}) for large
ranges of $M_1$ and $\widetilde{m}_1$ values.
Fig.~\ref{fig:TRH} shows
our numerical results for $T_N$, $T_\textrm{RH}$ and $T_R$.
The temperature $T_S$ is always only slightly larger than $T_{\textrm{RH}}$,
which is why it is not included in Fig.~\ref{fig:TRH}. 
For all of the four temperatures $T_N$, $T_S$, $T_{\textrm{RH}}$ and $T_R$,
we are able to derive fit formulas reproducing our numerical results to
very high precision.
These fit formulas can all be brought into the following form,
\begin{align}
T_i \simeq \min\left\{T_i^{(-)},T_i^{(+)}\right\} \simeq
\begin{cases}
T_i^{(-)} \,; & \:\: y \gtrsim c_i \:\Leftrightarrow\: y_i \gtrsim d_i \\
T_i^{(+)} \,; & \:\: y \lesssim c_i \:\Leftrightarrow\: y_i \lesssim d_i
\end{cases} \,, \quad
i = N,\, S,\, \textrm{RH},\, R \,,
\label{eq:Ti}
\end{align}
with the temperatures $T_i^{(-)}$ and $T_i^{(+)}$ being simple power laws,
\begin{align}
T_i^{(-)} \simeq & \:\: T_{i,0}^{(-)} \left(\frac{M_1}{10^{11}\,\textrm{GeV}}\right)^{p_i^{(-)}}
\left(\frac{\widetilde{m}_1}{10^{-2}\,\textrm{eV}}\right)^{q_i^{(-)}} \,, \\
T_i^{(+)} \simeq & \:\: T_{i,0}^{(+)} \left(\frac{M_1}{10^{11}\,\textrm{GeV}}\right)^{p_i^{(+)}}
\left(\frac{\widetilde{m}_1}{10^{-4}\,\textrm{eV}}\right)^{q_i^{(+)}} \,.
\end{align}
The quantity $y$ in Eq.~\eqref{eq:Ti} denotes the ratio of the two decay rates
$\Gamma_{N_1}^0$ and $\Gamma_S^0$,
\begin{align}
y = \frac{\Gamma_{N_1}^0}{\Gamma_S^0}
= \frac{8\,\widetilde{m}_1}{m_S} \left(\frac{v_{B-L}}{v_{\textrm{EW}}}\right)^2
\left[1 - 4\eta^2\right]^{-1/2} \simeq 2200 \left(\frac{\widetilde{m}_1}{10^{-2}\,\textrm{eV}}\right)
\left(\frac{10^{11}\,\textrm{GeV}}{M_1}\right) \,.
\end{align}
Similarly, $y_i$ is the ratio of the effective (s)neutrino decay rate $\Gamma_{N_1}^S$
and the \BmL Higgs decay rate $\Gamma_S^0$ evaluated at $a = a_i$,
\begin{align}
y_i = \left.\frac{\Gamma_{N_1}^S}{\Gamma_S^0}\right|_{a_i} = 
\gamma^{-1}\left(a_i\right) \frac{\Gamma_{N_1}^0}{\Gamma_S^0} = \gamma^{-1}\left(a_i\right) y\,.
\end{align}
The values of $c_i$ and $d_i$, $T_i^{(-)}$ and $T_i^{(+)}$
as well as of the powers $p_i^{(\pm)}$ and $q_i^{(\pm)}$
for each of the four temperatures $T_N$, $T_S$, $T_{\textrm{RH}}$
and $T_R$ are summarized in Tab.~\ref{tab:Tfit}.

%%%%%%%%%%%%%%%%%%%%%%%%%%%%%%%%%%%%%%%%%%%%%%%%%%%%%%%%%%%%%%%%%%%%%%%%%%%%%%

\medskip

Our numerical results lead us to two interesting observations:%
\footnote{For further useful observations concerning the temperatures $T_N$, $T_S$,
$T_{\textrm{RH}}$ and $T_R$, cf.\ Appendix~\ref{app:temperatures}.}

%%%%%%%%%%%%%%%%%%%%%%%%%%%%%%%%%%%%%%%%%%%%%%%%%%%%%%%%%%%%%%%%%%%%%%%%%%%%%%

\medskip\noindent\textbf{(1)}
In the case of fast decaying heavy (s)neutrinos, $y \gtrsim c_i$, the temperatures
$T_S$, $T_{\textrm{RH}}$ and $T_R$ barely vary with $\widetilde{m}_1$.
For large values of $\widetilde{m}_1$, they are, thus, proportional to the would-be
reheating temperature $\widetilde{T}_{\textrm{RH}}$
as well as the effective kink temperature $T_*$,
\begin{align}
T_* (M_1) \simeq \frac{1}{1.6} \,\widetilde{T}_{\textrm{RH}} (M_1) \simeq
\frac{1}{1.4} \, T_S^{(-)} (M_1) \simeq \frac{1}{1.2} \, T_{\textrm{RH}}^{(-)} (M_1)
\simeq  \frac{1}{1.2} \, T_R^{(-)} (M_1) \,.
\label{eq:Trelations}
\end{align}
These relations illustrate the numerical imprecision which one introduces when using
$\widetilde{T}_{\textrm{RH}}$, $T_S$, $T_{\textrm{RH}}$ or $T_R$ instead of $T_*$
in order to estimate $k_{\textrm{RH}}$.
Obviously, the imprecision always remains at a moderate level, which is mostly due
to the temperature plateau during reheating.
But Eq.~\eqref{eq:Trelations} also illustrates that $T_*$, $\widetilde{T}_{\textrm{RH}}$,
$T_S$, $T_{\textrm{RH}}$ and $T_R$ are certainly not equivalent to each other and
that it is important to distinguish between them from a conceptional point of view.

%%%%%%%%%%%%%%%%%%%%%%%%%%%%%%%%%%%%%%%%%%%%%%%%%%%%%%%%%%%%%%%%%%%%%%%%%%%%%%

\medskip\noindent\textbf{(2)}
The temperatures $T_N$, $T_S^{(+)}$, $T_{\textrm{RH}}^{(+)}$ and $T_R^{(+)}$
not only depend on $M_1$, but also on the effective neutrino mass $\widetilde{m}_1$.
A mere measurement of the position of the kink in the GW spectrum, i.e.\
a determination of the wavenumber $k_{\textrm{RH}}$, would hence not
suffice to pinpoint the numerical values of these temperatures.
Instead, all of them could still be tuned by varying the heavy (s)neutrino decay
rate.
This is an important feature of our reheating scenario that arises due to its
two-stage nature, distinguishing it from the ordinary scenario of reheating
via inflaton decay. 

%%%%%%%%%%%%%%%%%%%%%%%%%%%%%%%%%%%%%%%%%%%%%%%%%%%%%%%%%%%%%%%%%%%%%%%%%%%%%%
\medskip
As discussed in Ref.~\cite{Buchmuller:2012wn}, the fact that the temperature evolution
during reheating also depends on $\widetilde{m}_1$ leads to a series of non-trivial
relations between superparticle masses and neutrino parameters.
In combination with the possibility to determine $M_1$ by measuring $k_{\textrm{RH}}$,
these \textit{relations} partly turn into \textit{predictions} rendering our scenario
testable in future experiments.
Imagine, for instance, that dark matter consists of gravitinos which are thermally produced during reheating.
Assuming the mass of the gravitino were known, a measurement of $k_{\textrm{RH}}$
via the observation of GWs would then predict the value of $\widetilde{m}_1$.
Vice versa, given the value of $\widetilde{m}_1$, a determination of $k_{\textrm{RH}}$
would yield a prediction of the gravitino mass.

%%%%%%%%%%%%%%%%%%%%%%%%%%%%%%%%%%%%%%%%%%%%%%%%%%%%%%%%%%%%%%%%%%%%%%%%%%%%%%
\section{Observational Prospects \label{sec:observations}}

\begin{figure}[t!]
\begin{center}
\includegraphics[width=0.82\textwidth]{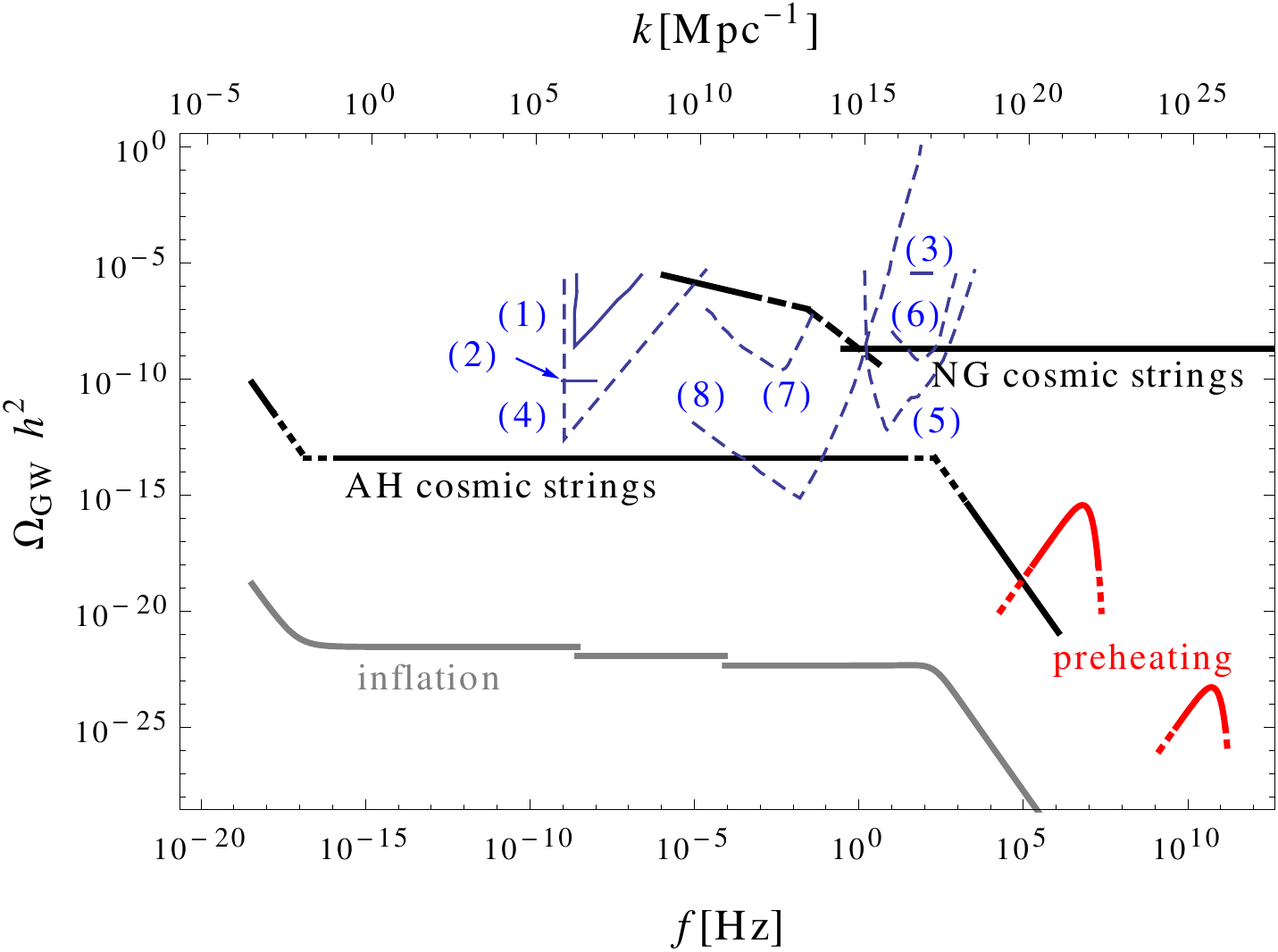}
\caption{Predicted GW spectrum and the (expected) sensitivity of current and upcoming experiments. The GW spectrum due to inflation (gray), preheating (red) as well as AH and NG cosmic strings (black) is shown for $v_{B-L} = 5 \times 10^{15}$~GeV, $M_1 = 10^{11}$~GeV, $m_S = 3 \times 10^{13}$~GeV, and $\alpha = 10^{-12}$, as in Figs.~\ref{fig_infl_vs_cs} and \ref{fig:comparison}. 
This corresponds to an effective kink temperature of $T_* = 3.9 \times 10^9$~Gev, which, for $\widetilde m_1 = 0.04$~eV, yields a reheating temperature of $T_{\text{RH}} = 4.9 \times 10^9$~GeV.
The current bounds on the stochastic GW spectrum from (1) millisecond pulsar timing (taken from Ref.~\cite{Smith:2005mm}, with (2) marking the update from EPTA~\cite{vanHaasteren:2011ni}) and (3) LIGO~\cite{Palomba:2012wn} are marked by solid blue lines. The dashed blue lines mark the expected sensitivity of some planned experiments: (4) SKA~\cite{Kramer:2004rwa}, (5) ET~\cite{etreport}, (6) advanced LIGO (taken from Ref.~\cite{etreport}), (7) eLISA~\cite{AmaroSeoane:2012km}, and (8) BBO and DECIGO~\cite{Alabidi:2012ex}. Note that with a correlation analysis ultimate DECIGO  has a
sensitivity down to $10^{-18}$.
}
\label{fig_obs}
\end{center}
\end{figure}

In Secs.~\ref{sec:inflation} to \ref{sec:comparison}, we discussed the GW background produced during the different stages of a $B$$-$$L$ phase transition in the early universe, namely during inflation, during tachyonic preheating, and from cosmic strings in the scaling regime. Fig.~\ref{fig_obs} summarizes the resulting GW spectrum of all these different sources for a representative parameter point, see Eq.~\eqref{eq:exampleparameterpoint}, and for $\alpha = 10^{-12}$. Additionally we show current bounds and the expected sensitivity of upcoming  GW experiments, depicted by the solid and dashed blue curves, respectively. These experiments can be grouped into three categories, see e.g.\ Ref.~\cite{Maggiore:1999vm} for a review: millisecond pulsar timing measurements (e.g.\ EPTA~\cite{vanHaasteren:2011ni}, PPTA~\cite{Manchester:2012za} and SKA~\cite{Kramer:2004rwa, Aharonian:2013av}) sensitive to GWs with a frequency of about $10^{-9}$~Hz, space-based interferometers (e.g.\ BBO~\cite{Crowder:2005nr}, DECIGO~\cite{Kawamura:2011zz} and eLISA~\cite{AmaroSeoane:2012km}) sensitive at about $10^{-1}$~Hz and ground-based interferometers (e.g.\ (advanced) LIGO~\cite{Abbott:2009ws, advligopaper}, ET~\cite{ethomepage}, and KAGRA~\cite{Kuroda:2010zzb}) which are most sensitive at about $10^2$~Hz. It should be noted that in particular for the ground-based detectors the sensitivity is typically given in terms of the strain $\tilde h_f$, which is related to the amplitude $\Omega_\text{GW}$ as~\cite{Maggiore:1999vm}
\begin{equation}
\Omega_\text{GW} h^2 = \frac{\text{SNR}}{F} \frac{4 \pi^2 f^3}{3 (H_0/h)^2} \tilde h_f^2 \,,
\end{equation}
with the sensitivity factor $F = 2/5$ for the ground-based interferometers and $\text{SNR}$ denoting the signal-to-noise ratio. Moreover, constraints on the effective number of relativistic DOFs from BBN and the CMB yield an upper bound of $\Omega_\text{GW} h^2 \lesssim 10^{-5}$ for $k > k_\text{eq}$, with a weak dependence on the origin of the GW background, cf.\ Refs.~\cite{Izotov:2010ca, Smith:2006nka}. For further ideas of how to possibly probe the parameter space of Fig.~\ref{fig_obs} in the future, see also Ref.~\cite{Maggiore:1999vm} and references therein.

A clear message of Fig.~\ref{fig_obs} is that GWs from a GUT-scale $U(1)$ phase transition will be seriously probed by a number of upcoming experiments. Here, the dominant contribution originates from cosmic strings in the scaling regime, cf.\ the black curves in Fig.~\ref{fig_obs}. However, at the same time, this is the contribution with the largest theoretical uncertainty. Fig.~\ref{fig_obs} shows the predictions based on the AH cosmic string model as well as on the NG model, cf.\ Secs.~\ref{sec:cosmicstrings} and \ref{sec:comparison}. In spite of the quite dramatic differences, it is worth noting that in both cases, upcoming experiments are expected to reach the sensitivity to probe the `plateau value' for GUT-scale cosmic strings. Moreover, in both cases it seems feasible that we might be able to measure a very interesting feature of the spectrum. In the case of NG strings, the bend-over marking the transition from radiation to matter domination, cf.\ Fig.~\ref{fig:NGstring10-6}, may be within range of 
planned satellite-based experiments, depending on the parameter $\alpha$. 
Note that for the large values obtained by the simulations in Ref.~\cite{BlancoPillado:2011dq} the model under consideration is in contradiction with 
millisecond pulsar timing measurements.
In the case of AH strings, the transition between an early matter dominated reheating phase and radiation domination has the right frequency to possibly be within reach of future ground-based GW detectors, cf.\ Sec.~\ref{sec:reheatingtemp}. Assuming that the actual signal from cosmic strings lies somewhere between the AH and the NG prediction (or maybe can even be obtained by interpolating between the AH model at early times and the NG model at late times), it indeed seems conceivable that future experiments will be able to probe one, or maybe even both, of these features.

The GW background from inflation, depicted in gray in Fig.~\ref{fig_obs} is probably the best understood cosmological source. However, as Fig.~\ref{fig_obs} shows, it is clearly subdominant compared to the GW background from cosmic strings in hybrid inflation models, which typically feature a very small tensor-to-scalar ratio, cf.\ Eq.~\eqref{eq:deltatsquared2}. Nevertheless, a precise understanding of the GW background from inflation is crucial for two reasons: First, although their origin is very different, both the AH string and the inflationary GW spectrum are governed by the respective Hubble-sized modes throughout the expansion history, see discussion at the end of Sec.~\ref{sec:cosmicstrings}. Hence the kink marking the transition between an early matter dominated reheating phase and radiation domination occurs at the same frequency in both spectra. So although the shape of this kink is modified in the case of the signal from AH cosmic strings due to the precise shape of $\widetilde C$, cf.\ Eq.~\eqref{eq_OmegaGW_2} and the discussion below Eq.~\eqref{eq_measurefRH}, all the conclusions we can draw from measuring the position of the kink in the inflationary spectrum hold just as well for the AH cosmic string spectrum. In particular, this means that we can directly read off the effective kink temperature $T_*$, determined by the parameters of the $B$$-$$L$ Higgs sector. Independent requirements on the reheating temperature from successful leptogenesis and dark matter production then allow to constrain the parameters of the neutrino sector, cf.\ Sec.~\ref{sec:reheatingtemp}.

Second, the prediction for the GW background from cosmic strings is plagued with large uncertainties. The discrepancy between the AH and the NG prediction is one, but there are also other mechanisms which might reduce or even eliminate the cosmic string contribution. For example, a coupling between the GUT and SM Higgs bosons can effectively reduce the string tension~\cite{Hindmarsh:2013xg}, or several e-folds of inflation occurring after the $U(1)$ phase transition could dilute cosmic strings in the NG model, thereby reducing
the GW background from cosmic strings, see e.g.\ Refs.~\cite{Shafi:1984tt, Yokoyama:1989pa, Freese:1995vp, Kamada:2012ag, Linde:2013aya}. Note also, that all simulations have been done for a bosonic Abelian Higgs model, whereas we are interested in a supersymmetric theory. Additional fermionic decay channels may further relax the cosmic string bound by a factor ${\cal{O}}(1)$. Last but not least, one has to worry about initial conditions. Clearly, strings cannot form until the causal horizon is larger than their characteristic width \cite{Bevis:2006mj}, and one should remember that tachyonic preheating proceeds very fast. In fact, the expectation value of the waterfall field  grows with time faster than exponentially \cite{Asaka:2001ez}. It is thus crucial to also take into account the second largest contribution, and over a wide range of frequencies, this is inflation.

Finally, the red curves in Fig.~\ref{fig_obs} depict the signals expected from tachyonic preheating, cf.\ Sec.~\ref{sec:preheating}. These are clearly at too high frequencies to be detectable in the near future. Consulting Eq.~\eqref{eq_pred_preheating}, it becomes clear that this is generically true for GUT-scale parameters. However, these signals show a distinctly different behaviour compared to the two before-mentioned contributions. It would thus be very interesting if one could probe such frequencies, since it would enable us to break parameter degeneracies and probe different models describing the preheating and reheating phase.

At this point it is worth mentioning, that although GWs yield an unique insight into the very early universe, there are other possible signals, in particular from cosmic strings, which are equally important: Cosmic strings distort the surface of last scattering of the CMB photons, as well as affect them on their path since then. Hence, CMB measurements yield upper bounds on cosmic string tension, cf.\ Ref.~\cite{Ade:2013xla} for a recent analysis. Moreover, if massive radiation is a significant energy loss mechanism also at late times, this leads to the production of ultra high-energetic cosmic rays and GeV-scale $\gamma$-rays. The non-observation of such signals imposes a (model-dependent) upper bound on the string tension in the AH model, cf.\ Refs.~\cite{Sigl:1995kk, Protheroe:1996pd, Bhattacharjee:1997in, Sigl:1998vz, Wichoski:1998kh}.

\section{Conclusions and Outlook \label{sec:conclusion}}

Cosmological $B$$-$$L$ breaking at the GUT scale can account for the initial
conditions of the hot early universe: the generation of entropy, baryogenesis
via leptogenesis and dark matter via thermal production of gravitinos. The 
temperature evolution during reheating is determined by the parameters of the
neutrino and $B$$-$$L$ Higgs sector. The false
vacuum of unbroken $B$$-$$L$ symmetry drives inflation, ending in tachyonic
preheating and thus with the production of $B$$-$$L$ Higgs bosons and cosmic
strings.

In the preceding sections we have calculated the various contributions to
the GW spectrum during this pre- and reheating process.
For the contribution from inflation we have discussed in detail the transfer
function, in particular the
departure from a flat spectrum at small and large frequencies and the
dependence on the reheating temperature.
Furthermore, we have studied the temperature evolution during the reheating process, 
which provided us with a clear understanding of its connection to the GW spectrum.
As a result of preheating, the scalar and vector degrees of
freedom lead to two peaks at very high frequencies, the positions and heights of which
we have estimated. The spectrum of GWs radiated from cosmic strings has
large theoretical uncertainties which are mostly due to different assumptions
on how strings lose their energy in the scaling regime. For Abelian Higgs
strings we find a GW spectrum which agrees in shape with the one obtained
for inflation. However, the normalization differs by about eight orders of
magnitude for typical parameter values.
This enhances the possibility of detecting a feature in the GW background associated with the
reheating temperature. Such a measurement could potentially allow for the determination of underlying
particle physics parameters, namely the \BmL Higgs mass and the heavy neutrino mass scale.
For Nambu-Goto strings, shape and normalization differ
significantly from AH strings. This is a consequence of the
energy loss of NG strings via GW radiation from
short string loops. The plateau of the spectrum is about five orders of 
magnitude higher than the one predicted by AH strings. In order
to make the parameter dependence of the GW spectrum from NG strings
transparent we have performed an approximate analytical calculation. 

As the model of cosmological $B$$-$$L$ breaking illustrates, the GW spectral
amplitude from cosmic strings can be many orders of magnitude larger than the
one from inflation. This is the main result of this paper, which opens new
perspectives for the observation of relic GWs. If the prediction of AH 
strings for the GW spectral amplitude is true, we have to wait for
BBO or DECIGO to detect relic GWs. NG strings
predict a much larger spectral amplitude which could be discovered
already with eLISA and advanced LIGO, as well as
with the Einstein Telescope. The true GW spectral amplitude for 
cosmic strings could of course lie in between the predictions of AH strings
and NG strings. We emphasize that several of the before-mentioned experiments are most
sensitive in frequency ranges where departures from a flat spectrum are expected. 
A discovery of such features in relic GWs would then contain valueable 
information about the energy loss mechanism of cosmic strings. Furthermore,
a determination of the reheating temperature of the early universe may be
feasible, a very exciting but clearly challenging goal!

\vspace{1cm}
\noindent \textbf{Acknowledgements} \newline
The authors thank D.~Figueroa, M.~Hindmarsh, T.~Hiramatsu, T.~Konstandin and A.~Shoda for helpful discussions and comments.
This work has been supported by the German Science Foundation (DFG)
within the Collaborative Research Center 676 ``Particles, Strings and the Early Universe'' as well as by the World Premier International Research Center Initiative (WPI Initiative) of the Ministry of Education, Culture, Sports, Science and Technology (MEXT) of Japan.

\newpage
\appendix
\newpage \section{Evolution of the Scale Factor During Reheating}
\label{app:scalefactor}

%%%%%%%%%%%%%%%%%%%%%%%%%%%%%%%%%%%%%%%%%%%%%%%%%%%%%%%%%%%%%%%%%%%%%%%%%%%%%%

In order to be able to precisely describe the connection between the reheating
process and the spectrum of GW modes re-entering the Hubble horizon
or being produced during reheating, we need to know the exact evolution of
the scale factor $a$ as a function of cosmic time $t$ during reheating.
This evolution is determined by the Friedmann equation in combination with
the Boltzmann equations for all non-relativistic and relativistic DOFs,
respectively, (cf.\ Ref.~\cite{Buchmuller:2011mw})
\begin{align}
H^2 = \left(\frac{\dot{a}}{a}\right)^2 = & \:
\frac{1}{3M_{\textrm{Pl}}^2} \left(\rho_S + \rho_{\textrm{rel}}\right) \,,
\label{eq:HBoltz1}\\
\left(\frac{\partial}{\partial t} - H p \frac{\partial}{\partial p}\right) f_S(t,p) = & \:
-\Gamma_S^0 \, f_S(t,p) \,, \label{eq:HBoltz2}\\
\left(\frac{\partial}{\partial t} - H p \frac{\partial}{\partial p}\right) f_{\textrm{rel}}(t,p) = & \:
\frac{2\pi^2}{g_{\textrm{rel}}} \frac{\delta(p-m_S/2)}{p^2} \,2 \,n_S(t)\, \Gamma_S^0 \,,
\label{eq:HBoltz3}
\end{align}
where all quantities labeled with an $S$ refer to the non-relativistic \BmL Higgs
bosons and their superpartners.
Similarly, the phase space distribution function (PSDF) $f_{\textrm{rel}}$ and the energy
density $\rho_{\textrm{rel}}$ subsume the contributions from all relativistic
DOFs, i.e.\ heavy (s)neutrinos and MSSM radiation,
\begin{align}
f_{\textrm{rel}} = f_N + f_{\tilde{N}} + f_r \,, \quad
\rho_{\textrm{rel}} = \rho_N + \rho_{\tilde{N}} + \rho_r \,.
\end{align}
In Eqs.~\eqref{eq:HBoltz2} and \eqref{eq:HBoltz3}, $p$ denotes the physical momentum
and $n$ the number density, which follows from the corresponding PSDF after integration
over the momentum phase space.
In this appendix, we solve Eqs.~\eqref{eq:HBoltz1} to \eqref{eq:HBoltz3}
for the scale factor $a$ and, based on our result for $a$, compute several
quantities required for our discussion in the main text.
In doing so, we will proceed as far as possible by means of analytical
calculations and only resort to numerical calculations if necessary.
This will provide us with a transparent picture,
allowing us to identify the parameter dependencies of all
quantities of interest.

%%%%%%%%%%%%%%%%%%%%%%%%%%%%%%%%%%%%%%%%%%%%%%%%%%%%%%%%%%%%%%%%%%%%%%%%%%%%%%

Putting the determination of the scale factor $a(t)$ on hold,
the Boltzmann equations in Eqs.~\eqref{eq:HBoltz2} and \eqref{eq:HBoltz3}
can be solved analytically for the PSDFs $f_S$ and $f_{\textrm{rel}}$,
\begin{align}
f_S(t,p) = & \: \frac{2\pi^2}{g_S}\left(\frac{a_{\textrm{PH}}}{a(t)}\right)^3
n_S \left(t_{\textrm{PH}}\right) \frac{\delta(p)}{p^2}
e^{-\Gamma_S^0\left(t - t_{\textrm{PH}}\right)} \,, \label{eq:fSres}\\
f_{\textrm{rel}}(t,p) = & \: \frac{2\pi^2}{g_{\textrm{rel}}} \int_{t_{\textrm{PH}}}^t dt'
\left(\frac{a(t')}{a(t)}\right)^3 \, \frac{2 \, n_S(t') \, \Gamma_S^0}{p^2} \ \ 
\delta\left(p-\frac{ a(t')}{a(t)} \, \frac{m_S}{2}\right) \,. \label{eq:frelres}
\end{align}
Here, we have used that the initial PSDF of the \BmL Higgs multiplet after preheating
is proportional to a delta function, $f_S(t_{\textrm{PH}},p) \propto \delta(p)/p^2$,
as well as that the initial PSDF of the relativistic DOFs vanishes,
$f_{\textrm{rel}}(t_{\textrm{PH}},p) = 0$.
Multiplying the PSDFs in Eqs.~\eqref{eq:fSres} and \eqref{eq:frelres} by
$E_S = \left(p^2 + m_S^2\right)^{1/2}$ and $E_{\textrm{rel}} = p$, respectively,
and integrating them over the momentum phase space, yields the energy densities
$\rho_S$ and $\rho_{\textrm{rel}}$,
\begin{align}
\rho_S (t) = & \: m_S \, n_S(t) \,, \quad n_S(t) = n_S \left(t_{\textrm{PH}}\right)
\left(\frac{a_{\textrm{PH}}}{a(t)}\right)^3
e^{-\Gamma_S^0 \left(t - t_{\textrm{PH}} \right)} \,, \label{eq:rhoSres} \\
\rho_{\textrm{rel}} (t) = & \: m_S \, \Gamma_S^0 \int_{t_{\textrm{PH}}}^t dt'
\left(\frac{a(t')}{a(t)}\right)^4 n_S(t')\,. \label{eq:rhorelres}
\end{align}
Given these results for $\rho_S$ and $\rho_{\textrm{rel}}$, it is easy to
see that they are also solutions to the following differential equations,
\begin{align}
\dot{\rho}_S + 3 H \rho_S = - \Gamma_S^0 \, \rho_S \,, \quad
\dot{\rho}_{\textrm{rel}} + 4 H \rho_{\textrm{rel}} = \Gamma_S^0 \, \rho_S \,.
\end{align}
This is a useful cross check, since the first equation obviously derives from the
Klein-Gordon equation for the \BmL Higgs bosons while the second equation is a direct
consequence of the covariant energy conservation.

%%%%%%%%%%%%%%%%%%%%%%%%%%%%%%%%%%%%%%%%%%%%%%%%%%%%%%%%%%%%%%%%%%%%%%%%%%%%%%

With the expressions for $\rho_S$ and $\rho_{\textrm{rel}}$ in
Eqs.~\eqref{eq:rhoSres} and \eqref{eq:rhorelres} at hand, we are now able
to solve the Friedmann equation (cf.\ Eq.~\eqref{eq:HBoltz1}) and hence determine
the scale factor.
Let us begin by introducing the equation of state of the reheating
phase, which relates the total pressure
$p_{\textrm{tot}} = p_S + p_{\textrm{rel}}$, where $p_S = 0$ and
$p_{\textrm{rel}} = \frac{1}{3} \rho_{\textrm{rel}}$, of the reheating phase to
its total energy density $\rho_{\textrm{tot}} = \rho_S + \rho_{\textrm{rel}}$,
\begin{align}
\omega = \frac{p_{\textrm{tot}}}{\rho_{\textrm{tot}}} =
\frac{1}{3}\frac{\rho_{\textrm{rel}}}{\rho_S + \rho_{\textrm{rel}}} \,.
\end{align}
Due to the permanent energy transfer from the non-relativistic
\BmL Higgs particles to the relativistic DOFs, the equation
of state coefficient $\omega$ continuously changes during reheating.
Starting out at a value $\omega = 0$, it monotonically grows until
it eventually reaches $\omega = 1/3$.
We can, hence, model the evolution of the scale factor by
approximating $\omega$ to be a piecewise constant function
of time taking successively larger
values $\omega_i \in \left[0,\frac{1}{3}\right]$ during consecutive short time intervals
$(t_i,t_{i+1}]$, where $a_{\textrm{PH}} \leq a(t_i) < a(t_{i+1})$.
During each such time interval, $a$ is then of the following form,
\begin{align}
a(t) = a(t_i) \left[1 + \frac{1}{p_i} H(t_i)\left(t-t_i\right)\right]^{p_i}
\,, \:\: H(t_i) = \left(\frac{\rho_{\textrm{tot}}(t_i)}
{3M_{\textrm{Pl}}^2}\right)^{1/2} \,, \:\: p_i = \frac{2}{3(1 + \omega_i)} \,,
\label{eq:FriedmannSol}
\end{align}
which is nothing but the standard solution of the Friedmann equation
assuming a flat universe and a constant equation of state,
$\omega_i = p_{\textrm{tot}} / \rho_{\textrm{tot}} = \textrm{const}$.
The coefficients $\omega_i$ can be determined iteratively by requiring
self-consistency of the Friedmann equation.
For all time intervals $(t_i,t_{i+1}]$, we solve the following
equation numerically, until we reach $\omega_i = 1/3$,
\begin{align}
\frac{\rho_{\textrm{tot}}(t_i)}{\rho_{\textrm{tot}}(t_{i+1})} =
\left(\frac{a(t_{i+1})}{a({t_i})}\right)^{3 (1+\omega_i)} \,,\quad
\rho_{\textrm{tot}}(t) = \rho_S(t) + \rho_{\textrm{rel}} (t)\,.
\end{align}

%%%%%%%%%%%%%%%%%%%%%%%%%%%%%%%%%%%%%%%%%%%%%%%%%%%%%%%%%%%%%%%%%%%%%%%%%%%%%%

\begin{figure}
\begin{center}
\includegraphics[width=12cm]{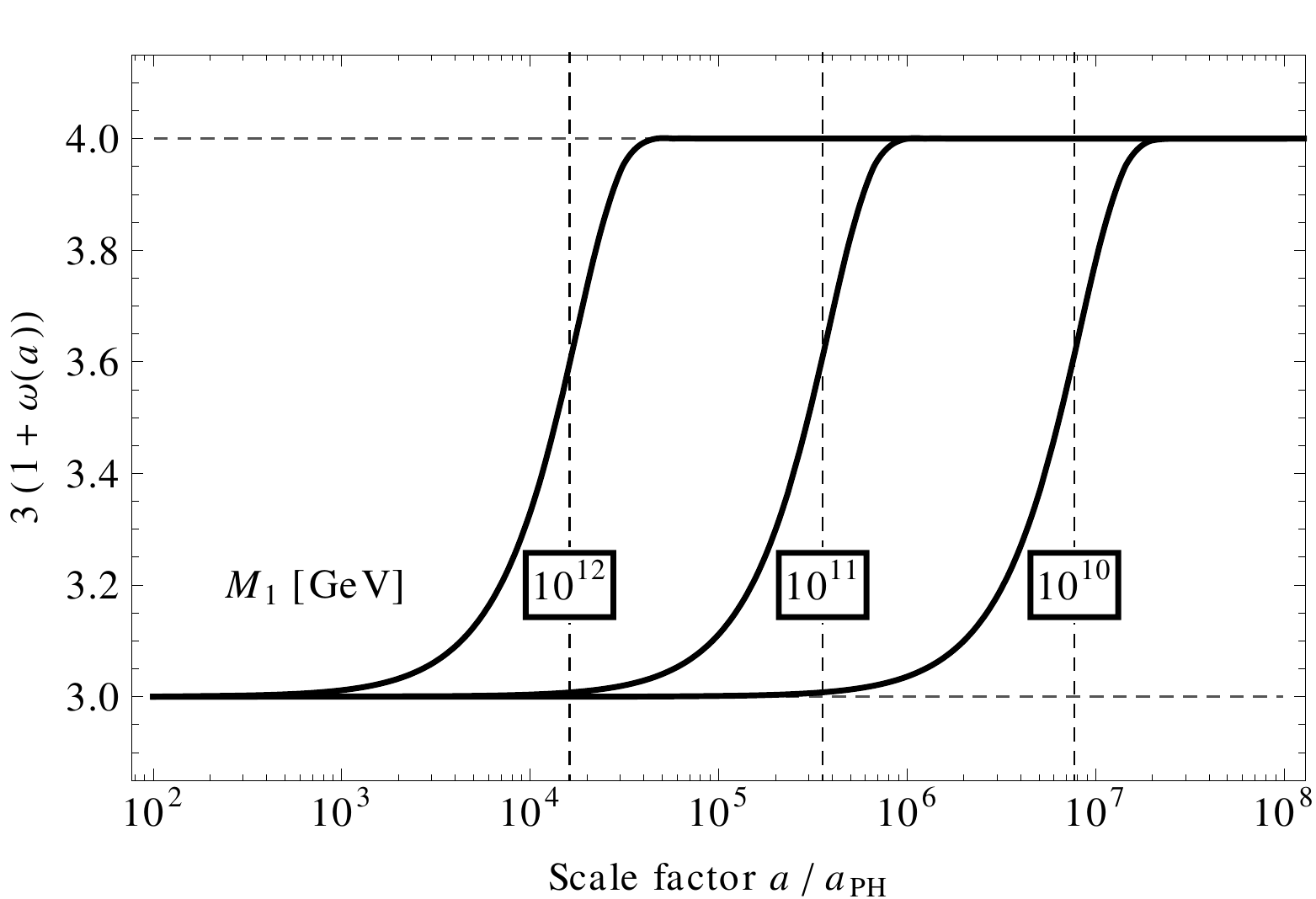}
\caption{Evolution of the equation of state coefficient $\omega$ as
function of the scale factor $a$ for three different values of $M_1 = \eta^2 m_S$.
The vertical dashed lines indicate the respective values of $a_{\textrm{RH}}$, i.e.\
the times when half of the non-relativistic \BmL Higgs particles have decayed
(cf.\ Eq.~\eqref{eq:defaR}).}
\label{fig:omega}
\end{center}
\end{figure}

%%%%%%%%%%%%%%%%%%%%%%%%%%%%%%%%%%%%%%%%%%%%%%%%%%%%%%%%%%%%%%%%%%%%%%%%%%%%%%

Fig.~\ref{fig:omega} presents the result of this calculation for
three different values of $m_S$, which, by means of our Froggatt-Nielsen
flavour model, may be translated into corresponding values of $M_1$.
Indeed, for all values of $M_1$, we see how $\omega$ gradually increases
from $\omega = 0$ to $\omega = 1/3$, with the transition between these two
asymptotic values always taking place at values of the scale factor around
$a_{\textrm{RH}}$.
But more than that, the three curves in Fig.~\ref{fig:omega}, in fact, all have
the same shape, differing from each other only by shifts along the horizontal
axis.
Moreover, as evident from the above calculation, the evolution of the
scale factor is insensitive to the distinction between nonthermal
heavy (s)neutrinos and thermal MSSM radiation.
As a consequence, the heavy (s)neutrino decay rate $\Gamma_{N_1}^0$ and
in particular the effective neutrino mass $\widetilde{m}_1$ do not enter
into the determination of the scale factor.
All quantities that may be directly derived from $a(t)$ as well as from the energy
densities $\rho_S$ and $\rho_{\textrm{rel}}$, and hence all quantities that we shall
compute in this appendix, thus, do not depend on $\widetilde{m}_1$.
In particular, we shall now discuss %calculate or comment on the calculation of
four important quantities required for our discussion in the main text:

%%%%%%%%%%%%%%%%%%%%%%%%%%%%%%%%%%%%%%%%%%%%%%%%%%%%%%%%%%%%%%%%%%%%%%%%%%%%%%

\medskip\noindent\textbf{(1)}
The evolution of the scale factor directly influences the GW spectrum through
its appearance in the equation of motion for the GW modes $\varphi_k^A$
(cf.\ Eqs.~\eqref{eq_einstein_k} and \eqref{eq_fouriermode}).
With regard to the GW background from inflation, $a(t)$ is even the \textit{only}
model-dependent ingredient to the GW mode equation, which,
in conformal coordinates, reads
\begin{align}
\frac{d^2}{d\tau^2}\tilde{\varphi}_k^A(\tau) + \left(k^2 - \frac{a''(\tau)}{a(\tau)}\right)
\tilde{\varphi}_k^A(\tau) = 0 \,, \quad
\tilde{\varphi}_k^A(\tau) = a(\tau)\, \varphi_k^A(\tau) \,.
\end{align}
Solving this mode equation for an appropriate range of $k$ values, thus,
allows us to determine the transfer function $T_2$ (cf.\ Eq.~\eqref{eq:T1T1result}).
Here, the fact that the shape of $a(t)$ is independent of any model parameter
(cf.\ Fig.~\ref{fig:omega}) directly
translates into the coefficients $c_2^{(1)}$ and $c_2^{(2)}$ being constants
across the entire parameter space.
In other words, the fixed shape of the kink
in the GW spectrum at $k = k_{\textrm{RH}}$ can be directly traced back to the
unchanging shape of the scale factor $a$ as a function of time.

%%%%%%%%%%%%%%%%%%%%%%%%%%%%%%%%%%%%%%%%%%%%%%%%%%%%%%%%%%%%%%%%%%%%%%%%%%%%%%

\medskip\noindent\textbf{(2)}
In Sec.~\ref{sec:inflation} we introduce the correction factor
$R = \rho_S\left(a_{\textrm{RH}}\right) / \tilde{\rho}_r^{\textrm{RH}}$ in order
to quantify the increase of $a^4\rho_{\textrm{rel}}$ after $a = a_{\textrm{RH}}$
(cf.\ Eq.~\eqref{eq:defR}).
To see that $R$ has exactly this meaning, note that both the \BmL Higgs energy density
$\rho_S\left(a_{\textrm{RH}}\right)$ as well as
the would-be radiation energy density $\tilde{\rho}_r^{\textrm{RH}}$ can be rewritten
in terms of the energy density $\rho_{\textrm{rel}}$,
\begin{align}
\rho_S\left(a_{\textrm{RH}}\right) = \rho_{\textrm{rel}}\left(a_{\textrm{RH}}\right) \,, \quad
\tilde{\rho}_r^{\textrm{RH}} = \left[\left(\frac{a}{a_{\textrm{RH}}}\right)^4 \rho_r
\right]_{a\gg a_{\textrm{RH}}} \hspace{-0.45cm} = \hspace{0.45cm}
\left[\left(\frac{a}{a_{\textrm{RH}}}\right)^4 \rho_{\textrm{rel}}
\right]_{a\gg a_{\textrm{RH}}} \,,
\end{align}
where we have neglected any possible changes in the number of relativistic DOFs in
thermal equilibrium during and shortly after reheating.
The factor $R$ is, hence, nothing but the ratio of two values of $a^4 \rho_{\textrm{rel}}$,
\begin{align}
R = \frac{\left(a^4 \rho_{\textrm{rel}}\right)_{a = a_{\textrm{RH}}}}
{\left(a^4 \rho_{\textrm{rel}}\right)_{a \gg a_{\textrm{RH}}}} =
\frac{I_R\left(t = t_{\textrm{RH}}\right)}{I_R\left(t \gg t_{\textrm{RH}}\right)}\,, \quad
I_R(t) = \int_{t_{\textrm{PH}}}^t dt' \, \frac{a(t')}{a_{\textrm{PH}}} \, e^{-\Gamma_S^0\left(t' - t_{\textrm{PH}}\right)} \,.
\label{eq_IR}
\end{align}
$I_R\left(t = t_{\textrm{RH}}\right)$ and
$I_R\left(t \gg t_{\textrm{RH}}\right)$ scale in exactly the same way
with the neutrino mass $M_1$, rendering the factor $R$ a constant
independent of all parameters of our model. Solving the Friedmann equation as outlined above and evaluating the expression for $I_R$ in Eq.~\eqref{eq_IR}, we find
\begin{align}
I_R\left(t = t_{\textrm{RH}}\right) \propto M_1^{-13/3} \,,\quad 
I_R\left(t \gg t_{\textrm{RH}}\right) \propto M_1^{-13/3} \,, \quad R = \textrm{const.} \,.
\end{align}
An explicit calculation then yields the value of $R$ used
throughout this paper, $R \simeq 0.41$.

%%%%%%%%%%%%%%%%%%%%%%%%%%%%%%%%%%%%%%%%%%%%%%%%%%%%%%%%%%%%%%%%%%%%%%%%%%%%%%

\medskip\noindent\textbf{(3)}
Besides $R$, we also introduce the correction factor
$C_{\textrm{RH}} = \big(a_{\textrm{PH}} H_{\textrm{PH}}^{2/3}\big) /
\big(a_{\textrm{RH}} H_{\textrm{RH}}^{2/3}\big)$ in Sec.~\ref{sec:inflation}
in order to account for the change in the equation of state between
$a = a_{\textrm{PH}}$ and $a = a_{\textrm{RH}}$.
The Hubble rate at the end of preheating, $H_{\textrm{PH}}$, is controlled by
the coupling constant of the \BmL Higgs particles, $H_{\textrm{PH}} \propto \sqrt{\lambda}$
(cf.\ Eq.~\eqref{eq_HPH}).
In the context of our Froggatt-Nielsen model, we estimate that $\lambda$
scales like $\left(M_1 / v_{B-L}\right)^2$, such that
$H_{\textrm{PH}} \propto M_1$.
Furthermore, based on our semi-analytical determination of the scale factor
we find that%
\footnote{In our Froggatt-Nielsen flavour framework, the \BmL Higgs decay rate
$\Gamma_S^0$ also scales like $M_1^{\,3}$, such that it is directly
proportional to $H_{\textrm{RH}}$.
Calculating $H_{\textrm{RH}}$ explicitly shows that $H_{\textrm{RH}} \simeq 0.58\, \Gamma_S^0$.}
\begin{align}
\frac{a_{\textrm{RH}}}{a_{\textrm{PH}}} \propto M_1^{-4/3} \,,\quad
H_{\textrm{RH}} \propto M_1^{\, 3} \,,
\end{align}
which results in $C_{\textrm{RH}}$ being also a constant independent of all
model parameters.
An explicit calculation leads to 
$C_{\textrm{RH}} \simeq 1.13$.

%%%%%%%%%%%%%%%%%%%%%%%%%%%%%%%%%%%%%%%%%%%%%%%%%%%%%%%%%%%%%%%%%%%%%%%%%%%%%%

\medskip\noindent\textbf{(4)}
Finally, we define the would-be reheating temperature $\widetilde{T}_{\textrm{RH}}$
in Sec.~\ref{sec:inflation}, allowing us to precisely describe the relation between
the wavenumber $k_{\textrm{RH}}$ and the temperature of the bath during reheating
(Eq.~\eqref{eq_Delta} and \eqref{eq:defTwb}).
Again neglecting any change in the number of relativistic DOFs during and after reheating,
we are able to write $\widetilde{T}_{\textrm{RH}}$ as
\begin{align}
\widetilde{T}_{\textrm{RH}} = \frac{1}{a_{\textrm{RH}}} \left(aT\right)_{a \gg a_{\textrm{RH}}} \,,\quad
\left(aT\right)_{a \gg a_{\textrm{RH}}} =
\left(\frac{30 \, a^4 \rho_{\textrm{rel}}}{\pi^2 g_*}\right)_{a \gg a_{\textrm{RH}}}^{1/4} \,.
\end{align}
Evaluated at late times, $(a/a_{\textrm{PH}})^4 \rho_{\textrm{rel}}$ scales like $M_1^{2/3}$.
Together with $a_{\textrm{RH}}/a_{\textrm{PH}} \propto M_1^{-4/3}$, we thus find that
$\widetilde{T}_{\textrm{RH}} \propto M_1^{3/2}$.
An explicit computation of $\widetilde{T}_{\textrm{RH}}$ then results in Eq.~\eqref{eq:Twbres}.

%%%%%%%%%%%%%%%%%%%%%%%%%%%%%%%%%%%%%%%%%%%%%%%%%%%%%%%%%%%%%%%%%%%%%%%%%%%%%%
\section{The GW Background from Nambu-Goto Strings}\label{app:GWBNG}

In this appendix we explain in some detail the approximations adopted
in evaluating the GW background in Sec.~\ref{sec:comparison}. According to Eq.~\eqref{eq:theintegral}, the spectrum is given by
\begin{align}
\Omega_{\rm GW}(f) = \frac{2 \pi^2 f^3}{3 H_0^2} 
\int_{0}^{h_*} dh \int_0^{z_{\rm PH}} dz  h^2 \frac{d^2R}{dz dh}\ , \nonumber
\end{align}
where the differential rate $d^2R/dz dh$ is defined in 
Eqs.~\eqref{eq:d2rdzdh} to \eqref{eq:hmax}. 
The GW background from NG strings can be obtained by calculating this
integral numerically. However, this does not shed much light on the underlying paramater dependencies.
Here, we clarify these by adopting some approximations for the integrand, which
allow us to evaluate the integral in Eq.~\eqref{eq:theintegral} analytically,
leading to the result given in Eq.~\eqref{eq:theresult}. 

The differential rate  $d^2R/dz dh$ depends on the Hubble parameter
$H(z)$ and the functions $\varphi_r(z)$ and $\varphi_t(z)$, i.e.\
distance to and cosmic time of the GW emission as functions of
redshift (cf. Eqs.~\eqref{eq_phi_r} and \eqref{eq_phi_t}), for which we use the approximations
\begin{align}
\frac{H(z)}{H_0}
&\simeq 
\begin{cases}
1 \ ,  & 0\phantom{^\Lambda}<z<z_\Lambda  \\
\Omega_{m}^{1/2} (1+z)^{3/2} , & z_\Lambda<z<z_{\rm eq} \\
\Omega_{r}^{1/2} (1+z)^2, & z_{\rm eq}<z<z_{\rm RH}  \\
\Omega_{r}^{1/2} (1+z_{\rm RH})^{1/2} (1+z)^{3/2}, \qquad & z_\text{RH} < z < z_\text{PH}
 \label{approx1}
\end{cases} \ ,\\
\rule{0pt}{11mm} \varphi_r(z)&\simeq \left\{
\begin{array}{ll}
\varphi_0 z \ , \hspace{2cm} & 0<z\leq 1 \  \\
\varphi_0 \ , & z>1 \  \label{approx2}
\end{array}\right. \ , \\
\rule{0pt}{15mm} \varphi_t(z)&\simeq
\begin{cases}
 \Omega_{m}^{-1/2}(1+z)^{-3/2}, & 0\phantom{_{\text{eq}}}<z<z_{\rm eq}\  \\
\Omega_{r}^{-1/2} (1+z)^{-2}, & z_{\rm eq}<z<z_{\rm RH} \  \\
 \Omega_{r}^{-1/2} (1+z_{\rm RH})^{-1/2} (1+z)^{-3/2}, \quad & z_\text{RH}<z<z_{\rm
   PH} \  \label{approx3} 
\end{cases} \ .
\end{align}
Here, $z_{\Lambda}$ denotes the boundary between matter and
$\Lambda$-domination,
\begin{align} 
z_\Lambda = \Omega_{m}^{-1/3}-1 \simeq 0.5 \ ,
\end{align}
and the constant $\varphi_0$ reads
 \begin{align}\label{eq:phi0}
\varphi_0 =
z_\Lambda+2 \, \Omega_{m}^{-1/2} (1+z_\Lambda)^{-1/2}  \ .
\end{align}
Note that the above approximations are rather rough around $z \sim z_{\Lambda}$,
a region not very important for our final result. 

With the approximations in Eqs.~\eqref{approx1} to \eqref{approx3}, the differential rate $hd^2R/dhdz$ is given by
\begin{align} \label{eq:appdrdhdz}
h \frac{d^2 R}{dz dh} &\simeq 
\frac{3 \pi H_0^2}{\gamma \alpha \Gamma hf^2} \Theta(1-\theta_m) \times 
\begin{cases}
\Omega_{m}^2  \varphi_0 z (1+z)\ , &  z<1    \\ 
\Omega_{m}^{3/2} \varphi_0 (1+z)^{-1/2} , \quad & 1 <z<z_{\rm eq}  \\ 
\Omega_{r}^{3/2} \varphi_0 (1+z) \ , & z_{\rm eq}<z<z_{\rm RH}    \\
\Omega_{r}^{3/2} \varphi_0 \dfrac{(1+z_{\rm RH})^{3/2}}{(1+z)^{1/2}}\  & z>z_{\rm RH} 
\end{cases} \ .
\end{align}
We can now perform the $z$-integration. The step function implies an
upper bound $z_m$ on the redshift, which dominates the integral,
\begin{align}
\frac{dR}{d \ln h} &\simeq \int^{z_m} dz h \frac{d^2R}{dzdh} \ .
\end{align}
 The upper boundary $z_m$ is a different function of $f$ and $h$ for each
 epoch and can be determined using Eqs.~\eqref{angle} to \eqref{amplitude}.
With this, we find for $dR/d\ln h$ the expressions
\begin{align}
\frac{dR}{d \ln h} \simeq \zeta f \left(\frac{H_0}{\alpha f}\right)^{5/3} \times \left\{
\begin{array}{ll}
\dfrac{1}{2} \Omega_{m}^{4/3} \left(\dfrac{h_0}{h}\right)^3 , & h_1 < h \\ 
2 \Omega_{m}^{11/8}  \left(\dfrac{h_0}{h}\right)^{5/3} , & h_2<h<h_1 \\ 
\dfrac{1}{2}\Omega_{r}^{11/10} \left(\dfrac{h_0}{h}\right)^{11/5}
 ,  &  h_3<h<h_2 \\ 
2 \Omega_{r}^{11/8} (1+z_{\rm RH})^{11/8}
\left(\dfrac{h_0}{h}\right)^{11/8}  ,  \quad &  h\phantom{_3}<h_3  
\end{array}\right. \label{eq:drdlnh} \ ,
\end{align}
where we have introduced
\begin{align}
\zeta = \left(\frac{3\pi^2 \varphi_0^2}{\gamma^2\Gamma G\mu}\right)^{3/5} \ .
\end{align}
The boundaries $h_1, h_2$ and $h_3$ separate the different
epochs, i.e.\ $\Lambda$-domination, matter domination, radiation domination
and reheating, respectively, with
\begin{equation}
h_1= \Omega_{m}^{-1/3} h_0\ , \quad 
h_2=\Omega_{m}^{-1/3} (1+z_{\rm eq})^{-4/3} h_0\ , \quad 
h_3= \Omega_{r}^{-1/3} (1+z_{\rm RH})^{-5/3} h_0\ , 
\end{equation}
where
\begin{equation}
h_0 = \frac{\alpha^{2/3} G \mu H_0^{1/3}}{\varphi_0 f^{4/3}}\ . 
\end{equation}

The GW background spectrum is obtained by finally integrating $h\,  dR/d \hspace{-1pt} \ln \hspace{-1pt} h$ over $h$.  For given $f$, there is an upper bound on the range of
integration, $h < h_*$, which follows from the requirement of including
only GW bursts which arrive within time intervals shorter than their
`oscillation period' \cite{Siemens:2006yp}. The step function in Eq.~\eqref{eq:appdrdhdz} implies the condition
$\theta_m(f,h,z_m)=((1+z_m)\, f \, l(f,h,z_m))^{-1/3}<1$, 
yielding a lower bound $h_c$ on the integration range,
which enforces the infrared cutoff  $l$ on the wavelength of the GWs
emitted at redshift $z_m$. Taking these boundaries into account, one can
perform the $h$-integration for the different epochs, which yields the 
result given in Eq.~\eqref{eq:theresult}:
\begin{equation}
\frac{\Omega_{\rm GW}(f)}{\Omega_{\rm GW}^{\rm pl}} \simeq  
\begin{cases}
\dfrac{\gamma_r^2}{\gamma_{\rm
    m}^2}\left(\dfrac{1}{5}\dfrac{\Omega_{m}^{5/3}}{\Omega_{\rm
      r}} +\dfrac{32}{25}\dfrac{\Omega_{m}^{7/6}}{\Omega_{\rm
      r}}\right)\left(\dfrac{H_0}{\alpha f}\right)^{1/3}, & \hspace{3mm}f_c\phantom{^\text{(NG)}}\hspace{-3mm} < f < f_1\\ 
\dfrac{32}{25} \left(\dfrac{6\pi \varphi_0^{2}}{\gamma_m^2\Gamma G
    \mu}
\right)^{5/11}\dfrac{\gamma_r^2}{\gamma_m^2}\dfrac{\Omega_{\rm
    m}^2}{\Omega_{r} }\left(\dfrac{H_0}{\alpha f}\right)^{12/11} ,
& \hspace{3mm}f_1\phantom{^\text{(NG)}}\hspace{-3mm} < f < f_2  \\
1 \ ,& f_{\rm eq}^{\rm (NG)} < f <  f_{\rm RH}^{\rm (NG)} \\ 
\left(1 + \dfrac{32}{25}\dfrac{\gamma_r^2}{\gamma_m^2}\right)\Omega_{\rm
  r}^{1/6}(1+z_{\rm RH})^{1/3} \left(\dfrac{H_0}{\alpha f}\right)^{1/3}, & 
f_{\rm RH}^{\rm (NG)} < f < f_3 \\ 
\dfrac{32}{25 } \left(\dfrac{6\pi \varphi_0^{2}}{\gamma_m^2\Gamma G
    \mu} \right)^{5/11}\dfrac{\gamma_r^2}{\gamma_m^2}
\dfrac{\Omega_{m}^2}{\Omega_{r}}(1+z_{\rm RH})^2
\left(\dfrac{H_0}{\alpha f}\right)^{12/11}, & \hspace{3mm}f_3\phantom{^\text{(NG)}}\hspace{-3mm} < f \ 
\end{cases} \nonumber
\end{equation}
with the height of the plateau given by
\begin{equation}
\Omega_{\rm GW}^{\rm pl}h^2 = 
\frac{5\pi^3\kappa^2 G\mu}{\Gamma\gamma_r^2}\Omega_r h^2 \ . \nonumber
\end{equation}

For $f_c < f < f_1$, the GW background is sourced by the vacuum and matter domination phases and is dominated by the region $h\sim h_1$, 
corresponding to GW emission around $z\sim 1$. In the frequency range 
$f_1 < f < f_2$, the GW background originates from the matter
dominated phase and is dominated by amplitudes close to the upper boundary in Eq.~\eqref{eq:theintegral},
\begin{equation}
h\sim h_*=\left(\frac{6 \pi}{\Gamma \gamma_m^2}\right)^{8/11}
\Omega_{m} \, \varphi_0^{5/11}
\left(\frac{G\mu}{\alpha^2}\right)^{3/11}
\frac{H_0^{17/11}}{f^{28/11}} < h_1 \ .
\end{equation}
The plateau for frequencies $f_{\rm eq}^\text{(NG)} < f < f_{\rm RH}^\text{(NG)}$ is produced
by GW bursts during the radiation dominated phase and is dominated by amplitudes 
\begin{equation}
h\sim h_c=\Omega_{r}^{1/2}\left(\frac{H_0}{\alpha f}\right)^{5/3} h_0 < h_2 \ . 
\end{equation} 
This means that the plateau is the envelope curve of GW bursts emitted
near $z_m$ in the radiation domination phase. 
The higher-frequency part, $f_{\rm RH} < f < f_3$, is due to the
radiation dominated phase and the first matter dominated phase and is dominated by  amplitudes around $h\sim h_3$, corresponding to GWs emitted near $z\simeq z_{\rm RH}$. 
The highest-frequency part, $f > f_3$, is the contribution from the
first matter dominated phase with amplitudes
\begin{equation}
h\sim h_*=\left(\frac{6 \pi}{\Gamma \gamma_m^2}\right)^{8/11} \Omega_{m} \, \varphi_0^{5/11} \left(\frac{G\mu}{\alpha^2}\right)^{3/11} 
(1+z_{\rm RH}) \frac{H_0^{17/11}}{f^{28/11}} <h_3 \ .
\end{equation} 

The approximate expression \eqref{eq:theresult} for the relic GW spectrum
is roughly consistent with the numerical evaluation in 
Ref.~\cite{Kuroyanagi:2012wm}. Additionally, it contains the effect of the
first matter dominated phase on the relic GW spectrum, which has
previously not been considered.

\section{Characteristic Temperatures during Reheating}
\label{app:temperatures}

%%%%%%%%%%%%%%%%%%%%%%%%%%%%%%%%%%%%%%%%%%%%%%%%%%%%%%%%%%%%%%%%%%%%%%%%%%%%%%

In Sec.~\ref{sec:reheatingtemp}, we introduced the neutrino
decay temperature $T_N$, the \BmL Higgs decay temperature $T_S$ as well as the
radiation domination temperature $T_R$ and discussed some of their properties.
In this appendix, we shall now provide the formal definitions of $T_N$, $T_S$ and $T_R$
and list a couple of further interesting results concerning these temperatures
which we are able to deduce from our numerical investigation of the reheating
process by means of Boltzmann equations.

%%%%%%%%%%%%%%%%%%%%%%%%%%%%%%%%%%%%%%%%%%%%%%%%%%%%%%%%%%%%%%%%%%%%%%%%%%%%%%

\medskip\noindent\textbf{Neutrino decay temperature:}
The neutrino decay temperature $T_N$ is defined such that at $T=T_N$ the Hubble
rate $H$ has just dropped to the value of the effective (s)neutrino decay rate
$\Gamma_{N_1}^S$,
\begin{align}
H\left(a_N\right) = \Gamma_{N_1}^S\left(a_N\right) \,,\quad T_N = T\left(a_N\right) \,.
\label{eq:TNdef}
\end{align}
As mentioned in Sec.~\ref{sec:model}, the reheating process is accompanied
by a characteristic plateau in the evolution of the radiation temperature
(cf.\ Fig.~\ref{fig:reheat}).
For $\Gamma_{N_1}^S \gg \Gamma_S^0$, the neutrino decay temperature
reflects best the typical temperature scale of this plateau (cf.\ Fig.~\ref{fig:Trhocompare})).
In our earlier works, we were particularly interested in exactly the
parameter region in which $\Gamma_{N_1}^S \gg \Gamma_S^0$, which is why
we referred to $T_N$ as the reheating temperature in these studies.

%%%%%%%%%%%%%%%%%%%%%%%%%%%%%%%%%%%%%%%%%%%%%%%%%%%%%%%%%%%%%%%%%%%%%%%%%%%%%%

\medskip\noindent\textbf{{\boldmath$B$$-$$L$} Higgs decay temperature:}
The equivalent of the inflaton decay temperature $T_\phi$ usually employed
in scenarios of reheating through inflaton decay (cf.\ Eq.~\eqref{eq:Tphidef})
for our reheating scenario is the \BmL Higgs decay temperature $T_S$, which is
defined as the temperature at the time when the Hubble rate $H$ equals the
\BmL Higgs decay rate $\Gamma_S^0$,
\begin{align}
H\left(a_S\right) = \Gamma_S^0\left(a_S\right) \,,\quad T_S = T\left(a_S\right) \,.
\label{eq:TSdef}
\end{align}

%%%%%%%%%%%%%%%%%%%%%%%%%%%%%%%%%%%%%%%%%%%%%%%%%%%%%%%%%%%%%%%%%%%%%%%%%%%%%%

\medskip\noindent\textbf{Radiation domination temperature:}
Due to the specifics of our reheating scenario, the transition from the cosmic
expansion being driven by non-relativistic DOFs to the cosmic expansion being
driven by relativistic DOFs does not coincide with the transition from
a phase dominated by nonthermal DOFs to a phase dominated by thermal DOFs.
While the former transition occurs at $a = a_{\textrm{RH}}$, the latter
takes place a bit later, at $a = a_R$, when enough nonthermal
(s)neutrinos have decayed into thermal MSSM radiation such that
half of the total energy is contained in the thermal plasma,
\begin{align}
\rho_{\textrm{tot}} \left(a_R\right) = 
2 \, \rho_r \left(a_R\right) =
2 \left[\rho_S + \rho_N + \rho_{\tilde{N}}  \right]_{a_R} \,,\quad
T_R = T\left(a_R\right) \,.
\label{eq:TRdef}
\end{align}
The temperature at this time defines what we shall refer to as the
radiation domination temperature $T_R$.
Among all physical temperatures introduced so far, i.e.\ $T_N$, $T_S$, $T_{\textrm{RH}}$
and $T_R$, the radiation domination temperature is always the smallest.
Using $T_R$ as `the reheating temperature' is particularly well motivated,
as, after all, it represents the highest temperature ever reached during
the thermal stage of the hot early universe.

%%%%%%%%%%%%%%%%%%%%%%%%%%%%%%%%%%%%%%%%%%%%%%%%%%%%%%%%%%%%%%%%%%%%%%%%%%%%%%

\medskip

Besides the observations already mentioned in Sec.~\ref{sec:reheatingtemp},
our numerical study of $T_N$, $T_S$, $T_{\textrm{RH}}$ and $T_R$
provides us with further interesting results:

%%%%%%%%%%%%%%%%%%%%%%%%%%%%%%%%%%%%%%%%%%%%%%%%%%%%%%%%%%%%%%%%%%%%%%%%%%%%%%

\medskip\noindent\textbf{(1)}
Up to some constant correction factor, the radiation domination temperature $T_R$ corresponds
to the minimum of the two decay temperatures $T_S$ and $T_N$,
\begin{align}
T_R \simeq 0.85 \times \min\left\{T_S, T_N\right\} \,.
\label{eq:TRrel}
\end{align}
This is, of course, expected since the radiation energy density $\rho_r$ can only
account for half of the total energy density $\rho_{\textrm{tot}}$ once both species,
the \BmL Higgs particles \textit{and} the heavy (s)neutrinos, have sufficiently decayed.
Making use of Eqs.~\eqref{eq:TNdef} and \eqref{eq:TSdef},
the relation in Eq.~\eqref{eq:TRrel} can be translated into a statement about
the Hubble rate at $a = a_R$, i.e.\ at the beginning of the thermal phase of the hot early universe,
\begin{align}
H\left(a_R\right) \simeq 0.72 \left(\frac{\alpha_R}{\min\{\alpha_S,\alpha_N\}}\right)^{1/2}
\min\left\{\Gamma_S^0,\Gamma_{N_1}^S\right\} \,.
\end{align}
Here, the factors $\alpha_S$, $\alpha_N$ and $\alpha_R$ are defined analogously to
the correction factor $\alpha_{\textrm{RH}}$,
\begin{align}
\alpha_i = \left.\frac{\rho_{\textrm{tot}}}{\rho_r}\right|_{a_i} \,,
\quad i = N,\,S,\,\textrm{RH},\,R\,.
\end{align}

%%%%%%%%%%%%%%%%%%%%%%%%%%%%%%%%%%%%%%%%%%%%%%%%%%%%%%%%%%%%%%%%%%%%%%%%%%%%%%

\medskip\noindent\textbf{(2)}
$\alpha_R$ is, by definition, given as $\alpha_R = 2$ (cf.\ Eq.~\eqref{eq:TRdef}).
$\alpha_N$ and $\alpha_S$ can be deduced from the following relations
in combination with our numerical results for $T_N$ and $T_S$,
\begin{align}
T_N = \left(\frac{90}{\alpha_N \pi^2 g_{*,\rho}^{\textrm{RH}}}\right)^{1/4}
\sqrt{\Gamma_{N_1}^S M_{\textrm{Pl}}} \,, \quad
T_S = \left(\frac{90}{\alpha_S \pi^2 g_{*,\rho}^{\textrm{RH}}}\right)^{1/4}
\sqrt{\Gamma_S^0 M_{\textrm{Pl}}} \,.
\label{eq:TNSalphaNS}
\end{align}
According to Eqs.~\eqref{eq:defTwb} and \eqref{eq:defR}, the factor
$\alpha_{\textrm{RH}}$ can meanwhile be obtained from the ratio of the would-be
reheating temperature and the actual reheating temperature,
\begin{align}
\alpha_{\textrm{RH}} = 2 \, R \, D^{4/3} = 2 \, R
\left(\frac{\widetilde{T}_{\textrm{RH}}}{T_{\textrm{RH}}}\right)^4 \,, \quad
D = \left(\frac{\widetilde{T}_{\textrm{RH}}}{T_{\textrm{RH}}}\right)^3 =
\left(\frac{\alpha_{\textrm{RH}}}{2\,R}\right)^{3/4} \,, \quad
R \simeq 0.41 \,.
\label{eq:alphaRHdelta}
\end{align}
With $\alpha_{\textrm{RH}}$ at hand, we are then immediately able to determine
the dilution factor $D$.
Solving the relations in Eq.~\eqref{eq:TNSalphaNS} for
$\alpha_N$ and $\alpha_S$ as well as computing $\alpha_{\textrm{RH}}$ and $D$
according to Eq.~\eqref{eq:alphaRHdelta}, we find that all four factors
can be brought into a similar form as the four temperatures in Eq.~\eqref{eq:Ti},
\begin{align}
\alpha_i \simeq \max\left\{\alpha_i^{(-)},\alpha_i^{(+)}\right\} \simeq
\begin{cases}
\alpha_i^{(-)} \,; & \:\: y \gtrsim e_i \:\Leftrightarrow\: y_i \gtrsim f_i \\
\alpha_i^{(+)} \,; & \:\: y \lesssim e_i \:\Leftrightarrow\: y_i \lesssim f_i
\end{cases} \,, \quad
i = N,\, S,\, \textrm{RH},\, D,
\label{eq:alphai}
\end{align}
where $\alpha_D \equiv D$ and
with the factors $\alpha_i^{(-)}$ and $\alpha_i^{(+)}$ being again simple power laws,
\begin{align}
\alpha_i^{(-)} \simeq & \:\: \alpha_{i,0}^{(-)} \left(\frac{M_1}{10^{11}\,\textrm{GeV}}\right)^{r_i^{(-)}}
\left(\frac{\widetilde{m}_1}{10^{-2}\,\textrm{eV}}\right)^{s_i^{(-)}} \,, \\
\alpha_i^{(+)} \simeq & \:\: \alpha_{i,0}^{(+)} \left(\frac{M_1}{10^{11}\,\textrm{GeV}}\right)^{r_i^{(+)}}
\left(\frac{\widetilde{m}_1}{10^{-4}\,\textrm{eV}}\right)^{s_i^{(+)}} \,.
\end{align}
Our numerical results for the values of $e_i$, $f_i$, $\alpha_i^{(\pm)}$, $r_i^{(\pm)}$
and $s_i^{(\pm)}$ are listed in Tab.~\ref{tab:alphafit}.

%%%%%%%%%%%%%%%%%%%%%%%%%%%%%%%%%%%%%%%%%%%%%%%%%%%%%%%%%%%%%%%%%%%%%%%%%%%%%%

\begin{table}
\begin{center}
\begin{tabular}{c||ccc|ccc|cc}
$\alpha_i$ & $\alpha_{i,0}^{(-)}$  & $r_i^{(-)}$ & $s_i^{(-)}$ &
$\alpha_{i,0}^{(+)}$ & $r_i^{(+)}$ & $s_i^{(+)}$ &
$e_i$ & $f_i$ \\\hline\hline
$\alpha_N$ & $280$ & $-1.0$ & $1.0$ & $5.6$ & $-0.1$ & $0.1$ & $50$ & $0.64$\\
$\alpha_S$ & $3.4$ & $0.0$ & $0.0$ & $37$ & $0.85$ & $-0.85$ & $300$ & $3.0$ \\
$\alpha_{\textrm{RH}}$ & $2.1$ & $0.0$ & $0.0$ & $14$ & $0.9$ & $-0.9$ & $180$ & $1.8$\\
$D$ & $2.0$ & $0.0$ & $0.0$ & $8.5$ & $0.65$ & $-0.65$ & $180$ & $1.8$
\end{tabular}
\caption{Numerical values to be used in the fit formulas for
$\alpha_N$, $\alpha_S$, $\alpha_{\textrm{RH}}$ and $\alpha_D \equiv D$
in Eq.~\eqref{eq:alphai}.}
\label{tab:alphafit}
\end{center}
\end{table}

%%%%%%%%%%%%%%%%%%%%%%%%%%%%%%%%%%%%%%%%%%%%%%%%%%%%%%%%%%%%%%%%%%%%%%%%%%%%%%

\medskip\noindent\textbf{(3)}
The fact that $T_S^{(-)}$ is proportional to $M_1^{3/2}$ can be easily understood
analytically.
In the framework of our Froggatt-Nielsen model, the \BmL Higgs decay rate
$\Gamma_S^0$ scales like $M_1^3$ (cf.\ Eq.~\eqref{eq:GammaSN1}).
Making use of the relation between $T_S$ and $\Gamma_S^0$ in Eq.~\eqref{eq:TNSalphaNS},
one then finds that $T_S$ exhibits exactly the same parameter dependence
as the square root of $\Gamma_S^0$ as long as $\alpha_S$ is a constant.
For fast decaying heavy (s)neutrinos, i.e.\ in the parameter region where
$T_S^{(-)} < T_S^{(+)}$, this is just the case, resulting in $T_S \propto M_1^{3/2}$.
Furthermore, since $T_{\textrm{RH}}$ is always only slightly smaller than $T_S$
and since $T_R \propto \min\left\{T_S,T_N\right\}$, this then explains
why $T_{\textrm{RH}}^{(-)}$ and $T_R^{(-)}$ are proportional to $M_1^{3/2}$ as well.
The scaling behaviour of $T_N^{(-)}$ is exactly the one already determined
in Ref.~\cite{Buchmuller:2012wn}.
Similarly, our result for $\alpha_N^{(-)}$ has already been derived in Ref.~\cite{Buchmuller:2012wn}.
In this sense, all other numerical results presented in this section may be regarded
as an extension of our previous findings.

%%%%%%%%%%%%%%%%%%%%%%%%%%%%%%%%%%%%%%%%%%%%%%%%%%%%%%%%%%%%%%%%%%%%%%%%%%%%%%

\end{document}